\documentclass[a4paper,11pt]{article}
\pdfoutput=1 

\usepackage{jheppub} 
\usepackage[T1]{fontenc}
\usepackage[utf8]{inputenc}

\usepackage{amssymb,amsfonts,amsmath,amsthm,bm,mathrsfs}
\usepackage{tikz}

\usepackage{pgfplots}
\usepackage{pdfpages}
\usepgfplotslibrary{dateplot}

\usetikzlibrary{snakes}
\usetikzlibrary{calc}
\usetikzlibrary{decorations}

\usepackage[T1]{fontenc} 
\usepackage{paralist}
\usepackage{soul}

\newcommand{\dif}{\mathrm{d}} 
\newcommand{\Li}{\operatorname{Li}}
\newcommand{\Res}{\operatorname{Res}}

\def\<{\langle}
\def\>{\rangle}

\renewcommand{\tilde}{\widetilde}

\definecolor{hblue}{rgb}{0,0,0.575}
\definecolor{hred}{rgb}{0.575,0.0,0.225}
\definecolor{hteal}{rgb}{0.0,0.545,0.7451}
\definecolor{optLegColour}{rgb}{0.5,0.5,0.5}

\def\figScale{0.835}\def\edgeLen{1*\figScale}\def\fScale{\small}\def\legLen{\edgeLen*0.65}\def\labelDist{\legLen*1.45}\def\lineThickness{(1pt)}\def\dotSize{(\figScale*10)}\def\ampSize{(1*\figScale*10pt)}
\tikzset{ddot/.style={fill=black,circle,minimum size=0.35*\dotSize,inner sep=0}}
\tikzset{int/.style={black,line width=\lineThickness,line cap=round,rounded corners=0.5pt}}\tikzset{ext/.style={black,line width=\lineThickness,line cap=round}}\tikzset{bdot/.style={fill=black,circle,minimum size=0.45*\ampSize,inner sep=0}}\tikzset{wdot/.style={draw=black,line width=\lineThickness,fill=white,circle,minimum size=0.50*\ampSize,inner sep=0}}
\newcommand{\leg}[3]{\draw[ext] #1--($#1+(#2:\legLen)$);\node at ($#1+(#2:\labelDist)$)[]{{\fScale #3}};}

\newcommand{\nonijkl}{\begin{tikzpicture}[scale=\figScale,baseline=-2.45]\useasboundingbox ($(-7.9/3*\figScale,-1.8)$) rectangle ($(7.9/3*\figScale,1.8)$);\draw[int,line width=0.1,red,draw=none] ($(-7.5/3*\figScale,-1.5)$) rectangle ($(7.5/3*\figScale,1.5)$);\coordinate(v1) at ($(0,0)+(90:\edgeLen/2)$);\coordinate(v2)at($(v1)+(18:\edgeLen)$);\coordinate(v3)at($(v2)+(18-72:\edgeLen)$);\coordinate(v4)at($(v3)+(18-2*72:\edgeLen)$);\coordinate(v5)at($(v4)+(18-3*72:\edgeLen)$);\coordinate(v6)at($(v5)+(198-0*72:\edgeLen)$);\coordinate(v7)at($(v6)+(198-1*72:\edgeLen)$);\coordinate(v8)at($(v7)+(198-2*72:\edgeLen)$);
\draw[int](v1)--(v2)--(v3)--(v4)--(v5)--(v6)--(v7)--(v8)--(v1);\draw[int](v1)--(v5);\leg{(v2)}{72}{$k$};
\leg{(v1)}{60}{};\leg{(v1)}{120}{};
\leg{(v2)}{72}{};
\leg{(v5)}{-60}{};
\leg{(v5)}{-120}{};
\leg{(v3)}{40}{\!};
\leg{(v3)}{-40}{\!};
\leg{(v7)}{140}{\!};
\leg{(v7)}{-140}{\!};
\node[fill=black,circle,minimum size=0.15*\dotSize,inner sep=0] at (-1.65,0) {};
\node[fill=black,circle,minimum size=0.15*\dotSize,inner sep=0] at (-1.6,0.13) {};
\node[fill=black,circle,minimum size=0.15*\dotSize,inner sep=0] at (-1.6,-0.13) {};
\node[fill=black,circle,minimum size=0.15*\dotSize,inner sep=0] at (1.65,0) {};
\node[fill=black,circle,minimum size=0.15*\dotSize,inner sep=0] at (1.6,0.13) {};
\node[fill=black,circle,minimum size=0.15*\dotSize,inner sep=0] at (1.6,-0.13) {};
\node[fill=black,circle,minimum size=0.15*\dotSize,inner sep=0] at (0,0.85) {};
\node[fill=black,circle,minimum size=0.15*\dotSize,inner sep=0] at (0.1,0.80) {};
\node[fill=black,circle,minimum size=0.15*\dotSize,inner sep=0] at (-0.1,0.80) {};
\node[fill=black,circle,minimum size=0.15*\dotSize,inner sep=0] at (0,-0.85) {};
\node[fill=black,circle,minimum size=0.15*\dotSize,inner sep=0] at (0.1,-0.80) {};
\node[fill=black,circle,minimum size=0.15*\dotSize,inner sep=0] at (-0.1,-0.80) {};
\draw[decorate, decoration=snake, segment length=6pt,segment amplitude=1pt,black,line width=\lineThickness] (v8) to[out=-60,in=60] (v6);
\draw[decorate, decoration=snake, segment length=6pt,segment amplitude=1pt,black,line width=\lineThickness] (v2)to[out=-120,in=120](v4);
\leg{(v4)}{72-2*72}{$l$};\leg{(v6)}{252-0*72}{$i$};\leg{(v8)}{252-2*72}{$j$};\foreach\a in {1,3,5,7}{\node at (v\a) [bdot]{};};\foreach\a in {2,4,6,8}{\node at (v\a) [wdot]{};};
\end{tikzpicture}}
\newcommand{\nonjkli}{\begin{tikzpicture}[scale=\figScale,baseline=-2.45]\useasboundingbox ($(-7.9/3*\figScale,-1.8)$) rectangle ($(7.9/3*\figScale,1.8)$);\draw[int,line width=0.1,red,draw=none] ($(-7.5/3*\figScale,-1.5)$) rectangle ($(7.5/3*\figScale,1.5)$);\coordinate(v1) at ($(0,0)+(90:\edgeLen/2)$);\coordinate(v2)at($(v1)+(18:\edgeLen)$);\coordinate(v3)at($(v2)+(18-72:\edgeLen)$);\coordinate(v4)at($(v3)+(18-2*72:\edgeLen)$);\coordinate(v5)at($(v4)+(18-3*72:\edgeLen)$);\coordinate(v6)at($(v5)+(198-0*72:\edgeLen)$);\coordinate(v7)at($(v6)+(198-1*72:\edgeLen)$);\coordinate(v8)at($(v7)+(198-2*72:\edgeLen)$);
\draw[int](v1)--(v2)--(v3)--(v4)--(v5)--(v6)--(v7)--(v8)--(v1);\draw[int](v1)--(v5);\leg{(v1)}{60}{};\leg{(v1)}{120}{};
\node[fill=black,circle,minimum size=0.15*\dotSize,inner sep=0] at (-1.65,0) {};
\node[fill=black,circle,minimum size=0.15*\dotSize,inner sep=0] at (-1.6,0.13) {};
\node[fill=black,circle,minimum size=0.15*\dotSize,inner sep=0] at (-1.6,-0.13) {};
\node[fill=black,circle,minimum size=0.15*\dotSize,inner sep=0] at (1.65,0) {};
\node[fill=black,circle,minimum size=0.15*\dotSize,inner sep=0] at (1.6,0.13) {};
\node[fill=black,circle,minimum size=0.15*\dotSize,inner sep=0] at (1.6,-0.13) {};
\node[fill=black,circle,minimum size=0.15*\dotSize,inner sep=0] at (0,0.85) {};
\node[fill=black,circle,minimum size=0.15*\dotSize,inner sep=0] at (0.1,0.80) {};
\node[fill=black,circle,minimum size=0.15*\dotSize,inner sep=0] at (-0.1,0.80) {};
\node[fill=black,circle,minimum size=0.15*\dotSize,inner sep=0] at (0,-0.85) {};
\node[fill=black,circle,minimum size=0.15*\dotSize,inner sep=0] at (0.1,-0.80) {};
\node[fill=black,circle,minimum size=0.15*\dotSize,inner sep=0] at (-0.1,-0.80) {};
\leg{(v2)}{72}{$j$};
\leg{(v5)}{-60}{};
\leg{(v5)}{-120}{};
\leg{(v3)}{40}{\!};
\leg{(v3)}{-40}{\!};
\leg{(v7)}{140}{\!};
\leg{(v7)}{-140}{\!};
\draw[decorate, decoration=snake, segment length=6pt,segment amplitude=1pt,black,line width=\lineThickness] (v8) to[out=-60,in=60] (v6);
\draw[decorate, decoration=snake, segment length=6pt,segment amplitude=1pt,black,line width=\lineThickness] (v2)to[out=-120,in=120](v4);
\leg{(v4)}{72-2*72}{$k$};\leg{(v6)}{252-0*72}{$l$};\leg{(v8)}{252-2*72}{$i$};\foreach\a in {1,3,5,7}{\node at (v\a) [bdot]{};};\foreach\a in {2,4,6,8}{\node at (v\a) [wdot]{};};
\end{tikzpicture}}

\pgfplotsset{compat=1.16}

\title{\boldmath The symbol and alphabet of two-loop NMHV amplitudes from $\bar{Q}$ equations}

\author[a,b,c,d]{Song He}
\author[b,d]{Zhenjie Li}
\author[b,e]{Chi Zhang}

\affiliation[a]{School of Fundamental Physics and Mathematical Sciences, Hangzhou Institute for Advanced Study, UCAS, Hangzhou 310024, China}
\affiliation[b]{CAS Key Laboratory of Theoretical Physics, Institute of Theoretical Physics, Chinese Academy of Sciences, Beijing 100190, China}
\affiliation[c]{ICTP-AP
International Centre for Theoretical Physics Asia-Pacific, Beijing/Hangzhou, China}
\affiliation[d]{School of Physical Sciences, University of Chinese Academy of Sciences, No.19A Yuquan Road, Beijing 100049, China}
\affiliation[e]{Niels Bohr International Academy, Niels Bohr Institute, Copenhagen University, Blegdamsvej 17, 2100 Copenhagen \O{}, Denmark}
\emailAdd{songhe@itp.ac.cn}
\emailAdd{lizhenjie@itp.ac.cn}
\emailAdd{chi.zhang@nbi.ku.dk}

\abstract{We study the symbol and the alphabet for two-loop NMHV amplitudes in planar ${\cal N}=4$ super-Yang-Mills from the $\bar{Q}$ equations, which provide a first-principle method for computing multi-loop amplitudes. Starting from one-loop N${}^2$MHV ratio functions, we explain in detail how to use $\bar{Q}$ equations to obtain the total differential of two-loop $n$-point NMHV amplitudes, whose symbol contains letters that are algebraic functions of kinematics for $n\geq 8$. We present explicit formula with nice patterns for the part of the symbol involving algebraic letters for all multiplicities, and we find $17-2m$ multiplicative-independent letters for a given square root of Gram determinant, with $0\leq m\leq 4$ depending on the number of particles involved in the square root. We also observe that these algebraic letters can be found as poles of one-loop four-mass leading singularities with MHV or NMHV trees. As a byproduct of our algebraic results, we find a large class of components of two-loop NMHV, which can be written as differences of two double-pentagon integrals, particularly simple and free of square roots. As an example, we present the complete symbol for $n=9$ whose alphabet contains $59\times 9$ rational letters, in addition to the $11 \times 9$ independent algebraic ones. We also give all-loop NMHV last-entry conditions for all multiplicities. }
\begin{document} 

\maketitle
\flushbottom

\section{Introduction}

Scattering amplitudes are central objects in fundamental physics: not only do they play a crucial role in bridging theory to high energy experiments such as Large Hadron Collider, but they also provide new insights into Quantum Field Theory (QFT) itself. Tremendous progress has been made in unravelling hidden mathematical structures of perturbative scattering amplitudes, especially in planar ${\cal N} = 4$ supersymmetric Yang-Mills theory (SYM) at the (all-loop) integrand level ({\it c.f.} \cite{ArkaniHamed:2010kv, ArkaniHamed:2012nw,Arkani-Hamed:2013jha}). Moreover, it has become an extremely fruitful playground for new methods of evaluating multi-loop Feynman integrals, which is a subject of enormous interests by itself ({\it c.f.} \cite{Bourjaily:2018aeq,Henn:2018cdp,Herrmann:2019upk} and references there in). Along these fascinating directions, we have discovered numerous new structures of the theory (and in many cases also for more general QFT), and it is conceivable that we will have more and more important understandings for both integrands and integrals. 

However, it is clear that there are major obstacles ahead when we go to sufficiently high loops and/or multiplicities. For example, we encounter transcendental functions that go beyond generalized polylogarithms~\cite{Bourjaily:2017bsb,Bourjaily:2018ycu}; even at two loops, it becomes more and more difficult to perform loop integration due to the rapid growth of complexity at higher multiplicity. More importantly, remarkable properties and simplicity of the final answer are sometimes obscured at integrand level or for individual integrals. It is thus tempting to ask for some alternative methods for computing multi-loop amplitudes, avoiding all the works of integrands and integrals, for higher loops and multiplicities.

One alternative is to use integrability of the theory~\cite{Beisert:2010jr}, which is believed to determine among other quantities scattering amplitudes (or equivalently light-like Wilson loops) for any value of the coupling~\cite{Basso:2013vsa}.  In principle all-loop amplitudes can be determined in this way, but in practice it is rather difficult to extract complete loop amplitudes as an analytic functions, and the state-of-art method is to use these results as powerful constraints on amplitudes. The most formidable progress for computing multi-loop amplitudes so far is through the hexagon and heptagon bootstrap program, which has exploited constraints from integrability very successfully \cite{Dixon:2011pw,Dixon:2014xca,Dixon:2014iba,Drummond:2014ffa,Dixon:2015iva,Caron-Huot:2016owq}. The first non-trivial amplitude in planar ${\cal N}=4$ SYM, the six-point amplitude (or hexagon) has been determined through seven and six loops for MHV and NMHV cases respectively~\cite{Caron-Huot:2019vjl}, and similarly the seven-point amplitude (or heptagon) has been determined through four loops for these cases respectively~\cite{Dixon:2016nkn,Drummond:2018caf}. This bootstrap method has been extremely powerful for $n=6,7$, but so far it has not been extended to $n\geq 8$ (octagons {\it etc.}), which are expected to be much more intricate. Despite all the progress on integrability and bootstrap, it is natural to ask the following question: is there an independent method for computing, at least in principle, scattering amplitudes of all loops, multiplicities and helicity sectors?

It is well known that scattering amplitudes in planar ${\cal N}=4$ SYM enjoy both superconformal and dual superconformal symmetries~\cite{Drummond:2006rz,Drummond:2008vq,Korchemsky:2010ut}, and they close into the infinite-dimensional Yangian symmetry which underpins integrability of the theory~\cite{Drummond:2009fd}. In~\cite{CaronHuot:2011kk}, based on the dual Wilson-loop picture, {\it exact} equations obeyed by the all-loop S-matrix were derived by determining the quantum corrections to the generators of Yangian symmetry acting on the Bern-Dixon-Smirnov (BDS)-renormalized S-matrix~\cite{Bern:2005iz}; the equations consist of $\bar{Q}$ equations for dual superconformal symmetry, and level-one generators $Q^{(1)}$ which are given by the parity-conjugate~\cite{CaronHuot:2011kk}. One answer to the question above is that by exploiting these anomaly equations, we have a first-principle method for recursively computing all-loop scattering amplitudes for all $n$ and $k$, which bypass loop integrands and integration altogether! Technically it is still very difficult to perform the recursive calculation explicitly for higher loops, but we would like to emphasize that there are no conceptual obstacles and everything boils down to solving first-order differential equations with sources from lower-loop amplitudes, as we will review shortly. 

The basic idea of computing amplitudes using anomaly equations is to convert the anomalies to total differentials, in terms of collinear integrals of lower-loop amplitudes with higher $n$ and $k$ ~\cite{CaronHuot:2011kk}. In all calculations using these anomaly equations so far, only half of them, namely the $\bar{Q}$ equations, have been used, which can already uniquely determine MHV and NMHV amplitudes given lower-loop ones. No serious attempts have yet been made to include both sets of equations, which would allow us to compute amplitudes with $k\geq 2$ and in turn MHV and NMHV amplitudes at higher loops, thus significantly push the limit of this method. Within the limitations of ${\bar Q}$ equations, the first application of the method has produced the complete symbol of two-loop MHV for all $n$, two-loop NMHV heptagon, and three-loop MHV hexagon~\cite{CaronHuot:2011kk}; for external kinematics in two dimensions,all two-loop NMHV and three-loop MHV amplitudes have been computed using $\bar{Q}$ equations~\cite{Caron-Huot:2013vda}. Moreover, the equations can provide all-loop constraints on scattering amplitudes which have proved to be very useful. As we will show shortly, without doing any loop-specific computations, ${\bar Q}$ equations provide the so-called last entry conditions for the {\it symbol}~\cite{Goncharov:2010jf, Duhr:2011zq, Duhr:2012fh} of not only MHV, but also NMHV amplitudes to all loops and all multiplicities~\footnote{Recall that MHV and NMHV amplitudes are expected to contain only generalized polylogarithms of weight $2L$ at $L$ loops~\cite{ArkaniHamed:2012nw}.}, and for $n=6,7$ they have been exploited in the hexagon and heptagon bootstrap. 

A crucial assumption for the hexagon and heptagon bootstrap is that the collection of letters entering the symbol, or the {\it alphabet}, consists of only $9$ and $42$ variables known as cluster coordinates~\cite{Golden:2013xva}, and the main challenge starting at $n=8$ is the lack of control for the symbol alphabet. The cluster algebra of $G_+(4,n)$ for $n\geq 8$ becomes infinite type and it is unclear which letters can appear for $L$-loop N${}^k$MHV amplitudes (see progress on analysis based on Landau equations~\cite{Dennen:2015bet,Prlina:2018ukf}). In addition, a new feature for $n\geq 8$ is the appearance of {\it algebraic letters} which can no longer be written as rational functions of momentum twistors~\cite{Hodges:2009hk}. It is of great interests even at two loops to understand the symbol alphabet and in particular what algebraic letters appear for $n\geq 8$. 

In our recent paper~\cite{Zhang:2019vnm}, we have computed the symbol of two-loop NMHV octagon using $\bar{Q}$ equations, as the first example of multi-loop amplitudes with algebraic letters. We have determined the alphabet of two-loop NMHV octagon, which consists of $180$ rational letters and $18$ algebraic ones that are independent under multiplicative relations. The $n=8$ alphabet of algebraic letters~\cite{Zhang:2019vnm} has been explained and conjectured to hold to higher loops using mathematical construction based on tropical Grassmannian~\cite{Drummond:2019cxm, Henke:2019hve} (see related ideas on positive configuration space~\cite{Arkani-Hamed:2019rds}). More recently, these $18$ algebraic letters have also been obtained from studying $n=8, k=2$ Yangian invariants or leading singularities~\cite{Mago:2020kmp,He:2020uhb} (see earlier works on ``cluster adjacency'' properties of rational letters based on poles of Yangian invariants~\cite{Drummond:2018dfd,Mago:2019waa}). It is tempting to ask if we can extend all these results on algebraic letters and the symbol to higher multiplicities. 

In this paper, we systematically derive such all-multiplicity results using $\bar{Q}$ equations, both at two-loop and all-loop orders, for NMHV amplitudes. Our results can be divided into three parts. First, since ${\bar Q}$ equations alone determine the total differential of MHV and NMHV amplitudes, the last entries of their symbols directly follow. The famous MHV last-entry conditions~\cite{CaronHuot:2011ky} follow from a simple residue computation on NMHV Yangian invariant~\cite{CaronHuot:2011kk}, without specifying loop orders. We will show that by the same residue computation on all possible N${}^2$MHV Yangian invariants, we obtain the complete last-entry conditions of $n$-point NMHV amplitudes, which we expect to be valid for all loops. 

As the main results of the paper, we present the $\bar{Q}$ calculation for the ``new'' part of two-loop $n$-point NMHV that contains algebraic symbol letters. By showing how to compute the action of collinear integrals on four-mass boxes of one-loop N${}^2$MHV, we determine the algebraic part of the symbol. All algebraic letters are grouped according to what ``four-mass'' square root $\Delta_{a,b,c,d}$ they contain: we find exactly $17-2 m$ multiplicative-independent algebraic letters for each $\Delta$, where $0\leq m\leq 4$ is the number of corners of the four-mass boxes that has only two particles. The most generic case is when all four corners contain more than two particles, {\it i.e.} $m=0$, we have $17$ independent letters which at least depend on $12$ particles, $a{-}1, a, a{+}1$, $\cdots$, $d{-}1, d, d{+}1$; other cases with $m>0$ are degenerate ones where some of the labels coincide. It nicely generalizes the $9$ independent algebraic letters for $n=8$~\cite{Zhang:2019vnm} with $m=4$. Moreover, we find that the symbol for this algebraic part can be written in a compact form: all new algebraic letters only appear in the third entry, and the first two-entry are the symbol for corresponding four-mass box; these weight-3 functions are also interlocked with specific (rational) last entries. 

Two interesting observations can be made about our results on the algebraic alphabet and words. First, we will show that all the algebraic letters can be interpreted as ``letters'' or poles of one-loop four-mass leading singularities with four trees that are either MHV or NMHV, which generalizes results for $n=8$ in~\cite{Mago:2020kmp, He:2020uhb}. Moreover, we find a large class of components of two-loop $n$-point NMHV amplitudes, which are absent of algebraic letters. They are the simplest NMHV components, which are coefficients of $\chi_i \chi_j \chi_k \chi_l$ for any non-adjacent $i, j, k, l$, and we show that they are completely free of square roots! Any such component can be written as the difference of two (cyclically related) double-pentagon integrals~\cite{ArkaniHamed:2010gh}, connecting our result to these important two-loop integrals. 

Finally, as the computation is straightforward but tedious for the remaining part which is independent of any algebraic letters,  we content ourselves by presenting explicit symbol for $n=9$. We find precisely $59 \times 9$ rational letters in addition to the $11 \times 9$ independent algebraic letters. Almost all rational letters are predicted by Landau equations except for one cyclic class, and we expect similar results for the rational alphabet extends to all multiplicities. We also show that all the algebraic letters are consistent with the rational letters in the sense of Landau analysis.

The paper is organized as follows. In sec. \ref{review}, after a quick review of $\bar{Q}$ equations, we move to list the complete collection of last-entry conditions for $n$-point NMHV amplitudes to all loop orders. 
In sec.~\ref{colintegralon4massbox}, we show how to compute the action of collinear integrals on four-mass boxes of the one-loop N${}^2$MHV amplitudes, which allows us to compute the algebraic part of the two-loop NMHV amplitudes. In sec.~\ref{algebraicletters}, we present the $17-2m$ independent algebraic letters after finding all the multiplicative relations they satisfied; we make a observation connecting them to one-loop four-mass leading singularities; we also give the nice patterns of how these algebraic letters appear in the symbol, and the implication for a large class of components. In sec.~\ref{exampleandcheck}, we present the full symbol of two-loop $n=9$ amplitudes including the complete alphabet. 

\section{Last-entry conditions for NMHV amplitudes from $\bar{Q}$ equations} \label{review}
\subsection{A lightning review of $\bar{Q}$ equations}

The infrared divergences of scattering amplitudes in planar $\mathcal{N}=4$ SYM exponentiate~\cite{Anastasiou:2003kj}, which can be captured by the famous BDS ansatz~\cite{Bern:2005iz}. For the $n$-point, N$^{k}$MHV amplitude $A_{n,k}$, we define an infrared-finite object, the so-called \emph{BDS-normalized amplitude}  $R_{n,k}=A_{n,k}/A_{n}^{\text{BDS}}$. $R_{n,k}$ is dual conformal invariant and enjoys a chiral half of the dual superconformal symmetries, but it is not invariant under the action of the other half~\cite{CaronHuot:2011kk}:
\begin{equation}
    \bar{Q}_{a}^{A}= \sum_{i=1}^n \chi_{i}^{A}\frac{\partial}{\partial Z_{i}^{a}} \:,
\end{equation}
where $Z_{i}$ and $\chi_{i}$ are the bosonic and fermionic part of the super momentum-twistor
\begin{align}
    \mathcal{Z}_{i}=(Z_{i}^{a}\vert \chi_{i}^{A}):=(\lambda_{i}^{\alpha},x_{i}^{\alpha\dot{\alpha}}\lambda_{i\alpha}\vert\theta_{i}^{\alpha A}\lambda_{i\alpha})
\end{align}
with dual super coordinates $(x\vert \theta)$. The remaining  unbroken $SL(4|4)$ dual superconformal generators include the ``good'' chiral half $Q_A^a=\sum_{i=1}^n Z_i^a \partial/\partial \chi_i^A$, the bosonic generators $K^a_b=\sum_{i=1}^n Z_i^a \partial/\partial Z_i^b$ and the R symmetry ones $R^A_B=\sum_{i=1}^n \chi_i^A \partial/\partial \chi_i^B$.

Thus super momentum-twistors make dual superconformal symmetry manifest, while the usual superconformal symmetry acts via level-one generators~\cite{Drummond:2010qh}. The most basic $\mathrm{SL}(4)$ invariants that one can build from bosonic momentum-twistors are the Pl\"{u}cker coordinates of $\mathrm{Gr}(4,n)$: $\langle ijkl\rangle:=\varepsilon_{abcd}Z_{i}^{a}Z_{j}^{b}Z_{k}^{c}Z_{l}^{d}$. Using supersymmetric momentum-twistors, one can build dual superconformal invariants or even Yangian invariants, which are most generally written in terms of contour integrals inside positive Grassmannian~ \cite{Drummond:2010uq,ArkaniHamed:2009vw,ArkaniHamed:2012nw}. For example, the most basic Yangian invariant, which were originally called the $R$ invariant~\cite{Drummond:2008vq,Mason:2009qx}, reads
\begin{equation}\label{Rinv}
    [i\,j\,k\,l\,m]:=\frac{\delta^{0\vert 4}(\chi_{i}^{A}\langle jklm\rangle+\text{cyclic})}{\langle ijkl\rangle\langle jklm\rangle
    \langle klmi\rangle\langle lmij\rangle\langle mijk\rangle} \:.
\end{equation}
It is antisymmetric in the five particle indices and satisfy the so-called six-term identity
\begin{equation}
    [ijklm]+[jklmn]+[klmni]+[lmnij]+[mnijk]+[nijkl]=0. \label{6ident}
\end{equation}
They are the most general Yangian invariants for $k=1$, and for general $k$, such Yangian invariants are leading singularities of loop amplitudes, including BCFW terms appearing in tree amplitudes. The NMHV tree amplitude, which we use shortly, is simply given by a sum of them: $R_{n,1}^{\rm tree}=\sum_{i<j} [1\,i\, i{+}1\,j\,j{+}1]$ (one can replace label $1$ by any other label here, which gives the same result). 

One of the main results of~\cite{CaronHuot:2011kk} is the following anomaly equation for the $\bar{Q}$ generators: it has been argued based on a Wilson-loop analysis that $\bar{Q}$ of the amplitude is given in terms of an integral of higher-point one with fermion insertion (which increases $k$) in the collinear limit, and by taking into account $\bar{Q}$ of the BDS ansatz we have:
\begin{align}
\bar{Q}_{a}^{A} R_{n,k} = \frac{1}{4}\Gamma_{\rm cusp}~\operatorname{Res}_{\epsilon=0}\int_{\tau=0}^{\tau=\infty}\Bigl(\mathrm{d}^{2\vert 3}\mathcal{Z}_{n+1}\Bigr)_{a}^{A} \:\bigl[ R_{n+1,k+1}
-R_{n,k}R_{n+1,1}^{\text{tree}} \bigr]+ \text{cyclic}  \label{Qbar} \:, 
\end{align}
where $\Gamma_{\rm cusp}$ is the cusp anomalous dimension~\cite{Beisert:2006ez} and ``cyclic'' denotes the $n{-}1$ images of the foregoing term under the rotation $1\to2\to\cdots \to n \to 1$. Eq. \eqref{Qbar} is conjectured to hold non-perturbatively~\cite{CaronHuot:2011kk}. On the RHS, we have shown the term with particle $n{+}1$ added in collinear limit with $n$, and its (super-) momentum-twistor $\mathcal{Z}_{n+1}$ is parameterized by $\epsilon, \tau$:
\begin{equation}
    \mathcal{Z}_{n+1}= \mathcal{Z}_{n}- \epsilon \mathcal{Z}_{n-1} + C \epsilon \tau \mathcal{Z}_{1} + C'\epsilon^{2} \mathcal{Z}_{2} \:,
    \label{colpara}
\end{equation}
with $C=\frac{\langle n{-}1\,n\,2\,3\rangle}{\langle n\,1\,2\,3\rangle}$ and $C'=\frac{\langle n{-}2\,n{-}1\,n\,1\rangle}{\langle n{-}2\,n{-1}\,2\,1\rangle}$. The integral measure $\int (\mathrm{d}^{2\vert 3}\mathcal{Z}_{n+1})_{a}^{A}$ consists of the bosonic part $(\mathrm{d}^{2}Z_{n+1})_{a}:=\varepsilon_{abcd}Z_{n+1}^{b}\mathrm{d}Z_{n+1}^{c}\mathrm{d}Z_{n+1}^{d}$ and the fermionic part $(\mathrm{d}^{3}\chi_{n+1})^{A}$; using (\ref{colpara}) the bosonic measure can be written as 
\begin{equation}C(\bar{n})_{a}\operatorname{Res}_{\epsilon=0}\int\epsilon\mathrm{d}\epsilon\int_{0}^{\infty} \mathrm{d}\tau
\end{equation}
with $(\bar{n})_{a}:=(n{-}1\,n\,1)_{a}$. Thus computing the $\bar{Q}$ anomaly is straightforward: after performing the 3-fold fermionic integration over $\chi_{n+1}$, taking the collinear parametrization \eqref{colpara} for both $Z_{n+1}$ and the remaining $\chi_{n+1}$, the notation $\operatorname{Res}_{\epsilon=0}$ means to extract the coefficient of $\mathrm{d}\epsilon/\epsilon$ under the collinear limit of $\epsilon\to 0$, and finally we integrate over ``momentum fraction'' $\tau$ from $0$ to $\infty$. As shown in~\cite{CaronHuot:2011kk}, precisely the difference of the two terms in the bracket ensures that the RHS of \eqref{Qbar} is finite: not only do possible $\log \epsilon$ divergences cancel, the combination is also free of endpoint divergences for the $\tau$-integral, which serve as important consistency checks of the calculation.

In practice, we can make enormous progress by expanding \eqref{Qbar} perturbatively: it relates the anomaly of $L$-loop amplitude, $R_{n,k}^{(L)}$, to lower-loop ones such as $ R_{n+1,k+1}^{(L-1)}$ and $R_{n,k}^{(L-1)}$. The next question is if we can use the anomaly to determine the full amplitude, and this amounts to solve  \eqref{Qbar} which can be viewed as a collection of first-order differential equations. From a ``modern'' perspective (c.f. \cite{He:2018okq}), the right thing to do is to replace the Grassmann variables by differentials of momentum twistors (which are also anti-commuting)~\cite{Arkani-Hamed:2017vfh} $\chi^A_i \to \dif Z^A_i$ for $i=1,2,\cdots, n$, and we have identified the index of $\dif Z$ with the R-symmetry index; by taking the trace of the operator $\sum_{i=1}^n \dif Z_i^A \partial/\partial Z_i^a$ we have that $\dif R_{n,k}^{(1)}$ is given by the RHS of  \eqref{Qbar} with the same replacement $\chi_i \to \dif Z_i$~\footnote{Note that when we take the differential of the usual super-amplitude, it is understood that the differential $\dif$ only acts on transcendental functions, not on the coefficients which are Yangian (thus $\bar{Q}$) invariants. This is consistent since after we do the replacement $\chi \to \dif Z$ for these Yangian invariants, they become differential forms that are closed.}. 

The remaining task in solving the differential equations is just to determine the ``kernel'' of $\bar{Q}$ (or $\dif$ if we make the replacement). As shown in~\cite{CaronHuot:2011kk}, for N$^{k}$MHV amplitude with $k\geq 2$, the kernel of $\bar{Q}$ does involve non-trivial dual conformal functions, thus $\bar{Q}$-equation can not determine the result uniquely without supplement with the parity-conjugate, $Q^{(1)}$ equations (with both $\bar{Q}$ and $Q^{(1)}$ equations, the kernel must be linear combination of Yangian invariants which can be determined in turn by collinear limits {\it etc.}). However, as we will see that for NMHV (and MHV) amplitudes, \eqref{Qbar} is very powerful as the kernel is essentially trivial, and it alone allows us to compute the differential of $R_{n, k}^{(L)}$ with $k=0,1$. 

For MHV amplitude, which is a function of bosonic momentum twistors (no Grassmann part), the only function that can be annihilated by $\bar{Q}$ or $\dif$ is a constant.  For NMHV cases, the $\bar{Q}_{a}^{A}$ operator has a non-trivial kernel since 
\begin{equation}
    \bar{Q}_{a}^{A}\biggl([12345]\log\frac{\langle 1234\rangle}{\langle 1235\rangle} \biggr)=0\,. 
           \label{qbarker}      
\end{equation}
Nevertheless, one can prove that the $\bar{Q}_{a}^{A}$ can never annihilate a function of form $[i\,j\,k\,l\,m] F(Z)$ where $F(Z)$ is a conformal invariant function of bosonic momentum-twistors~\cite{CaronHuot:2011kk}. Thus, $\bar{Q}$ equations \eqref{Qbar}, with the supplement of dual conformal invariance, can determine the differential of NMHV amplitudes on their own.

\subsection{Last-entry conditions for all-loop NMHV amplitudes}

It is expected that MHV or NMHV BDS-normalized amplitudes admit a schematic form 
\begin{equation}
    R_{n,k=0,1}^{(L)}=\sum_{\alpha} Y_{n,k=0,1}^{\alpha}\,\mathcal{I}^{(2L)}_{\alpha} \:, 
\end{equation}
where $Y_{n,k}$ denote the loop-independent Yangian invariants (recall $Y_{n,k=0}= 1$ and all $Y_{n,k=1}$ are given by R invariants of the form~\eqref{Rinv}), and $\mathcal{I}^{(2L)}$'s are linear combinations of generalized polylogarithms of weight $2L$. They can be defined by $2L$-fold iterated integrals~\cite{goncharov2005galois}
\begin{equation}
    G(a_{1},\ldots,a_{2L};z)=\int_{0}^{z}\frac{\dif t}{t-a_{1}}\, G(a_{2},\ldots,a_{2L};t), \label{Gpolylot} 
\end{equation} 
with the starting point $G(z):=1$. It is straightforward to see that the differential of a generalized polylogarithm $\mathcal{I}^{(2L)}_{\alpha}$ of weight $2L$ satisfy
\begin{equation}
    \dif \mathcal{I}_{\alpha}^{(2L)} = \sum_{\beta} \mathcal{I}_{\alpha,\beta}^{(2L-1)}\dif \log s_{\beta}
\end{equation}
where $\mathcal{I}^{(2L-1)}$ are some generalized polylogarithm of weight $2L{-}1$. Then, one can introduce a symbol map for generalized polylogarithms by recursively defining 
\begin{equation}
    \mathcal{S}(\mathcal{I}_{\alpha}^{(2L)}) =\sum_{\beta} \mathcal{S}(\mathcal{I}_{\alpha,\beta}^{(2L-1)})\otimes s_{\beta}
\end{equation}
with $\mathcal{S}(\log a):= a$. We call $s_{\beta}$'s generated in this way the {\it symbol letters}, each tensor product consisted of letters the {\it{word}}, and the collection of all letters the {\it symbol alphabet}~\cite{Goncharov.A.B.:2009tja,Goncharov:2010jf}.

Now, it is natural to write the differential of NMHV or MHV $L$-loop amplitudes as
\begin{equation} \label{devofR}
\mathrm{d} R_{n,k=0,1}^{(L)}=\sum_{\alpha,\beta} Y_{n,k=0,1}^{\alpha}~\mathrm{d}\log(s_\beta)~{\cal I}^{(2L{-}1)}_{\alpha,\beta}\,.
\end{equation}
To derive~\eqref{devofR} from $\bar{Q}$ equations \eqref{Qbar}, we use the fact that the RHS of \eqref{Qbar} consists of terms of the form $Y_{n{+}1, k{+}1} F^{(2L{-}2)}_{n+1}$~\footnote{For $k>0$, we do not assume the transcendental functions $F$ to be generalized polylogarithms, but we still schematically put a ``weight'' $2L{-}2$ for such functions at $L{-}1$ loops. In turn, just from $\bar{Q}$ equations we cannot prove NMHV amplitudes must be generalized polylogarithms without inspecting structures of N${}^2$MHV amplitudes on the RHS, though we expect this to be true.}.
A remarkable feature of the integral $\operatorname{Res}_{\epsilon=0}\int \mathrm{d}^{2\vert3}\mathcal{Z}_{n+1}$ is that the fermionic integral $\int\mathrm{d}^{3}\chi_{n+1}$ and residue $\operatorname{Res}_{\epsilon=0}$ part can be performed on Yangian invariants $Y_{n+1,k{+}1}$ independent of transcendental functions $F$'s. One may worry about the $\log^{L-1}\epsilon$ divergences arising from the collinear limit of but the divergences are always canceled after integrating over $\tau$, as shown in~\cite{CaronHuot:2011kk}. For MHV ($k=0$), the effect of $\operatorname{Res}_{\epsilon=0}\int \epsilon\mathrm{d}\epsilon\int \mathrm{d}^{3}\chi_{n+1}$ yields terms of the form 
\begin{equation}
\bar{Q}\log\frac{\langle\bar{n} i\rangle}{\langle\bar{n}j \rangle} \times \int \mathrm{d}\log f_{i,j}(\tau)~F_{n{+}1} (\tau, \epsilon\to 0)
\end{equation}
with some rational function $f_{i,j} (\tau)$ for each term. The last step is trivial for MHV: we can simply replace $\chi_{i}^{A}$ in $\bar{Q}_{a}^{A}$ with $\mathrm{d}Z_{i}^{A}$ then take the trace to obtain the external derivative $\mathrm{d}:=\sum_{i,a}\mathrm{d}Z_{i}^{a}\,\partial/\partial Z_{i}^{a}$, which reproduces the well-known MHV final entries $\mathrm{d}\log \langle i{-}1\, i\, i{+}1\, j\rangle$ after collecting all cyclic terms. The one-dimensional integrals for $F^{(2L{-}2)}_{n{+}1}$ gives weight-$(2L{-}1)$ functions in (\ref{devofR}). 

For NMHV ($k=1$), the effect of $\operatorname{Res}_{\epsilon=0}\int \epsilon\mathrm{d}\epsilon\int \mathrm{d}^{3}\chi_{n+1}$ on N${}^2$MHV Yangian invariants, $Y_{n{+}1,2}$ on the RHS of \eqref{Qbar} gives a list of possible $Y_{n,1}$ in \eqref{Rinv} times final entries as
\begin{equation}\label{NMHVfinal}
Y_{n,1}^\alpha~{\bar Q} \log (a_\alpha) \in \biggl\{  [i\,j\,k\,l\,m]~\bar{Q}\log\frac{\langle\bar{n} I\rangle}{\langle\bar{n} J \rangle} \biggr\}
\end{equation}
where $I, J$ can generally be intersections of momentum twistors of the form {\it e.g.} $(i j) \cap (k l m)$ (see~\cite{ArkaniHamed:2010kv,ArkaniHamed:2010gh}). To obtain the differential of $R_{n,1}$ in this case, the naive replacement above has an ambiguity due to the existence of non-trivial kernel of $\bar{Q}$, which always take the form \eqref{qbarker}.
%
Nevertheless, since the kernel of $\bar{Q}$ can not contain non-trivial functions of \emph{dual conformal invariants} (DCI) in this case, the replacement $\chi_{i}\to \mathrm{d}Z_{i}$ has no ambiguity once we convert the arguments of $\bar{Q}\log$ to DCI 
by adding ``0'' of the form~\eqref{qbarker}. It is a straightforward but tedious algorithm to arrive at such a manifestly DCI form, which gives the final answer for $\mathrm{d} R_{n,1}$.

To obtain all possible last entries for NMHV amplitudes, one needs to consider the action of $\,\int\dif^{2|3}\mathcal{Z}_{n+1}\,$ on all possible N$^{2}$MHV Yangian invariants. Unlike the unique type of NMHV Yangian invariant \eqref{Rinv}, there are $14$ distinct N$^{2}$MHV Yangian invariants up to cyclic rotations~\cite{ArkaniHamed:2012nw}. Apart from the algebraic leading singularities of four-mass boxes, which we will discuss below, the other $13$ of them are all rational, for which the effect of $\operatorname{Res}_{\epsilon=0}\int \epsilon\mathrm{d}\epsilon\int \mathrm{d}^{3}\chi_{n+1}$ is given in Appendix \ref{appa}. Now we list all the NMHV last-entry conditions by applying the operation all these $14$ types of invariants. 

We have obtained three types of last entries (dressed with NMHV Yangian invariants). First, we obtain last entries of the form
\begin{equation}
    [abcde]\bar{Q}\log\frac{\<\bar{n}i\>}{\<\bar{n}j\>}\,.  \label{lastentry1}
\end{equation}
By considering six-term identities~\eqref{6ident} and relations of the form \eqref{qbarker} we see that there are $(n-4)\binom{n-1}{4}-\binom{n-3}{2}$ such last entries.\footnote{This counting comes from the fact that the number of independent $R$-invariants is $\binom{n-1}{4}$, the number of independent $\bar{Q}\log\frac{\langle\bar{n}i\rangle}{\langle\bar{n}j\rangle}$ is $n{-}4$, and there are $\binom{n-3}{2}$ vanishing combinations of form $[n{-}1\,n\,1\,i\,j]\bar{Q}\log\frac{\langle\bar{n}i\rangle}{\langle\bar{n}j\rangle}$.}\\

The second type of last entries we obtain are
\begin{align}
    &[1\,i_{1}\,i_{2}\,i_{3}\,i_{4}]\,\bar{Q}\log\frac{\<1(n{-}1\,n)(i_{1}\,i_{2})(i_{3}\,i_{4})\>}{\<\bar{n}i_{1}\>\<1i_{2}i_{3}i_{4}\>} \:,  \nonumber \\
    &[i_{1}\,i_{2}\,i_{3}\,i_{4}\,n{-}1]\,\bar{Q}\log\frac{\<n{-}1(n\,1)(i_{1}\,i_{2})(i_{3}\,i_{4})\>}{\<\bar{n}i_{1}\>\<n{-}1\,i_{2}i_{3}i_{4}\>} \label{lastentry2} \\
    &[i_{1}\,i_{2}\,i_{3}\,i_{4}\,n]\,\bar{Q}\log\frac{\<n(1\,n{-1})(i_{1}\,i_{2})(i_{3}\,i_{4})\>}{\<\bar{n}i_{1}\>\<n\,i_{2}i_{3}i_{4}\>}  \nonumber
\end{align}
where $\,1<i_{1}<i_{2}<i_{3}<i_{4}<n{-}1$, and we abbreviate $\<a(bc)(de)(fg)\>:=\<abde\>\<acfg\>-\<acde\>\<abfg\>$. There are \,$3\binom{n-3}{4}\,$ such last entries. \\

Finally, we have the third type of last entries
\begin{equation} \label{lastentry3}
    [i_{1}\,i_{2}\,i_{3}\,i_{4}\,i_{5}]\,\bar{Q}\log\frac{\<\bar{n}(i_{1}i_{2})\cap(i_{3}i_{4}i_{5})\>}{\<\bar{n}i_{1}\>\<i_{2}i_{3}i_{4}i_{5}\>} \:, \:\: [i_{1}\,i_{2}\,i_{3}\,i_{4}\,i_{5}]\,\bar{Q}\log\frac{\<\bar{n}(i_{1}i_{2}i_{3})\cap(i_{4}i_{5})\>}{\<\bar{n}i_{1}\>\<i_{2}i_{3}i_{4}i_{5}\>} \:,
\end{equation}
where $\,1<i_{1}<i_{2}<i_{3}<i_{4}<i_{5}<n{-}1$, and we have $\,2\binom{n-3}{5}\,$ of them.

By considering cyclic rotations, we see that altogether there are $\,42\binom{n}{6}\,$ last entries. 
However, since $\bar{Q}$ has a non-trivial kernel, we need to turn the arguments of $\,\bar{Q}\log\,$ into dual conformal invariants (DCI): this has already been done in the second and third cases, but we need to do it for the first case. This is realized by expanding the last entries on the basis which has the minimal number of the basis vectors that are not DCI. After then, the coefficients of these basis vectors automatically vanish due to the dual conformal invariance of amplitudes. 
To see this, we temporarily introduce equivalence relations $\,Y_{n,1}\bar{Q}\log v\sim 0$, where $\,v\,$ is a DCI. Since $\,[c\,d\,e\,f\,g]\bar{Q}\log \frac{\<pqra\>}{\<pqrb\>}\sim [c\,d\,e\,f\,g]\bar{Q}\log\frac{\<ijka\>}{\<ijkb\>}$, let's introduce the abbreviation
\begin{equation}
    [ab;cdefg]:=[c\,d\,e\,f\,g]\bar{Q}\log\frac{\<pqra\>}{\<pqrb\>} \:.
\end{equation}
It is easy to see that $\,[ab;abcde]\sim 0$ by adding ``0'' of the form \eqref{qbarker},
\[
[a\,b\,c\,d\,e]\bar{Q}\log\frac{\<pqra\>}{\<pqrb\>} 
=[a\,b\,c\,d\,e]\bar{Q}\log\frac{\<pqra\>\langle bcde\rangle}{\<pqrb\>\< acde\>} \sim 0\:,
\]
and
\begin{align*}
    [ab;cdefg] &\sim [ac;cdefg]+[cb;cdefg] \:,  \\
    [ab;bcdef] &\sim [ab;hcdef]-[ab;bhdef]+[ab;cdhef] \\
    &\quad -[ab;cdehf]+[ab;cdefh]\:,
\end{align*}
where the six-term identity \eqref{6ident} has been used in the second relation, especially $[ab;bcdef]\sim [ac;bcdef]\,$ and $\,[ab;bcdef]\sim[ab;acdef]$. By repeated use of such relations, all $\,[\ast;\ast]\,$ can be expanded on a basis $\,\{[ab;bcden],1\leq a<b<c<d<e\leq n{-}1\}$. 
As mentioned, all the coefficients of these basis vectors, which are weight-$(2L{-}1)$ functions, must vanish. After removing them, we obtain $\,42\binom{n}{6}-\binom{n-1}{5}\,$ last-entry conditions for all-loop NMHV BDS-normalized amplitudes. This reduces to the last-entry conditions given in \cite{Zhang:2019vnm} for $n=8$.

\section{The action of collinear integrals on the four-mass boxes} \label{colintegralon4massbox}

\subsection{A quick review of box expansions in \texorpdfstring{$\mathcal{N}=4$}{N=4} SYM}

According to $\bar{Q}$ equations \eqref{Qbar}, we need the one-loop N$^{2}$MHV amplitudes as the input in the computation of 2-loop NMHV amplitudes. Such data are available from the familiar \emph{box expansion}~\cite{Bern:1992em,Bern:1993kr}. Let us quickly review this result and setup some notations.

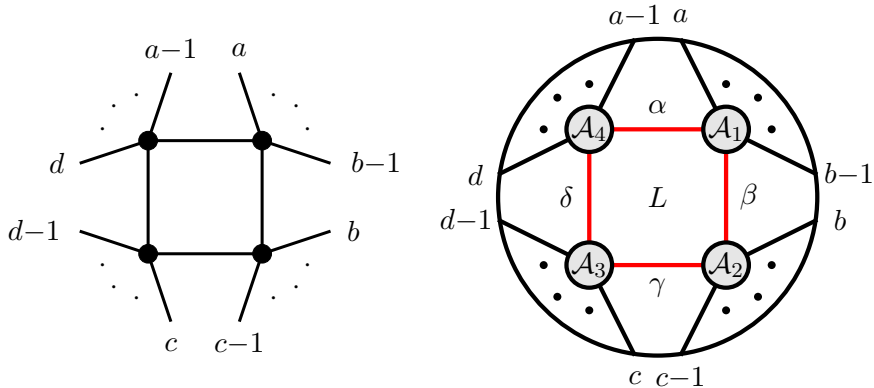
\begin{figure}
    \centering
    \begin{tikzpicture}[scale=1.5,baseline={([yshift=-.5ex]current bounding box.center)}]
        \node[fill=black,circle,draw=black, inner sep=0pt,minimum size=7pt] at (-2,0) {};
        \node[fill=black,circle,draw=black, inner sep=0pt,minimum size=7pt] at (-2,-1) {};
        \node[fill=black,circle,draw=black, inner sep=0pt,minimum size=7pt] at (-1,-1) {};
        \node[fill=black,circle,draw=black, inner sep=0pt,minimum size=7pt] at (-1,0) {};
        \draw[line width=0.4mm] (-2,0) -- (-2,-1) -- (-1,-1) -- (-1,0) -- cycle;
        \draw[line width=0.4mm] (-2.6,-0.8) -- (-2,-1);
        \draw[line width=0.4mm] (-2,-1) -- (-1.8,-1.6);
        \draw[line width=0.4mm] (-1,-1) -- (-0.4,-0.8);
        \draw[line width=0.4mm] (-1,-1) -- (-1.2,-1.6);
        \draw[line width=0.4mm] (-2.6,-0.2) -- (-2,0);
        \draw[line width=0.4mm] (-1.8,0.6) -- (-2,0);
        \draw[line width=0.4mm] (-1.2,0.6) -- (-1,0);
        \draw[line width=0.4mm] (-1,0) -- (-0.4,-0.2);
        \node at (-1.2,0.8) {$a$};
        \node at (0,-0.2) {$b{-}1$};
        \node at (-0.2,-0.8) {$b$};
        \node at (-1.2,-1.8) {$c{-}1$};
        \node at (-1.8,-1.8) {$c$};
        \node at (-3,-0.8) {$d{-}1$};
        \node at (-2.8,-0.2) {$d$};
        \node at (-1.8,0.8) {$a{-}1$};
        \node at (-2.4,0.1) {$\cdot$};
        \node at (-2.3,0.3) {$\cdot$};
        \node at (-2.1,0.4) {$\cdot$};
        \node at (-0.9,-1.4) {$\cdot$};
        \node at (-0.7,-1.3) {$\cdot$};
        \node at (-0.6,-1.1) {$\cdot$};
        \node at (-0.6,0.1) {$\cdot$};
        \node at (-0.7,0.3) {$\cdot$};
        \node at (-0.9,0.4) {$\cdot$};
        \node at (-2.4,-1.1) {$\cdot$};
        \node at (-2.3,-1.3) {$\cdot$};
        \node at (-2.1,-1.4) {$\cdot$};
    \end{tikzpicture}
        \def\mydotintikz{\node[circle,fill=black, inner sep=0pt,minimum size=3pt]}
    \begin{tikzpicture}[scale=0.6,baseline={([yshift=-.5ex]current bounding box.center)}]
        \draw[ultra thick]  (-1.5,0.5) node{$L$} ellipse (3.5 and 3.5);
        \draw[ultra thick,red] (-2.5,2) -- (-0.5,2); \draw[ultra thick,red] (0,-0.5) -- (0,1.5);
        \draw[ultra thick,red] (-2.5,-1) -- (-0.5,-1); \draw[ultra thick,red] (-3,1.5) -- (-3,-0.5);
        \draw[ultra thick] (0,-1) -- (-1,-3); \draw[ultra thick] (0,-1) -- (2,0);
        \draw[ultra thick] (0,2) -- (2,1); \draw[ultra thick] (0,2) -- (-1,4);
        \draw[ultra thick] (-3,-1) -- (-5,0); \draw[ultra thick] (-3,-1) -- (-2,-3);
        \draw[ultra thick] (-3,2) -- (-5,1); \draw[ultra thick] (-3,2) -- (-2,4);
        \filldraw[ultra thick, fill=gray!20]  (-3,2) node {$\mathcal A_4$} ellipse (0.5 and 0.5);
        \filldraw[ultra thick, fill=gray!20]  (0,2) node {$\mathcal A_1$} ellipse (0.5 and 0.5);
        \filldraw[ultra thick, fill=gray!20]  (0,-1) node {$\mathcal A_2$} ellipse (0.5 and 0.5);
        \filldraw[ultra thick, fill=gray!20]  (-3,-1)node {$\mathcal A_3$} ellipse (0.5 and 0.5);
        \node at (-2,4.5) {$a{-}1$}; \node at (-1,4.5) {$a$};
        \node at (2.7,1) {$b{-}1$}; \node at (2.5,0) {$b$};
        \node at (-1,-3.5) {$c{-}1$}; \node at (-2,-3.5) {$c$};
        \node at (-5.7,0) {$d{-}1$}; \node at (-5.5,1) {$d$};
        \mydotintikz at (-4,-1) {}; \mydotintikz at (-3.7,-1.7) {};
        \mydotintikz at (-3,-2) {}; \mydotintikz at (0,3) {};
        \mydotintikz at (0.7,2.7) {}; \mydotintikz at (1,2) {};
        \mydotintikz at (-4,2) {}; \mydotintikz at (-3.7,2.7) {};
        \mydotintikz at (-3,3) {}; \mydotintikz at (1,-1) {};
        \mydotintikz at (0.7,-1.7) {}; \mydotintikz at (0,-2) {};
        \node at (-3.5,0.5) {$\delta$}; \node at (-1.5,2.5) {$\alpha$};
        \node at (0.5,0.5) {$\beta$}; \node at (-1.5,-1.5) {$\gamma$};
    \end{tikzpicture}
    \caption{The four mass box and four mass leading singularity}
    \label{fig:1}
\end{figure}
    Because the scattering amplitudes in $\mathcal{N}=4$ are free of UV divergence, one-loop amplitudes can be expanded in a basis of box integrals $I_{a,b,c,d}$ involving four inverse loop momentum propagators $x_{0a}^{2}$, $x_{0b}^{2}$, $x_{0c}^{2}$ and $x_{0d}^{2}$:
\begin{equation}
    \mathcal{A}_{n,k}^{1\text{-loop}} = \sum_{1\leq a<b<c<d} \mathcal{L}_{a,b,c,d} I_{a,b,c,d} \label{scalarboxexpansion}
\end{equation}
with the \emph{leading singularities} $\mathcal{L}_{a,b,c,d}$ as the coefficients. The most generic terms are the so-called ``four-mass'' box, {\it{i.e.}}, all four mass corners $\{\{a,\ldots,b{-}1\},\{b,\ldots,c{-}1\},\{c,\ldots,d{-}1\},$ $\{d,\ldots ,a{-}1\}\}$ involve two or more particles, (see figure \ref{fig:1}). For such terms, the box integrals $I_{a,b,c,d}$ is free of any divergence and can be evaluated to weight-2 polylogarithms:
\begin{align}
    I_{a,b,c,d}:=\int \dif^{4} x_{0} \frac{-x_{ac}^{2}x_{bd}^{2}\Delta}{x_{0a}^{2}x_{0b}^{2}x_{0c}^{2}x_{0d}^{2}} =
    -\Li_{2}(z)+\Li_{2}(\bar{z})-\frac{1}{2}\log(z\bar{z}) \log\biggl(\frac{1-z}{1-\bar{z}}\biggr) \label{boxintegral}
\end{align}
where
\begin{align}
    u&:=\frac{x_{ab}^{2}x_{cd}^{2}}{x_{ac}^{2}x_{bd}^{2}}\:,\quad v:=\frac{x_{bc}^{2}x_{da}^{2}}{x_{ac}^{2}x_{bd}^{2}} \:,\quad  \Delta:=\sqrt{(1-u-v)^{2}-4uv} \:. \label{uvdel} \\
    z&=\frac{1}{2}\bigl(1+u-v+\Delta\bigr)\:, \qquad \bar{z}=\frac{1}{2}\bigl(1+u-v-\Delta\bigr)\:. \label{zzbar}
\end{align}
The subscript $a,b,c,d$ will be restored to indicate the specific box when necessary, otherwise suppressed. The coefficient $\mathcal{L}_{a,b,c,d}$ for each four-mass box is the sum of products of 4 tree amplitudes and the N$^{2}$MHV ``four-mass'' Yangian invariant~\cite{ArkaniHamed:2012nw}
\begin{align}
    \mathcal{L}_{a,b,c,d}=&\sum \sum_{\pm}\left(\begin{array}{r}
        \quad \mathcal{A}_{k_1}(\alpha_{\pm}, a, \cdots, b{-}1, \beta_{\pm}) \\
        \times \mathcal{A}_{k_2}(\beta_{\pm}, b, \cdots, c{-}1, \gamma_{\pm}) \\
        \times \mathcal{A}_{k_3}(\gamma_{\pm}, c, \cdots, d{-}1, \delta_{\pm}) \\
        \times \mathcal{A}_{k_4}(\delta_{\pm}, d, \ldots, a{-}1, \alpha_{\pm})
        \end{array}\right) \nonumber \\
        &\quad \times \frac{1-u-v\pm \Delta}{2\Delta}[\alpha_{\pm},b{-}1,b,c{-}1,c][\gamma_{\pm},d{-}1,d,a{-}1,a]\label{4massLS}
\end{align}
where the first sum is over all sets of four tree amplitudes satisfying $\sum_{i=1}^4 k_i=k{-}2$, 
and the second sum is over the two solutions of the Schubert problem
\begin{align}
    \alpha&= (a{-}1\,a)\cap (d\,d{-1}\,\gamma)\:,\qquad  \gamma= (c{-}1\,c)\cap (b\,b{-}1\, \alpha) \:,\\
    \beta&=(b\,b{-1})\cap (c{-}1\,c\,\delta) \:,\qquad   \,  \:\:\delta=(d\,d{-1})\cap (a{-}1\,a\,\beta) \:.
\end{align}
The other terms in the box expansion \eqref{scalarboxexpansion} can be obtained from the general four-mass boxes by taking one or more mass corners massless (say, $b\to a{+}1$). The coefficients $\mathcal{L}_{a,b,c,d}$ vary smoothly in this limit.  However, the box integrals become divergent and must be regulated. There are several regularization schemes, say dimensional regularization~\cite{Bern:1993kr} and Higgs regularization~\cite{Alday:2009zm}. Here we follow a dual conformal invariant regularization scheme introduced in~\cite{Bourjaily:2013mma}. In this regularization, the infrared finite and regulator-independent BDS-subtracted $S$-matrix $R_{n,k}$ at 1-loop reads 
\begin{eqnarray}
    R_{n,k}^{(1)} = \sum_{1\leq a<b<c<d} \bigl(\mathcal{L}_{a,b,c,d}-\mathcal{A}_{n,k}^{\text{tree}}\mathcal{L}_{a,b,c,d}^{\text{MHV}}\bigr) I_{a,b,c,d}^{\text{fin}}
\end{eqnarray}
where $I_{a,b,c,d}^{\text{fin}}$ denote the finite part of DCI-regulated box integrals; $\mathcal{L}_{a,b,c,d}^{\text{MHV}}=0,1$ are 1-loop MHV box coefficients. The reader who is interested in the other coefficients $\mathcal{L}_{a,b,c,d}$ and DCI-regulated box integrals is urged to see \cite{Bourjaily:2013mma} for a detailed discussion and a complete list. 

For our purpose, we only need $k=2$ thus the 4 tree amplitudes in \eqref{4massLS} are all MHV with ${\cal A}_{k=0}=1$, and we are left with the last line, which we denote by $f^{\pm}_{a,b,c,d}$ for the two solutions. An important point we want to emphasize here is that all boxes other than four-mass ones are totally free of the square root $\Delta$, since the corresponding $u$ and/or $v$ vanish. The $\dif^{2\vert 3}\mathcal{Z}_{n+1}$ integration for these boxes can thus be easily performed without any obstacle. However, the $\dif^{2\vert 3}\mathcal{Z}_{n+1}$ integration for four-mass boxes are non-trivial due to the existence of the square root $\Delta$. In the rest of this section, we will work out on the prescription for four-mass boxes and obtain algebraic part of the two-loop answer. 


\subsection{The prescription for four-mass boxes}

Now we consider how $\dif^{2\vert 3}\mathcal{Z}_{n+1}$ acts on N$^{2}$MHV four-mass boxes with coefficients, and it is easy to see that only two kinds of $f^{\pm}_{a,b,c,d}$ survive: $f_{1,b,c,n}^{\pm}I_{1,b,c,n}$ with corners
\begin{equation}
    \{\{1,\ldots,b{-}1\},\{b,\ldots,c{-}1\},\{c,\ldots,n{-}1\},\{n,n{+}1\}\} \:,
\end{equation}
and $f^{\pm}_{a,b,c,n+1}I_{a,b,c,n{+}1}$ with corners
\begin{equation}
    \{\{a,\ldots,b{-}1\},\{b,\ldots,c{-}1\},\{c,\ldots,n\},\{n{+1},\ldots,a{-}1\}\}.
\end{equation}
Since $\Delta_{1,b,c,n}$ become rational in terms of momentum-twistors under the collinear limit $Z_{n+1}\to Z_{n}$, it is straightforward to see that no square root remains, and we have
\begin{align}
    &\sum_{\pm}\Res_{\epsilon=0}\int \dif^{2\vert 3}\mathcal{Z}_{n+1} \: f_{1,b,c,n}^{\pm} \nonumber  \\
    &\qquad \qquad =\int \dif\log\tau \:\bar{Q}\log \frac{\langle \bar{n}2\rangle}{\langle \bar{n}(b{-}1\,b)\cap(c{-1}\,c\,n) \rangle}[b{-1}\,b\,c{-1}\,c\,n] 
\end{align}
and 
\begin{equation*}
    \lim_{Z_{n+1}\to Z_{n}} I_{1,b,c,n} = {-}\Li_{2}\biggl(1-\frac{x_{1b}^{2}x_{cn}^{2}}{x_{1c}^{2}x_{bn}^{2}}\biggr) + \frac{1}{2}\Biggl(\log\biggl(\frac{x_{2n}^{2}x_{1\,n{-1}}^{2}x_{bc}^{2}}{x_{2\,n{-1}}^{2}x_{1c}^{2}x_{bn}^{2}}\biggr)+2\log\epsilon\Biggr)\log\biggl(\frac{x_{1b}^{2}x_{cn}^{2}}{x_{1c}^{2}x_{bn}^{2}}\biggr) 
\end{equation*}
Note that the divergence of $\log\epsilon$ and $\tau$-integration will be cancelled in the final answer.

However, for the second kind of boxes, after taking the collinear limit $Z_{n+1}\to Z_{n}$, the square root $\Delta$ remains and we do not have a rational functions of $\tau$. 
To perform the $\tau$-integration, we need rationalizing these $\tau$ integrands first. In other words, we need to find a variable substitution $t(\tau)$ such that $\Delta^{2}$ in terms of $t$ is a perfect square. Since $\Delta^{2}$ is a quadratic polynomial in $\tau$ (after factoring out a perfect-squared denominator), this is just a classical problem to find a rational parameterization of a quadratic curve. For the rational curve defined by
\begin{equation}
y^{2}=x^{2}+ax+b \:, \label{raexp}
\end{equation}
If there is a rational point $(x_{\ast},y_{\ast})$ on this curve\footnote{By a rational points $(x_{\ast},y_{\ast})$ we mean 
$x_{\ast},y_{\ast}\in \mathbb{Q}(a,b)$.}, then we can insert $y=y_{\ast}+t(x-x_{\ast})$ in eq. (\ref{raexp}) to work out the rational parameterization $x(t)$ and hence $y(t)$. For a more comprehensive treatment of rationalizing roots in Feynman integrals, we refer the reader to \cite{Besier:2018jen,Besier:2019kco}. 

For our problem, there are two kinds of obvious rational points, one with $u(\tau_{\ast})=0$ and the other with $v(\tau_{\ast})=0$. In what follows, we will denote these two points as $\tau_{u}$ and $\tau_{v}$. Depending on the values of $\tau_{u}$ and $\tau_{v}$, the second kind of 4-mass boxes can be decomposed into 4 classes further: 
\begin{enumerate}[(i)]
    \item $a>2$ and $c<n-1$: \qquad $
        \tau_{u}=-\frac{x_{cn}^{2}x_{13}^{2}}{x_{1c}^{2}x_{3n}^{2}}\:$ , \qquad $
        \tau_{v}=-\frac{x_{an}^{2}x_{13}^{2}}{x_{1a}^{2}x_{3n}^{2}}$ \:.
     \item $a>2$ and $c=n{-}1$: \qquad \:\:$\tau_{u}=0\:$, \qquad \qquad \:\:\:\:
           $\tau_{v}=-\frac{x_{an}^{2}x_{13}^{2}}{x_{1a}^{2}x_{3n}^{2}}$ \:.
    \item $a=2$ and $c<n{-}1$: \qquad \,\!  $\tau_{u}=-\frac{x_{cn}^{2}x_{13}^{2}}{x_{1c}^{2}x_{3n}^{2}},$ \qquad \:
        $\tau_{v}=\infty$ \:
      \item $a=2$ and $c=n{-1}$:\qquad  \,\: $\tau_{u}=0\:,$ \qquad\qquad\quad \,\! $\tau_{v}=\infty\:.$       
\end{enumerate}
Case (i) is the generic case,  which first appears for one-loop 10-point N$^{2}$MHV. Case (ii) and (iii) are two special cases of Case (i), both of which first appear for 9 points,  and Case (iv) is the most special case which first appears for 8 points. Let us consider Case (i) first:

\subsection*{Case (i): $2<a<a{+1}<b<b{+1}<c<n{-1}$}
The $\dif^{2\vert 3}\mathcal{Z}_{n+1}$ integration for such boxes will introduce two square roots
\begin{align}
    \Delta_{1}=\sqrt{(1-u_{1}-v_{1})^{2}-4u_{1}v_{1}}\:,\qquad 
    \Delta_{n}=\sqrt{(1-u_{n}-v_{n})^{2}-4u_{n}v_{n}} \label{twodel}
\end{align}
where
\begin{align}
    u_{1}=\frac{x_{1 a}^{2} x_{b c}^{2}}{x_{1 b}^{2} x_{a c}^{2}}, \quad v_{1}=\frac{x_{a b}^{2} x_{c 1}^{2}}{x_{1 b}^{2} x_{a c}^{2}}\:,\quad u_{n}=\frac{x_{a b}^{2} x_{c n}^{2}}{x_{a c}^{2} x_{b n}^{2}}, \quad v_{n}=\frac{x_{b c}^{2} x_{n a}^{2}}{x_{a c}^{2} x_{b n}^{2}} \:.
\end{align}
Similarly, we have $z_{1},\bar{z}_{1}$ and $z_{n},\bar{z}_{n}$ as in eq.\eqref{zzbar}.
In terms of these new variables, there are two rational parameterizations
\begin{equation}
    \tau=\tau_{v} \frac{\left(t-u_{1}\left(1-u_{n}-\Delta_{n}\right)\right)\left(t-u_{1}\left(1-u_{n}+\Delta_{n}\right)\right)}{\left(t-v_{n}\left(1-v_{1}-\Delta_{1}\right)\right)\left(t-v_{n}\left(1-v_{1}+\Delta_{1}\right)\right)}  \label{rationpara1}
\end{equation}
and
\begin{equation}
\tau=\tau_{u} \frac{\left(t-v_{1}\left(1-v_{n}-\Delta_{n}\right)\right)\left(t-v_{1}\left(1-v_{n}+\Delta_{n}\right)\right)}{\left(t-u_{n}\left(1-u_{1}-\Delta_{1}\right)\right)\left(t-u_{n}\left(1-u_{1}+\Delta_{1}\right)\right)} 
\label{rationpara2}
\end{equation}
based on the rational points $\tau_{u}$ and $\tau_{v}$, respectively. By using the first parameterization \eqref{rationpara1}, the $\dif^{2\vert 3}\mathcal{Z}_{n+1}$ integration for the sum of $f_{a,b,c,n{+}1}^{\pm}$ gives
\begin{align}
    &\Res_{\epsilon=0}\int\epsilon\dif\epsilon\dif\tau \int\dif\chi_{n+1}^{3}\:\bigl(f_{a,b,c,n{+}1}^{+}+f_{a,b,c,n{+}1}^{-}\bigr) =  \nonumber \\
    &\quad \int\mathrm{d} \log \frac{t-u_{1} v_{n}\left(1+2 \bm{x}_{c}\right)}{t-u_{1} v_{n}\left(1-2 \bm{y}_{c}\right)} \,[a{-}1\, a\, b{-}1\, b\, c] \bar{Q} \log \bm{x}_{c}  - (c\leftrightarrow c{-}1)\nonumber  \\
    &\qquad + \mathrm{d} \log \frac{t-u_{1} v_{n}\left(1+2 \bm{x}_{b}\right)}{t-u_{1} v_{n}\left(1-2 \bm{y}_{b}\right)} \,[a{-}1\, a\, b\, c{-}1\, c] \bar{Q} \log \bm{x}_{b} -(b\leftrightarrow b-1)\nonumber \\
    &\qquad +\mathrm{d} \log \frac{t-u_{1} v_{n}\left(1+2 \bm{x}_{a}\right)}{t-u_{1} v_{n}\left(1-2 \bm{y}_{a}\right)} \, [a\, b{-}1\, b\, c{-}1\, c] \bar{Q} \log \bm{x}_{a} -(a\leftrightarrow a{-}1) \nonumber \\
    &\qquad +\mathrm{d} \log \frac{t-u_{1} v_{n}}{t+u_{1} v_{n}(1-2 \mu)} \bar{Q} \log \frac{\langle\bar{n}\, c{-}1\rangle}{\langle\bar{n} \,c\rangle} \,[a{-}1\, a\, b{-}1\, b\,(c{-}1\, c) \cap(\bar{n})] \nonumber  \\
    &\qquad +\mathrm{d} \log \left(t-u_{1} v_{n} \frac{\mu+\nu}{\mu-\nu}\right) \bar{Q} \log \frac{\langle\bar{n}\, a{-}1\rangle}{\langle\bar{n} \,a\rangle} \,[(a{-}1\, a) \cap(\bar{n})\, b{-}1\, b\, c{-}1\, c] \:, \label{fabcn}
\end{align}
where $(i\leftrightarrow j)$ denote the exchange of particle labels $i$ and $j$ of the foregoing terms, and
\begin{align}
    &\mu=\frac{\langle n(b{-}1\,b)(c{-}1\,c)(n{-}1\,1)\rangle\langle a{-}1\,a\,c{-}1\,c\rangle}{\langle n(a{-}1\,a)(c{-}1\,c)(n{-}1\,1)\rangle\langle b{-}1\,b\,c{-1}\,c\rangle} \:,\nonumber \\
    &\nu=\frac{\langle n(b{-}1\,b)(c{-}1\,c)(n{-}1\,1)\rangle\langle a{-}1\,a\,b{-}1\,b\rangle}{\langle n(a{-}1\,a)(b{-}1\,b)(n{-}1\,1)\rangle\langle b{-}1\,b\,c{-1}\,c\rangle}  \:,\nonumber \\
    & \bm{x}_{a}=\frac{\langle\bar{n} (c{-}1\,c)\cap(a\,b{-}1\,b) \rangle}{\langle \bar{n}\,a\rangle \langle b{-}1\,b\,c{-}1\,c\rangle} \:,\quad \bm{y}_{a}=\frac{\langle a{-}1\,a\,b{-}1\,b\rangle\langle a\,c{-}1\,c\,n \rangle}{\langle a(b{-}1\,b)(c{-}1\,c)(n\,a{-}1)\rangle}\:,  \nonumber \\
    &\bm{x}_{b}=\frac{\langle \bar{n} (c{-}1\,c)\cap (a{-}1\,a\,b)\rangle}{\langle \bar{n} (a{-}1\,a)\cap(b\,c{-}1\,c)\rangle} \:, \quad
    \bm{y}_{b}= \frac{\langle a{-}1\,a\,b{-}1\,b\rangle\langle b\,c{-}1\,c\,n \rangle}{\langle a{-}1\,a\,b\,n\rangle\langle b{-}1\,b\,c{-}1\,c\rangle} \:,\nonumber \\
    &\bm{x}_{c}=\frac{\langle \bar{n}\,c\rangle\langle a{-}1\,a\,b{-}1\,b\rangle}{\langle \bar{n}(a{-}1\,a)\cap(b{-}1\,b\,c)\rangle} \:,\quad 
    \bm{y}_{c}=\frac{\langle c(a{-1}\,a)(b{-}1\,b)(c{-}1\,n)\rangle}{\langle a{-}1\,a\,c\,n\rangle\langle b{-}1\,b\,c{-}1\,c\rangle} \:. \label{xabc}
\end{align}
Note that, the $R$ invariant $[(a{-}1\,a)\cap(\bar{n})\,b{-1}\,b\,c{-1}\,c]$ in eq.\eqref{fabcn} should be understood as the R invariant $[I\,b{-1}\,b\,c{-1}\,c]$ with $I=\mathcal{Z}_{a-1}\langle a\bar{n}\rangle+\mathcal{Z}_{a}\langle\bar{n}\,a{-}1 \rangle$ rather than $I=\mathcal{Z}_{n-1}\langle n\,1\,a{-}1\,a\rangle+\mathcal{Z}_{n}\langle 1\,a{-}1\,a\,n\rangle+\mathcal{Z}_{1}\langle a{-}1\,a\,n{-1}\,n\rangle$, similarly for the $R$ invariant $[a{-}1\, a\, b{-}1\, b\,(c{-}1\, c) \cap(\bar{n})]$. This kind of $R$ invariants can be expressed as the $R$ invariants without any intersection by using of the six-term identity \eqref{6ident} and 
\begin{equation}
    [a_{1} a_{2} a_{3} b_{1}(b_{1} b_{2}) \cap (c_{1} c_{2} c_{3})]=\biggl(1+\frac{\langle b_{1} a_{1} a_{2} a_{3}\rangle\langle b_{2} c_{1} c_{2} c_{3}\rangle}{\langle c_{1} c_{2} c_{3} b_{1}\rangle\langle b_{2} a_{1} a_{2} a_{3}\rangle}\biggr)^{-1} [a_{1} a_{2} a_{3} b_{1} b_{2}] \:. \label{rinvid1}
\end{equation}
The box integral under this collinear limit in terms of $t$ becomes
\begin{equation}
    \lim_{\mathcal{Z}_{n+1}\to\mathcal{Z}_{n}}I_{a,b,c,n+1}= \Li_{2}(\zeta)-\Li_{2}(\bar{\zeta})+\frac{1}{2}
    \log(\zeta\bar{\zeta})\log\frac{1-\zeta}{1-\bar{\zeta}} \label{boxint1}
\end{equation}
where
\begin{equation}
    \zeta=\frac{t-u_{1} v_{n}}{t+u_{1} v_{n}}, \qquad \bar{\zeta}=\left(\frac{\nu}{\mu}\right) \frac{t+u_{1} v_{n}(1-2 \mu)}{t-u_{1} v_{n}(1+2 \nu)} \:.
\end{equation}
Note that although the $\tau$ integrand becomes rational in terms of $t$, the integration region for $t$ is either $[u_{1}(1-u_{n}+\Delta_{n}),v_{n}(1-v_{1}+\Delta_{1})]$ or $[u_{1}(1-u_{n}-\Delta_{n}),v_{n}(1-v_{1}-\Delta_{1})]$ from which these two square roots enter the final result, as in \cite{Zhang:2019vnm}.

For future use, here we also give the result under the second parameterization \eqref{rationpara2}:
\begin{align}
    &\Res_{\epsilon=0}\int\epsilon\dif\epsilon\dif\tau \int\dif\chi_{n+1}^{3}\:\bigl(f_{a,b,c,n{+}1}^{+}+f_{a,b,c,n{+}1}^{-}\bigr) =  \nonumber \\
    &\quad \int \mathrm{d} \log \frac{t-u_{n} v_{1}(1+2 \bm{x}_{c}^{-1})}{t-u_{n} v_{1}(1-2 \bm{y}^{-1}_{c})} \,[a{-}1\, a\, b{-}1\, b\, c] \bar{Q} \log \bm{x}_{c}^{-1}  - (c\leftrightarrow c{-}1)\nonumber  \\
    &\qquad + \mathrm{d} \log \frac{t-u_{n} v_{1}(1+2 \bm{x}_{b}^{-1})}{t-u_{n} v_{1}(1-2 \bm{y}^{-1}_{b})} \,[a{-}1\, a\, b\, c{-}1\, c] \bar{Q} \log \bm{x}_{b}^{-1} -(b\leftrightarrow b-1)\nonumber \\
    &\qquad +\mathrm{d} \log \frac{t-u_{n} v_{1}(1+2 \bm{x}_{a}^{-1})}{t-u_{n} v_{1}(1-2 \bm{y}^{-1}_{a})} \, [a\, b{-}1\, b\, c{-}1\, c] \bar{Q} \log \bm{x}_{a}^{-1} -(a\leftrightarrow a{-}1) \nonumber \\
    &\qquad +\mathrm{d} \log \left(t-u_{n} v_{1} \frac{\tilde{\mu}+\tilde{\nu}}{\tilde{\mu}-\tilde{\nu}}\right)  \bar{Q} \log \frac{\langle\bar{n}\, c{-}1\rangle}{\langle\bar{n} \,c\rangle} \,[a{-}1\, a\, b{-}1\, b\,(c{-}1\, c) \cap(\bar{n})] \nonumber  \\
    &\qquad +\mathrm{d} \log \frac{t-u_{n} v_{1}}{t+u_{n} v_{1}(1-2 \tilde{\mu})}\bar{Q} \log \frac{\langle\bar{n}\, a{-}1\rangle}{\langle\bar{n} \,a\rangle} \,[(a{-}1\, a) \cap(\bar{n})\, b{-}1\, b\, c{-}1\, c] \:, \label{fabcn2}
\end{align}
where
\begin{equation}
\begin{split}
    \tilde{\mu}&=\frac{\langle n(a{-}1\,a)(b{-1}\,b)(n{-1}\,1)\rangle\langle a{-}1\,a\,c{-}1\,c\rangle}{\langle n(a{-1}\,a)(c{-}1\,c)(n{-}1\,1)\rangle\langle a{-1}\,a\,b{-1}\,b\rangle}  \\
    \tilde{\nu} &= \frac{\langle n(a{-}1\,a)(b{-1}\,b)(n{-1}\,1)\rangle\langle b{-}1\,b\,c{-}1\,c\rangle}{\langle n(b{-1}\,b)(c{-}1\,c)(n{-}1\,1)\rangle\langle a{-1}\,a\,b{-1}\,b\rangle} 
\end{split}
\end{equation}
and $\bm{x}_{a},\bm{y}_{a}$ {\it etc}. are the same as in eq.\eqref{xabc}. The box integral under this parameterization has the same form as eq.\eqref{boxint1} but now with   
\begin{equation}
    \zeta=1-\biggl(\frac{\tilde{\nu}}{\tilde{\mu}}\biggr) \frac{t+u_{n} v_{1}(1-2 \tilde{\mu})}{t-u_{n} v_{1}(1+2 \tilde{\nu})}\qquad \bar{\zeta}=1-\frac{t-u_{n}v_{1}}{t+u_{n}v_{1}}\:.
\end{equation}
Now the integration region for $t$ is either $[v_{1}(1-v_{n}-\Delta_{n}),u_{n}(1-u_{1}-\Delta_{1})]$ or $[v_{1}(1-v_{n}+\Delta_{n}),u_{n}(1-u_{1}+\Delta_{1})]$.

\subsection*{Case (ii): $2<a<a{+1}<b<b{+1}<c=n{-1}$}

Unlike the previous case, the $\dif^{2\vert 3}\mathcal{Z}_{n+1}$ integration for these boxes only give the square root $\Delta_{1}$ in eq. \eqref{twodel}, since now 
\begin{equation}
    u_{1}=\frac{x_{1 a}^{2} x_{b\,n-1}^{2}}{x_{1 b}^{2} x_{a \,n-1}^{2}} \:, \quad 
    v_{1}=\frac{x_{a b}^{2} x_{n-1\, 1}^{2}}{x_{1 b}^{2} x_{a \,n-1}^{2}} \:, \quad 
    u_{n}=0\:,\quad 
    v_{n}=\frac{x_{b\,n-1}^{2}x_{an}^{2}}{x_{a\,n-1}^{2}x_{bn}^{2}} \:.
\end{equation}
In this case, the first rational parameterization \eqref{rationpara1} is still available while the second one \eqref{rationpara2} becomes singular. Thus, the $\dif^{2\vert 3}\mathcal{Z}_{n+1}$ integration for the sum of $f_{a,b,n-1,n{+}1}^{\pm}$ gives almost the same result as in \eqref{fabcn} with $c=n{-1}$ except that the terms with $\bar{Q}\log \langle \bar{n}c\rangle$ have to be modified since now $\langle \bar{n}c\rangle =0$. A straightforward calculation shows that
\begin{align}
    &\quad \mathrm{d} \log \frac{t-u_{1} v_{n}\left(1+2 \bm{x}_{c}\right)}{t+u_{1} v_{n}\left(1-2 \bm{y}_{c}\right)} \,[a{-}1\, a\, b{-}1\, b\, c] \bar{Q} \log \frac{\langle \bar{n}\,c\rangle\langle a{-}1\,a\,b{-}1\,b\rangle}{\langle \bar{n}(a{-}1\,a)\cap(b{-}1\,b\,c)\rangle}  \nonumber \\
    &+\mathrm{d} \log \frac{t-u_{1} v_{n}}{t+u_{1} v_{n}(1-2 \mu)} \,[a{-}1\, a\, b{-}1\, b\,(c{-}1\, c) \cap(\bar{n})] 
    \bar{Q}\log \frac{\langle\bar{n}\,c{-}1\rangle}{\langle\bar{n}\,c\rangle}\nonumber \\
    &\xrightarrow[]{\mathcal{Z}_{c}\to\mathcal{Z}_{n-1}}\dif\log\frac{t-u_{1}v_{n}}{t-u_{1}(2-v_{n})} [a{-1}\,a\,b{-1}\,b\,n{-}1] \bar{Q}\log \frac{\langle \bar{n}\,n{-}2\rangle\langle a{-}1\,a\,b{-}1\,b\rangle}{\langle \bar{n}(a{-}1\,a)\cap(b{-}1\,b\,n{-}1)\rangle} \:. \label{modxc}
\end{align}
Note that, the 1-D integral over $t$ for this term now is divergent since the poles of \eqref{modxc} meet with one endpoint of both integration regions. The divergence of this $t$ integration can be easily removed by subtracting  
\begin{equation}
\int \dif\log\frac{\tau+1}{\tau} \: \Li_{2}(1-v_{n})+\frac{1}{2}
\biggl(\log\tau+\log\frac{x_{ab}^{2}x_{1\,n-1}^{2}x_{3n}^{2}}{x_{13}^{2}x_{a\,n-1}^{2}x_{bn}^{2}}\biggr)\log v_{n}\:,
\end{equation}
which is cancelled in the final result.

\subsection*{Case (iii): $2=a<a{+1}<b<b{+1}<c<n{-1}$}

This case is very similar to the Case (ii). The $\dif^{2\vert 3}\mathcal{Z}_{n+1}$ integration for these boxes only give the square root $\Delta_{n}$ in eq.\eqref{twodel} with
\begin{equation}
    u_{1}=0 \:, \quad 
    v_{1}=\frac{x_{2 b}^{2} x_{c1}^{2}}{x_{1 b}^{2} x_{2c}^{2}} \:, \quad 
    u_{n}=\frac{x_{2b}^{2}x_{cn}^{2}}{x_{2c}^{2}x_{bn}^{2}}\:,\quad 
    v_{n}=\frac{x_{bc}^{2}x_{2n}^{2}}{x_{2c}^{2}x_{bn}^{2}} \:.
\end{equation}
Now, the second rational parameterization \eqref{rationpara2} is available while the first one \eqref{rationpara1} is not. Again, the $\dif^{2\vert 3}\mathcal{Z}_{n+1}$ integration for the sum of $f_{2,b,c,n{+}1}^{\pm}$ gives almost the same result as in \eqref{fabcn2} with $a=2$ except that the terms with $\bar{Q}\log \langle \bar{n}\,a{-}1\rangle$ have to be modified since now $\langle \bar{n}\,a{-}1\rangle =0$. A straightforward calculation shows that
\begin{align}
    &\quad \mathrm{d} \log \frac{t-u_{n} v_{1}(1+2 \bm{x}_{a-1}^{-1})}{t-u_{n} v_{1}(1-2 \bm{y}_{a-1}^{-1})} \,[1\, b{-}1\, b\, c{-}1\,c] \bar{Q} \log\frac{\langle\bar{n} (c{-}1\,c)\cap(a{-}1\,b{-}1\,b) \rangle}{\langle \bar{n}\,a{-}1\rangle \langle b{-}1\,b\,c{-}1\,c\rangle}  \nonumber \\
    &+\mathrm{d} \log \frac{t-u_{n} v_{1}}{t+u_{n} v_{1}(1-2 \tilde{\mu})}\,[(a{-}1, a) \cap(\bar{n})\, b{-}1\, b\, c{-}1\, c] \bar{Q} \log \frac{\langle\bar{n}\, a{-}1\rangle}{\langle\bar{n} \,a\rangle}  \nonumber \\
    &\xrightarrow[]{\mathcal{Z}_{a-1}\to\mathcal{Z}_{1}}\dif\log\frac{t-u_{n}(2-v_{1})}{t-u_{n}v_{1}} [1\,b{-1}\,b\,c{-}1\,c] \bar{Q}\log \frac{\langle \bar{n}\,2\rangle\langle b{-}1\,b\,c{-}1\,c\rangle}{\langle \bar{n}(c{-}1\,c)\cap(1\,b{-}1\,b)\rangle} \:. \label{modxa}
\end{align}
Again, the 1-D integral over $t$ for this term now is divergent since the poles of \eqref{modxa} meet with one endpoint of both integration regions. The divergence of this $t$ integration can be easily removed by subtracting  
\begin{equation}
-\int \dif\log(\tau+1)\: \Li_{2}(1-v_{1})+\frac{1}{2}
\biggl(\log\frac{x_{13}^{2}x_{2n}^{2}x_{bc}^{2}}{x_{2c}^{2}x_{1b}^{2}x_{3n}^{2}}-\log\tau\biggr)\log v_{1}\:,
\end{equation}
which is cancelled in the final result.

\subsection*{Case (vi): $2=a<a{+1}<b<b{+1}<c=n{-1}$}
We remark that in fact the last case does not introduce any square root since 
\begin{equation}
    u_{1}=0 \:, \quad 
    v_{1}=\frac{x_{2 b}^{2} x_{n-1\,1}^{2}}{x_{1 b}^{2} x_{2\,n-1}^{2}} \:, \quad 
    u_{n}=0\:,\quad 
    v_{n}=\frac{x_{b\,n-1}^{2}x_{2n}^{2}}{x_{2\,n-1}^{2}x_{bn}^{2}} \:,
\end{equation}
and we do not need it for the algebraic part. However, for completeness, we present the result for this case in Appendix \ref{appc}.

Having rationalized the $\tau$ integrand, one can easily perform the $\tau$ integration for the above four-mass boxes and obtain generalized polylogarithms of weight 3. This is done by using, say {\tt PolylogTools}~\cite{Duhr:2019tlz} or {\tt HyperInt}~\cite{Panzer:2014caa}, or the algorithm provided in the Appendix A of~\cite{CaronHuot:2011kk} if one only needs the symbol. 


\section{Algebraic letters and words for two-loop NMHV amplitudes} \label{algebraicletters}

\subsection{Algebraic letters and their multiplicative relations}

The full computation including contributions from lower-mass boxes becomes tedious for large $n$, and for now we will be interested in a part of the answer depending on algebraic letters, which turns out to be quite neat. Before spelling out the general result for this part, we present all the algebraic letters appeared in the result. We write them in the form
\[
\frac{a+\Delta}{a-\Delta}    
\]
where $a$ is a rational function of Pl\"{u}cker coordinates, and $\Delta$ is a square root for one of the four-mass boxes. The nice thing about this representation is that the multiplicative relations of algebraic letters do not involve rational ones, as shown in Appendix \ref{appb}. In terms of notations we introduced in {\bf{Case(i)}}, we find the following algebraic letters from the integral of $\int \dif^{2\vert 3}\mathcal{Z}_{n+1}\sum_{\pm}f^{\pm}_{a,b,c,n+1}I_{a,b,c,n+1}$:
\begin{equation}
    \frac{z_{1}}{\bar{z}_{1}}\:,\qquad \frac{1-z_{1}}{1-\bar{z}_{1}}\:,\qquad 
    \frac{z_{n}}{\bar{z}_{n}}\:,\qquad \frac{1-z_{n}}{1-\bar{z}_{n}}\:,
\end{equation}
which are letters consisting of the symbol of the 4-mass box integral \eqref{boxintegral}, and {{\it new}} symbol letters of the form (with $\ast$ denotes possible superscripts)
\begin{equation}
    \frac{(\bm{x}^{\ast}_{a,b,c,n}+1)^{-1}-z_{1,a,b,c}}{(\bm{x}^{\ast}_{a,b,c,n}+1)^{-1}-\bar{z}_{1,a,b,c}}\:,\qquad
    \frac{(\bm{x}_{a,b,c,n}^{\ast}+1)^{-1}-\bar{z}_{n,a,b,c}}{(\bm{x}_{a,b,c,n}^{\ast}+1)^{-1}-z_{n,a,b,c}}  \label{newalgebraicletter}
\end{equation}
where $\bm{x}^{\ast}_{a,b,c,n}$ are simply $\bm{x}_{a},\bm{x}_{b},\bm{x}_{c}$ defined in \eqref{xabc} as well as $\bm{x}_{a-1},\bm{x}_{b-1},\bm{x}_{c-1}$ differing by exchanges of particle labels. Here we restore the subscript to indicate the specific boxes. Since {\bf{Case (ii)}} and {\bf{Case (iii)}} can be viewed as degenerations of {\bf{Case (i)}}, the new algebraic letters produced by them have the same form as in \eqref{newalgebraicletter}.

All new algebraic letters are generated by cyclic rotations of eq.\eqref{newalgebraicletter}. Collect all new algebraic letters, we find they can be filled into two classes:
\begin{equation}\label{defiofchi}
    \mathcal{X}_{a,b,c,d}^{\ast}:=\frac{(\bm{x}^{\ast}_{a,b,c,d}+1)^{-1}-\bar{z}_{d,a,b,c}}{(\bm{x}^{\ast}_{a,b,c,d}+1)^{-1}-z_{d,a,b,c}}\:, \qquad     
    \widetilde{\mathcal{X}}_{a,b,c,d}^{\ast}:=\frac{(\bm{x}^{\ast}_{a,b,c,d-1}+1)^{-1}-z
    _{d,a,b,c}}{(\bm{x}^{\ast}_{a,b,c,d-1}+1)^{-1}-\bar{z}_{d,a,b,c}}
\end{equation}
with 6 choices $a{-}1,a,b{-}1,b,c{-}1,c$ of the superscript $\ast$, where
\begin{align}
    & \bm{x}^{a}_{a,b,c,d}=\frac{\langle\overline{d} (c{-}1\,c)\cap(a\,b{-}1\,b) \rangle}{\langle \overline{d}\,a\rangle \langle b{-}1\,b\,c{-}1\,c\rangle} \:, \nonumber \\
    &\bm{x}_{a,b,c,d}^{b}=\frac{\langle \overline{d} (c{-}1\,c)\cap (a{-}1\,a\,b)\rangle}{\langle \overline{d} (a{-}1\,a)\cap(b\,c{-}1\,c)\rangle} \:, 
  \label{xabcd1}\\
    &\bm{x}_{a,b,c,d}^{c}=\frac{\langle \overline{d}\,c\rangle\langle a{-}1\,a\,b{-}1\,b\rangle}{\langle \overline{d}(a{-}1\,a)\cap(b{-}1\,b\,c)\rangle} \:,  \nonumber  
\end{align}
and $\bm{x}_{a,b,c,d}^{a-1},\bm{x}_{a,b,c,d}^{b-1},\bm{x}_{a,b,c,d}^{c-1}$ differing by exchanges of particle labels $a\leftrightarrow a{-}1$ {\it etc.}. Note that $\mathcal{X}_{a,b,c,d}^{\ast}$, $\mathcal{X}_{b,c,d,a}^{\ast}$, $\mathcal{X}_{c,d,a,b}^{\ast}$ and $\mathcal{X}_{d,a,b,c}^{\ast}$ involving the same square root $\Delta_{a,b,c,d}$, and so do $\widetilde{\mathcal{X}}$'s. Naively, there are $4\times 2\times 6+2=50$ algebraic letters involving the square root $\Delta_{a,b,c,d}$. However, some new algebra letters reduce to $z/\bar{z}$ or $(1-z)/(1-\bar{z})$ when some mass corners only contain 2 particles, for instance, 
\[
\mathcal{X}_{d+2,b,c,d}^{d+1}=\frac{\bar{z}_{d,d+2,b,c}}{z_{d,d+2,b,c}}\:,\qquad 
\widetilde{\mathcal{X}}_{a,b,d-2,c}^{d-2}=\frac{1-z_{d,a,b,d-2}}{1-\bar{z}_{d,a,b,d-2}} \:.
\]
It is straightforward to show that there are $50-2m$ algebraic letters involve the same square root if the corresponding four-mass box has $0\leq m\leq 4$ corners that contain only two particles. Note that $m$ also signifies the number of momentum twistors involved in such a square root. For the most generic case with $m=0$, it involves at least $12$ momentum twistors, $a{-}1, a, a{+}1$, $\cdots$, $d{-}1, d, d{+}1$. For degenerate cases with $m>0$, some particles become coincide, and the most degenerate case with $m{=}4$ we have only $8$ momentum twistors. These $\mathcal{X}$'s and $\widetilde{\mathcal{X}}$'s, together with $z/\bar{z}$ and $(1-z)/(1-\bar{z})$ give a set of algebraic letters whose logarithms are invariant up to a sign under the cyclic rotation $i\to i{+1}$ and the reflection $i\to n{-}i{+}1$. 
Algebraic letters involving different $\Delta$'s are manifestly multiplicatively independent, while algebraic letters involving the same $\Delta$ are \emph{not}. A rather remarkable observation we have is that there are precisely $33$ multiplicative relations among them. In the most general form, these relations read:
\begin{gather}
\frac{\mathcal{X}_{a,b,c,d}^{a-1}}{\mathcal{X}_{a,b,c,d}^{a}} =\frac{\mathcal{X}_{d,a,b,c}^{a}}{\mathcal{X}_{d,a,b,c}^{a-1}}\:,\quad \frac{\mathcal{X}_{d,a,b,c}^{a-1}}{\mathcal{X}_{d,a,b,c}^{a}} =\frac{\mathcal{X}_{c,d,a,b}^{a}}{\mathcal{X}_{c,d,a,b}^{a-1}}\:, \nonumber \\ 
\frac{\widetilde{\mathcal{X}}_{a,b,c,d}^{a-1}}{\widetilde{\mathcal{X}}_{a,b,c,d}^{a}} =\frac{\widetilde{\mathcal{X}}_{d,a,b,c}^{a}}{\widetilde{\mathcal{X}}_{d,a,b,c}^{a-1}}\:,\quad \frac{\widetilde{\mathcal{X}}_{d,a,b,c}^{a-1}}{\widetilde{\mathcal{X}}_{d,a,b,c}^{a}} =\frac{\widetilde{\mathcal{X}}_{c,d,a,b}^{a}}{\widetilde{\mathcal{X}}_{c,d,a,b}^{a-1}}  \:, \label{productrealtion1}\\
\frac{\mathcal{X}_{a,b,c,d}^{a-1}}{\mathcal{X}_{a,b,c,d}^{a}}=\frac{\widetilde{\mathcal{X}}_{c,d,a,b}^{a}}{\widetilde{\mathcal{X}}_{c,d,a,b}^{a-1}} \:,\quad \frac{\mathcal{X}_{a,b,c,d}^{a}}{\mathcal{X}_{a,b,c,d}^{b}}=\frac{\widetilde{\mathcal{X}}_{a,b,c,d}^{b}}{\widetilde{\mathcal{X}}_{a,b,c,d}^{a}}\:,\qquad 
\frac{\mathcal{X}_{a,b,c,d}^{b}}{\mathcal{X}_{a,b,c,d}^{c}}=\frac{\widetilde{\mathcal{X}}_{a,b,c,d}^{c}}{\widetilde{\mathcal{X}}_{a,b,c,d}^{b}} \nonumber
\end{gather}
and 21 images under the rotations of $a\to b\to c\to d\to a$, as well as 
\begin{align}
    \frac{\mathcal{X}_{a,b,c,d}^{a}\mathcal{X}_{b,c,d,a}^{d}\mathcal{X}_{c,d,a,b}^{d}\mathcal{X}_{d,a,b,c}^{a}}{\mathcal{X}_{a,b,c,d}^{b}\mathcal{X}_{b,c,d,a}^{c}\mathcal{X}_{c,d,a,b}^{c}\mathcal{X}_{d,a,b,c}^{b}} &=1 \:, \nonumber \\
    \frac{\mathcal{X}_{a,b,c,d}^{a}\mathcal{X}_{b,c,d,a}^{d}\mathcal{X}_{d,a,b,c}^{a}}{\mathcal{X}_{a,b,c,d}^{c}\mathcal{X}_{b,c,d,a}^{c}\mathcal{X}_{d,a,b,c}^{d}} &=1 \:, \label{productrelation2}\\
    \frac{\mathcal{X}_{b,c,d,a}^{b}\mathcal{X}_{c,d,a,b}^{a}\mathcal{X}_{d,a,b,c}^{a}}{\mathcal{X}_{b,c,d,a}^{c}\mathcal{X}_{c,d,a,d}^{c}\mathcal{X}_{d,a,b,c}^{b}} &=1 \:, \nonumber
\end{align}
and 
\begin{equation}\label{productrelation3}
\begin{split}
    \frac{\mathcal{X}_{c,d,a,b}^{a}\mathcal{X}_{d,a,b,c}^{a}}{\mathcal{X}_{c,d,a,b}^{d}\mathcal{X}_{d,a,b,c}^{d}} &=\frac{z_{a,b,c,d}}{\bar{z}_{a,b,c,d}} \:, \\
    \frac{\mathcal{X}_{b,c,d,a}^{c}\mathcal{X}_{c,d,a,b}^{c}}{\mathcal{X}_{b,c,d,a}^{d}\mathcal{X}_{c,d,a,b}^{d}} &= \frac{1-z_{a,b,c,d}}{1-\bar{z}_{a,b,c,d}} \:.
\end{split}
\end{equation}
Note that eqs.\eqref{productrelation2} and \eqref{productrelation3} can also be written in terms of $\widetilde{\mathcal{X}}$'s by using eq.\eqref{productrealtion1}. These relations leave us $17-2m$  multiplicatively independent algebraic letters for where $\Delta$ involves $m$ corners that contain only two particles.

Let us complete this subsection by commenting on the consistency of these algebraic letters with rational letters. These algebraic letters can be rewritten in terms of $(a\pm\sqrt{a^{2}-4b})/2$. The discriminants $a^{2}-4b$ are always proportional to $\Delta^{2}$ in eq.\eqref{uvdel}, which are the square-root branch points from the Landau analysis~\cite{Prlina:2017tvx}, while the branch points $b=0$ correspond to zero locus of some rational letters since $\log(a-\sqrt{a^{2}-4b})= \log b + O(b)$. It is straightforward to show that:
\begin{equation}
    \begin{aligned}
    \bigl|(\bm{x}^{c}_{a,b,c,d-1}+1)^{-1}-z
    _{d,a,b,c}\bigr|^{2}&\propto \langle c(A)(B)(D_{-})\rangle\,\langle(A) \cap(\overline{d{-}1}) B(C) \cap(\overline{d{-}1})\rangle\langle A B\rangle \:, \\
    \bigl|(\bm{x}^{c}_{a,b,c,d}+1)^{-1}-\bar{z}_{d,a,b,c}\bigr|^{2}&\propto\langle c(A)(B)(D_{+})\rangle\langle(A) \cap(\overline{d}) B(C) \cap(\overline{d})\rangle\langle A B\rangle \:,\\
    \bigl|(\bm{x}^{b}_{a,b,c,d-1}+1)^{-1}-z
    _{d,a,b,c}\bigr|^{2} &\propto \langle b(A)(C)(D_{-})\rangle\langle(A) \cap(\overline{d{-}1}) B(C) \cap(\overline{d{-}1})\rangle \:,\\
    \bigl|(\bm{x}^{b}_{a,b,c,d}+1)^{-1}-\bar{z}_{d,a,b,c}\bigr|^{2}&\propto \langle b(A)(C)(D_{+})\rangle\langle(A) \cap(\overline{d}) B(C) \cap(\overline{d})\rangle \:, \\
    \bigl|(\bm{x}^{a}_{a,b,c,d-1}+1)^{-1}-\bar{z}_{d,a,b,c}\bigr|^{2}&\propto \langle a(B)(C)(D_{-})\rangle\langle(A) \cap(\overline{d{-}1}) B(C) \cap(\overline{d{-}1})\rangle\langle B C\rangle \:, \\
    \bigl|(\bm{x}^{a}_{a,b,c,d}+1)^{-1}-\bar{z}_{d,a,b,c}\bigr|^{2}&\propto \langle a(B)(C)(D_{+})\rangle\langle(A) \cap(\overline{d}) B(C) \cap(\overline{d})\rangle\langle B C\rangle \:,
    \end{aligned} \label{algebraiclandu}
\end{equation}
where $\lvert a-z\rvert^{2}$ stands for $(a-z)(a-\bar{z})$, and we introduce 5 lines $A=(a{-1}\,a)$, $B=(b{-1}\,b)$, $C=(c{-}1\,c)$, $D_{-}=(d{-}2\,d{-}1)$ as well as $D_{+}=(d{-}1\,d)$. The expressions for $\vert(
\bm{x}^{a-1}_{a,b,c,d}+1)-z_{d,a,b,c}\vert^{2}$ {\it etc.} can be easily obtained by the exchanges of $a{-1}$ and $a$, {\it etc.}.  The consistent alphabets for two-loop NMHV amplitudes thus require the appearance of factors on right-hand side of eq.\eqref{algebraiclandu}. 
As we will see in the next section, these rational letters indeed appear the alphabet for the 2-loop 9-point NMHV amplitudes.

\subsection{Algebraic words of the symbol and a large class of simple components}
 
Once the algebraic letters are expressed in terms of $(a+\Delta)/(a-\Delta)$, we can separate the words involving algebraic letters from the symbol unambiguously, and we call such words {\it algebraic words}.
As noted in~\cite{Zhang:2019vnm} for $n=8$, these algebraic words, although \emph{not} integrable, follow a very simple pattern where the first two entries consist of the symbol of the four-mass box integral $I_{a,b,c,d}$,
\begin{equation}
    \mathcal{S}(I_{a,b,c,d})= \frac{1}{2}\biggl( u\otimes \frac{1-\bar{z}}{1-z}+v\otimes \frac{z}{\bar{z}}\biggr) \label{boxsymbol} \:,
\end{equation}
while the third entry would be an arbitrary algebraic letter in the alphabet. Our calculation shows that not only this is true for all $n$, but there is a much \emph{stronger} result, which we present now. 
In fact, the final entries and the accompanied $R$ invariants are also completely fixed after knowing the first three entries of the algebraic words, as indicated in eqs.\eqref{fabcn} and \eqref{fabcn2}. In summary, when the third entry is non-degenerate $\mathcal{X}_{a,b,c,d}^{\ast}$'s, the algebraic words become extremely simple:
\begin{align}
    &\quad \mathcal{S}(I_{a,b,c,d})\otimes \mathcal{X}_{a,b,c,d}^{c-1}\otimes \bm{x}_{a,b,c,d}^{c-1} \,[a{-}1\,a\,b{-1}\,b\,c-1] \nonumber \\
    &-\mathcal{S}(I_{a,b,c,d})\otimes \mathcal{X}_{a,b,c,d}^{c}\otimes \bm{x}_{a,b,c,d}^{c} \,[a{-}1\,a\,b{-1}\,b\,c] \nonumber \\
    &+ \mathcal{S}(I_{a,b,c,d})\otimes \mathcal{X}_{a,b,c,d}^{b-1}\otimes \bm{x}_{a,b,c,d}^{b-1} \,[a{-}1\,a\,b{-1}\,c-1\,c] \nonumber \\
    &-\mathcal{S}(I_{a,b,c,d})\otimes \mathcal{X}_{a,b,c,d}^{b}\otimes \bm{x}_{a,b,c,d}^{b} \,[a{-}1\,a\,b\,c{-1}\,c] \nonumber \\
    &+\mathcal{S}(I_{a,b,c,d})\otimes \mathcal{X}_{a,b,c,d}^{a-1}\otimes \bm{x}_{a,b,c,d}^{a-1} \,[a{-}1\,b{-1}\,b\,c-1\,c] \nonumber \\
    &-\mathcal{S}(I_{a,b,c,d})\otimes \mathcal{X}_{a,b,c,d}^{a}\otimes \bm{x}_{a,b,c,d}^{a} \,[a\,b{-1}\,b\,c{-}1\,c]  \:,\label{Xalgebraicword}
\end{align}
and likewise for $\widetilde{\mathcal{X}}$'s. We see that the $\bm{x}^{\ast}_{a,b,c,d}$ variables, which we have used to define $\mathcal{X}$ and $\tilde{\mathcal{X}}$'s, exactly appear as the last entries for the corresponding third entries, and they also determine the accompanying $R$ invariants.

When the third entry is $z/\bar{z}$ or $(1-z)/(1-\bar{z})$, one can directly show from eq.\eqref{fabcn} or eq.\eqref{fabcn2} that, for general $a,b,c,d$ which are non-adjacent, the algebraic words take the form
\begin{align} \label{algwordsz}
    & \mathcal{S}(I_{a,b,c,d}) \otimes \frac{\bar{z}_{a,b,c,d}}{z_{a,b,c,d}}  \otimes \biggl(\frac{\langle \overline{d{-}1} \,c{-}1\rangle\langle \overline{d}\, c\rangle}{\langle \overline{d{-}1} \,c\rangle\langle \overline{d}\, c{-}1\rangle} \,[a{-}1\,a\,b{-}1\,b\,c] \nonumber \\
        &+\frac{\langle\overline{d{-}1}\,(c{-}1\,c)\cap(a{-}1\,a\,b{-}1)\rangle \langle\overline{d}\, c{-}1\rangle}{\langle\overline{d}\,(c{-}1\,c)\cap(a{-}1\,a\,b{-}1)\rangle\langle\overline{d{-}1} \,c{-1}\rangle} [a{-}1\,a\,b{-}1\,c{-}1\,c] - (b{-}1\leftrightarrow b) \nonumber \\
        &+\frac{\langle\overline{d{-}1}\,(c{-}1\,c)\cap(a{-}1\,b{-}1\,b)\rangle \langle\overline{d}\, c{-}1\rangle}{\langle\overline{d}\,(c{-}1\,c)\cap(a{-}1\,b{-}1\,b)\rangle\langle\overline{d{-}1} \,c{-}1\rangle} [a{-}1\,b{-}1\,b\,c{-}1\,c] - (a{-}1\leftrightarrow a) \biggr)
\end{align}
and
\begin{align}\label{algwords1-z}
    &    \mathcal{S}(I_{a,b,c,d})\otimes \frac{1-z_{a,b,c,d}}{1-\bar{z}_{a,b,c,d}} \otimes \biggl( \frac{\langle\overline{d{-1}}\,a{-}1 \rangle\langle\overline{d}\,a \rangle}{\langle\overline{d{-1}}\,a \rangle\langle\overline{d}\,a{-}1\rangle} \, [a{-}1\,b{-}1\,b\,c{-}1\,c] \nonumber \\
    &+\frac{\langle \overline{d{-}1} (a{-1}\,a)\cap(b{-}1\,c{-1}\,c) \rangle\langle\overline{d}\, a\rangle}{\langle \overline{d}\, (a{-1}\,a)\cap(b{-}1\,c{-1}\,c) \rangle\langle \overline{d{-}1} \,a\rangle} [a{-}1\,a\,b{-1}\,c{-1}\,c] - (b{-}1\leftrightarrow b) \nonumber \\
    &+\frac{\langle \overline{d{-}1} (a{-1}\,a)\cap(b{-}1\,b\,c{-1}) \rangle\langle\overline{d} \,a\rangle}{\langle \overline{d}\, (a{-1}\,a)\cap(b{-}1\,b\,c{-1}) \rangle\langle \overline{d{-}1} \,a\rangle} [a{-}1\,a\,b{-1}\,b\,c{-1}] - (c{-}1\leftrightarrow c) \biggr)
\end{align}
as well as their cyclic images under the rotation $a\to b\to c\to d \to a$. Again, the final entries have to be modified as in eqs.\eqref{modxc} and \eqref{modxa} when $c=d{-}2$ or $a=d{+}2$. More precisely, the final entries in the first lines of \eqref{algwordsz} and \eqref{algwords1-z} are modified by
\[
    \frac{\langle \overline{d{-}1} \,c{-}1\rangle\langle \overline{d}\, c\rangle}{\langle \overline{d{-}1} \,c\rangle\langle \overline{d}\, c{-}1\rangle} \xrightarrow{c\to d-2}  \frac{\langle \overline{d{-}1}\,d{-}3\rangle\langle \overline{d}\,d{-}2\rangle\langle a{-}1\,a\,b{-}1\,b\rangle}{\langle \overline{d}\,d{-}3\rangle\langle d{-}2\,(a{-1}\,a)(b{-1}\,b)(d{-1}\,d)\rangle} 
\]
and
\[
    \frac{\langle \overline{d{-}1} \,a{-}1\rangle\langle \overline{d}\, a\rangle}{\langle \overline{d{-}1} \,a\rangle\langle \overline{d}\, a{-}1\rangle} \xrightarrow{a\to d+2} \frac{\langle \overline{d{-}1}\,d{+}1\rangle\langle\overline{d}\,d{+}2\rangle\langle b{-}1\,b\,c{-}1\,c\rangle}{\langle\overline{d{-}1}\,d{+2}\rangle\langle d{+}1\,(b{-}1\,b)\,(c{-}1\,c)(d{-}1\,d) \rangle}  \:,
\]
respectively. This concludes our result for the algebraic words of two-loop $n$-point NMHV. 

Finally, let us remark on an obvious but interesting corollary from the pattern of algebraic words. Let's consider the  $\chi_{i}\chi_{j}\chi_{k}\chi_{l}$ components with non-adjacent $i,j,k,l$ of the two-loop NMHV amplitudes (recall the MHV tree amplitudes are stripped off), or Wilson loops. Given that in the algebraic words, all $R$ invariants always contain two pairs of adjacent particles, {\it i.e.} $[a, a{+}1, b, b{+}1, c]$, no such components can be extracted, thus any such component is simply free of square roots! Note that when $n$ is large enough, we have ${\cal O}(n^4)$ such component, which are the majority of all NMHV components. 

Qualitatively we do expect these to be the simplest components of NMHV amplitudes, since they not only vanish at tree and one-loop level (which means they are finite at two loops), but each of them can be written as a combination of only two integrals! These facts are clear in the representation of NMHV amplitudes (up to two loops) in~\cite{ArkaniHamed:2010gh}, but even more invariantly follow from the super-Wilson-loop picture~\cite{CaronHuot:2010ek}. As noted in~\cite{ArkaniHamed:2010gh}, it is straightforward to show that the component $\chi_i \chi_j \chi_k \chi_l$ is simply given by the difference of double-pentagon integral $I_{\rm{dp}}(i,j,k,l)$ and its cyclic rotation $I_{\rm{dp}}(l,i,j,k)$:
\begin{equation}
    \nonijkl-\nonjkli
\end{equation}
and we record the definition of non-adjacent double pentagon integral $I_{\rm{dp}}(i,j,k,l)$:
\begin{align}
 \int \frac{\dif^4 \ell_1 \dif^4 \ell_2~\langle \ell_1 \bar{i}\cap \bar{j}\rangle \langle \ell_2 \bar{k}\cap \bar{l} \rangle \langle i j k l\rangle}{\langle \ell_1\, i{-}1 i\rangle \langle \ell_1 \,i i{+}1\rangle \langle \ell_1\, j{-}1 j\rangle \langle \ell_1 \,j j{+}1 \rangle \langle \ell_1 \ell_2 \rangle \langle \ell_2\, k{-}1 k\rangle \langle \ell_2\, k k{+}1\rangle\langle \ell_2\, l{-}1 l\rangle \langle \ell_2\, l l{+}1\rangle}\nonumber
\end{align}
where $\ell_1$ and $\ell_2$ denote the two bi-twistors for the loop momenta. 

Remarkably, in terms of these integrals, what we find, rather indirectly through two-loop NMHV amplitudes, is that for any non-adjacent $i,j,k,l$, this difference is free of algebraic letters! We have also obtained the complete symbol of the differences for $n=8,9$, which depends on relatively small number of rational letters, and we expect the simplicity continues to all $n$ (note the difference depends on at most $12$ twistors).

Of course, these integrals themselves are important ingredients of two-loop amplitudes and it would be fantastic to study them individually. The symbol of such integrals are currently unknown, and individual integral does contain algebraic letters involving $\Delta$'s, as shown in~\cite{Bourjaily:2019igt} by evaluating it at a specific kinematic point. We leave the comprehensive study of both the differences and the integrals themselves to the future.

\subsection{Comments on algebraic letters from leading singularities}

Following~\cite{Mago:2020kmp,He:2020uhb}, we make a simple observation that all our algebraic symbol letters for two-loop $n$-point NMHV amplitudes, \eqref{defiofchi}, are ``letters'', or simply singularities, of one-loop leading singularities for the four-mass boxes. As reviewed in sec.~\ref{colintegralon4massbox}, these are leading singularities (LS) gluing together four tree amplitudes, each with at least $4$ legs. If the number of legs for them are $n_1, n_2, n_3, n_4$ respectively, we have $n=\sum_{i=1}^4 n_i-8$. For details of ``letters'' of leading singularities, or Yangian invariants, please refer to~\cite{He:2020uhb}. For these algebraic functions, we do not need to compute all the ``letters'', and it suffices to list the poles of such leading singularities. 

The simplest examples are $8$-point N${}^2$MHV leading singularities (with $n_i=4$-point MHV amplitudes for $i=1,2,3,4$), which was the primary example in~\cite{Mago:2020kmp, He:2020uhb}. Quite nicely, we find that there are $9$ independent letters associated with such a leading singularity, all containing the square root {\it e.g.} $\Delta_{2,4,6,8}$, and similarly $9$ letters with square root $\Delta_{1,3,5,7}$. We denote such leading singularities as ${\cal L}^{k=2}_{2,4,6,8}$ and ${\cal L}^{k=2}_{1,3,5,7}$. By just using these two leading singularities, we obtain exactly the $18$-dim space of algebraic symbol letters for our $n=8$ case~\footnote{To compare directly to our symbol letters which are in the form of $(a+\Delta)/(a-\Delta)$, we can take the ratio of each pole/letter of leading singularities evaluated at two solutions $\pm$.}. Encouraged by this success, now we move to general 1-loop four-mass leading singularity, which was given in \eqref{4massLS}. The independent algebraic letters/poles of such a leading singularity are given by the $9$ independent ones of ${\cal L}^{k=2}_{a,b,c,d}$ and those from the four tree amplitudes at the corners. 

To be concrete, we focus on a particularly simple sub-class of these leading singularities, where each corner has either MHV or NMHV degree ($k_i=0, 1$ for $i=1,2,3,4$). N${}^2$MHV leading singularities correspond to all $k_i=0$, and now we allow some (or all) of the corners to have $k_i=1$. We start with the case where one corner, say, the first one, has $k_1=1$, and without loss of generality we consider $(a,b,c,d)=(1,4,6,8)$ for $n=9$ ({\it i.e.} $3$ external legs only at the first corner). In this case we have ${\cal A}_{n_{1}=5,k_1=1}=[\alpha, 1,2,3,\beta]$ (and the other three ${\cal A}=1$). Now in addition to the $9$ independent algebraic letters of ${\cal L}_{k=2} (1,4,6,8)$, we have $5$ letters/poles from $[\alpha, 1,2,3, \beta]$:
\begin{equation}
\langle \alpha \beta 1 2 \rangle, \langle \alpha \beta 2 3 \rangle, \langle \alpha \beta 1 3 \rangle, \langle \alpha 1 2 3 \rangle, \langle \beta 1 2 3 \rangle\,.
\end{equation}
Note that $\alpha=(9 1) \cap (8 7 \gamma)$ and $\beta=(3 4) \cap (5 6 \delta)$, thus $ \langle \alpha 1 2 3 \rangle$, $\langle \beta 1 2 3 \rangle$ and $\langle \alpha \beta 1 3 \rangle$ are in fact rational functions of Pl\"{u}cker coordinates, so the only two algebraic letters are $\langle \alpha \beta 1 2 \rangle$ and $\langle \alpha \beta 2 3 \rangle$. It is straightforward to check that they are multiplicatively independent with the $9$ letters above, thus we conclude that there are $9+2=11$ independent algebraic letters for this leading singularity. By taking the ratio of two solutions $\pm$, they span precisely the same space as the $11$ algebraic symbol letters for $n=9$ which are associated with $\Delta_{1,4,6,8}$.  

More generally, it turns out that the complete algebraic alphabet of two-loop $n$-point NMHV, \eqref{defiofchi}, can be obtained from one-loop four-mass leading singularities with $k_i=0, 1$. The correspondence works for each four-mass configuration $(a,b,c,d)$ individually: for the generic case, the $17$ independent symbol letters of two-loop NMHV amplitude with square root of $\Delta_{a,b,c,d}$ can be obtained from a single leading singularities in \eqref{4massLS} with 4 NMHV tree amplitudes, {\it i.e.} $k_i=1$ for $i=1,2,3,4$. One can check that in addition to the $9$ algebraic letters for ${\cal L}^{k=2}_{a,b,c,d}$, each tree amplitudes at least contain two new algebraic letters similar to those in the $n=9$ example.  Altogether this means we can generate the $17$ independent algebraic letters for the generic case, which we first encounter at $n=12$. Note that as we have seen before~\cite{He:2020uhb}, the correspondence between letters from leading singularities and symbol letters does no preserve $k$: for NMHV (two-loop) amplitudes, we need leading singularities with up to $k=6$.

\section{The complete symbol and alphabet: \texorpdfstring{$n=9$}{n=9} example} \label{exampleandcheck}

Now that we have the algebraic part of the symbol, we can finish the calculation by including the rational part which also receive contribution from all lower-mass boxes. This part has been automatized, which produces the complete symbol of two-loop $n$-point NMHV amplitude. However, the length of the symbol (especially the rational data) grows rapidly when $n$ increases, and we content ourselves by presenting the result for $n=9$ as an example.

\subsection{The symbol of two-loop 9-point NMHV amplitude}

The differential of two-loop NMHV amplitude from $\bar Q$ equation can be written as 
\[
\dif R_{9,1}^{(2)}=\sum_\alpha [i_\alpha\,j_\alpha\,k_\alpha\,l_\alpha\,m_\alpha](F_\alpha \dif\log x_\alpha),
\]
where $F_\alpha$ are the weight-$3$ generalized polylogarithms arising from 1-D integrals. Here, we only compute their 
symbol $S(F_{\alpha})$ by using the algorithm in~\cite{CaronHuot:2011kk}.
We record the symbol of $R_{9,1}^{(2)}$,
\[
\mathcal S(R_{9,1}^{(2)})=\sum_\alpha [i_\alpha\,j_\alpha\,k_\alpha\,l_\alpha\,m_\alpha]\bigl(\mathcal{S}(F_\alpha) \otimes x_\alpha\bigr),
\]
explicitly in the supplementary material of this paper as well as the repository \cite{datasite}. The data is organized as follows: we choose a basis of 
$R$ invariants $[i\,j\,k\,l\,9]$ for $1\leq i<j<k<l<9$, and record ${8 \choose 4}=70$ coefficients ${\cal S}_{i,j,k,l}$. 
Then, we find $59\times 9=531$ rational letters in the alphabet, and they are:
\begin{itemize}
\item 13 cyclic classes of $\langle 12kl\rangle$ for $3\leq k<l\leq 8$ but $(k,l)\neq (6,7), (7,8)$;
\item 7 cyclic classes of $\langle 12(ijk)\cap (lmn)\rangle$ for $3\leq i<j<k<l<m<n\leq 9$;
\item 8 cyclic classes of \\
$\langle \bar 2 \cap (245) \cap \bar 6 \cap (691) \rangle,
\langle \bar 2  \cap (346) \cap \bar 6 \cap (892)\rangle,
\langle \bar 2  \cap (346) \cap \bar 6 \cap (782)\rangle,$\\
$\langle \bar 2 \cap (245) \cap \bar 7 \cap (791) \rangle,
\langle \bar 2  \cap (245) \cap (568) \cap \bar 8\rangle,
\langle \bar 2 \cap (245) \cap (569) \cap \bar 9 \rangle,$\\
$\langle \bar 2\cap (245) \cap (679) \cap \bar 9  \rangle,
\langle \bar 2\cap (256) \cap (679) \cap \bar 9 \rangle;$
\item 

10 cyclic classes of 
$\langle 1(i\,i{+}1)(j\,j{+}1)(k\,k{+}1)\rangle$ for $2\leq i<i+1<j$, $j+1<k\leq 8$,


6 cyclic classes of $\langle 1(2i)(j\, j{+}1)(k9)\rangle$ for $3\!\leq\! i\!<\!j$, $j+1\!<\!k\!\leq\! 8$ but $(i,k)\neq (3,8), (4,7)$,

14 cyclic classes of $\langle 1(29)(ij)(k\,k{+}1)\rangle$ for $3\!\leq\! i\!<\!j\!\leq\! 8$, $3\!\leq\! k\!\leq\! i{-}2$ or $j{+}1\!\leq\!k\!\leq\!7$;

\item 1 cyclic class of $\langle 1\, (56)\cap (\bar 3)\, (78)\cap (\bar 3)\, 9\rangle$.
\end{itemize}

A few comments are in order. First of all, by combining these $59\times 9$ rational letters with $11 \times 9$ independent algebraic letters, we have the complete alphabet of $630$ letters for $n=9$. We see that the alphabet is consistent as we have mentioned in the last section: for each algebraic letter of the form $a \pm \sqrt{a^2-4b}$, $b$ is indeed a rational letter. We expect this to hold for all multiplicities. Furthermore, we have found some discrepancies with the predictions from Landau analysis~\cite{Prlina:2017tvx}: not only some rational letters predicted there do not appear in our alphabet, but more importantly exactly the last class, {\it i.e.} cyclic rotations of $\langle 1\, (56)\cap (\bar 3)\, (78)\cap (\bar 3)\, 9\rangle$, are absent in the Landau analysis of~\cite{Prlina:2017tvx}.

\subsection{Consistency checks}

Even before obtaining the final result, the $\bar{Q}$ calculation is very rigid: it cannot be carried through till the end unless various tests have been passed. For example, in the collinear integral, all the $\log \epsilon$ divergence must be accompanied by vanishing $\tau$-integrals; also it is highly non-trivial that we are able to convert the arguments of $\bar{Q}\log $ into DCI combinations. All that being said, to make sure our result is correct, we have performed various consistency checks including easy ones such as cyclicity, dual conformal invariance and the condition of physical first entries. Let's present details for the more non-trivial checks, such as integrability, collinear limits and absence of spurious poles.

In these checks, it's usually difficult to determine whether a symbol with algebraic letters vanishes or not before imposing the multiplicative relations of algebraic letters. We leave this technical problem in the Appendix \ref{appb}.

\paragraph*{Integrability}

It is a non-trivial but crucial check that our symbol is integrable. We expand the symbol of the total differential on a basis of ${n{-}1 \choose 4}$ $R$ invariants, and we check that each coefficient 
can be integrated to a function. 
%
%
These coefficients have the form
\[
	\sum_{\alpha_1,\alpha_2,\alpha_3,\alpha_4}c_{\alpha_1,\alpha_2,\alpha_3,\alpha_4}
	l_{\alpha_1}\otimes l_{\alpha_2}\otimes l_{\alpha_3} \,\,\dif\log l_{\alpha_4},
\]
where symbols in the coefficients of $d\log$ comes from polylogarithms, so
it's integrable if and only if (see \cite{Chen:1977oja,brown2009multiple})
\[
	\sum_{\alpha_1,\alpha_2,\alpha_3,\alpha_4}c_{\alpha_1,\alpha_2,\alpha_3,\alpha_4}
	l_{\alpha_1}\otimes l_{\alpha_2} \,\,
	\dif \log l_{\alpha_3} \wedge \dif\log l_{\alpha_4}=0.
\]

In order to calculate $\dif \log l_i$, we choose a positive parameterization of 
$\mathrm{Gr}_+(4,9)/T$~\cite{speyer2005tropical} which makes all arguments of square roots positive 
\[
\begin{pmatrix} 
-1 & 0 & 0 & 0 & 1 & p_{1,6} & p_{1,7} & p_{1,8} & p_{1,9} \\
 0 & -1 & 0 & 0 & -1 & -p_{2,6} & -p_{2,7} & -p_{2,8} & -p_{2,9} \\
 0 & 0 & -1 & 0 & 1 & p_{3,6} & p_{3,7} & p_{3,8} & p_{3,9} \\
 0 & 0 & 0 & -1 & -1 & -1 & -1 & -1 & -1 
\end{pmatrix},
\]
where we have the polynomials of face variables
\[
	p_{i,j}(x)=\sum_{q:i\to j}\,\, \prod_{\text{face $f$ under the path $q$}}x_f
\]
naturally defined on the following network:
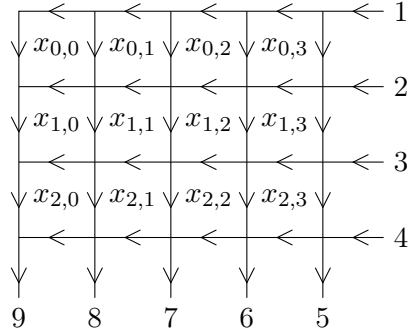
\begin{center}
\begin{tikzpicture}
\draw (3.3,2.5) node[right]{1} -- (-1.5,2.5) -- (-1.5,-1.3)  node[below]{9};
\draw (-0.5,2.5) -- (-0.5,-1.3) node[below]{8};
\draw (0.5,2.5) -- (0.5,-1.3) node[below]{7};
\draw (1.5,2.5) -- (1.5,-1.3) node[below]{6};
\draw (2.5,2.5) -- (2.5,-1.3) node[below]{5};
\draw (3.3,1.5) node[right]{2}-- (-1.5,1.5);
\draw (3.3,0.5) node[right]{3}-- (-1.5,0.5);
\draw (3.3,-0.5) node[right]{4}-- (-1.5,-0.5);

\foreach \x in {0,1,2,3,4} {
\foreach \y in {0,1,2,3}{
	\draw (3.1-\x,2.6-\y) -- (2.9-\x,2.5-\y) -- (3.1-\x,2.4-\y);
}}

\foreach \x in {0,1,2,3,4} {
\foreach \y in {0,1,2,3}{
	\draw (-1.6+\x,2.1-\y) -- (-1.5+\x,1.9-\y) -- (-1.4+\x,2.1-\y);
}}

\foreach \x in {0,1,2,3} {
\foreach \y in {0,1,2}{
	\node at (-1+\x,2-\y) {$x_{\y,\x}$};
}}
\end{tikzpicture}
\end{center}
and the unlabeled face variables are fixed to be $1$. 

We use this positive parametrization to check the integrability of the coefficients
of all linear-independent $R$ invariants, and we find that they are indeed integrable.

\paragraph*{Collinear limits.} 

We check that the NMHV $9$-point amplitude reduces to NMHV and MHV $8$-point amplitude upon 
taking the $k$-preserving and $k$-decreasing collinear limits respectively.  
We consider the limit $9||8$ by sending 
 \begin{equation*}
     Z_9\to Z_8+
       \epsilon \frac{\langle 1258\rangle }{\langle 1257\rangle } Z_7 + 
       \epsilon \tau \frac{\langle 2568\rangle }{\langle 1256\rangle } Z_1 + 
       \eta \frac{\langle 1568\rangle }{\langle 1256\rangle } Z_2,
 \end{equation*}
for fixed $\tau$ then taking the limit $\eta\to 0$ before $\epsilon\to 0$. 
Under the $k$ preserving limit, $R$ invariants behave as  $[a b c 8 9]\to 0$ 
and $[a b c d 9] \to [abcd8]$, while under the $k$ decreasing limit, 
the $R$ invariants behave as $[1a789]\to 1$ with the others vanishing.
After taking such limits and keeping leading terms of $\eta$ and $\epsilon$, 
it is highly non-trivial that the limits do not depend on the parameters $\eta$,
$\epsilon$ and $\tau$, {\it i.e.} it has smooth limits, and then we find that these two limits are exactly the known symbols of NMHV $8$-point amplitude~\cite{Zhang:2019vnm} and MHV $8$-point amplitude~\cite{CaronHuot:2011ky}.

\paragraph*{Cancellation of spurious poles.} 

Finally, we check that the residue on any spurious pole of $R$ invariants vanishes. For instance, $\langle 1235\rangle=0$ is a non-physical pole of $[1 2 3 5 9]$, thus the residue as $\langle 1235\rangle \to 0$ should vanish. Thanks to cyclicity, we only need to check the following poles
 \[
\langle 1235\rangle, \langle 1236\rangle, \langle 1237\rangle, \langle 1238\rangle, 
 \langle 1246\rangle, \langle 1247\rangle, \langle 1248\rangle, \langle 1257\rangle, 
 \langle 1258\rangle, \langle 1268\rangle, \langle 1357\rangle
 \]
each of which is of the form $\langle 1abc\rangle$ and belongs to exactly one $R$ invariant in our basis. The cancellation of the pole $\langle 1abc\rangle$ means that the coefficient of the corresponding $R$ invariant vanishes as $\langle 1abc\rangle \to 0$. To see this, we send $Z_{1} \to \alpha Z_{a}+\beta Z_{b}+\gamma Z_{c} +\delta Z_{9}$
for fixed $\alpha,\beta,\gamma$, and verified numerically that the coefficient of $[1abc9]$ vanishes under the limit of $\delta \to 0$.

\section{Discussions}

In this paper, following the $n=8$ result~\cite{Zhang:2019vnm}, we have systematically studied NMHV amplitudes to all multiplicities based on the recursive method of $\bar{Q}$ equations~\cite{CaronHuot:2011kk}. In addition to the first all-loop results for last-entry conditions of $n$-point NMHV amplitudes~\eqref{lastentry1}--\eqref{lastentry3}, we have focused on the computation of two-loop NMHV amplitudes. The main results we have presented are the symbol and alphabet of the non-trivial, algebraic words, derived using relevant four-mass boxes for one-loop N${}^2$MHV amplitudes. For a generic square root which involves four corners with at least $3$ particles, we find $50$ algebraic letters \eqref{defiofchi} satisfying exactly $33$ multiplicative relations, \eqref{productrealtion1}-\eqref{productrelation3}, thus resulting in $17$ independent algebraic letters (for degenerate cases the number reduces to $17-2m$ with $1\leq m\leq 4$ corners containing $2$ particles). The symbol has a nice pattern where the R-invariant and last-entry are directly correlated with the algebraic letters on the third entry (while the first two entries being the symbol of four-mass boxes). Moreover, we have computed for the first time the complete symbol for $n=9$, and obtained the full alphabet with $59 \times 9$ rational letters, in addition to $11 \times 9$ algebraic ones. Our results have passed various consistency checks, and interestingly the rational letters for $n=9$ raise tensions with Landau analysis though the majority of them are consistent with it. 

One of the motivations here is to extend the $n=8$ alphabet~\cite{Zhang:2019vnm} to higher $n$, namely the algebraic letters for all $n$ and the full alphabet for at least $n=9$. It is straightforward but tedious to compute the full alphabet for higher $n$, which would provide a new family of data points besides $n$-point MHV alphabet~\cite{CaronHuot:2011ky}. It is then highly desirable to ``explain'' such alphabets from certain mathematical structures~\cite{Drummond:2019cxm, Henke:2019hve, Arkani-Hamed:2019rds}. We have provided a simple explanation by listing the letters/poles of one-loop leading singularities with MHV/NMHV corners, and it would be interesting to pursue that direction further. For example, even when restricted to quadratic ones, we find many ``new'' algebraic letters/poles (most of which with new $\Delta$'s) of higher-loop leading singularities already for $n=9,10$, and it would be interesting to see which of them appear as symbol letters. Moreover, the remarkable simplicity of the algebraic words suggests a deeper structures, and it is worth studying properties such as cluster adjacency/extended Steinmann~\cite{Drummond:2017ssj,Caron-Huot:2019bsq,Caron-Huot:2020bkp}, now for the part involving algebraic letters. It is also highly desirable to ``complete'' such algebraic words into integrable ones, which would allow us to write down weight-$4$ functions for the algebraic part. 

Regarding computation of loop amplitudes, the most pressing question is to compute the long-sought-after symbol of three-loop $n=8$ MHV, from our two-loop $n=9$ NMHV results. As is familiar from ${\bar Q}$ computations before, the computation of MHV amplitudes from NMHV ones can be completely automatized, though again we need to rationalize all the square roots as we have done for two-loop NMHV in this paper. This is a tedious but straightforward exercise, and we expect to report the result in the near future~\cite{toapp}, which would add a data point of the alphabet as well as give the ``lost symbol'' for the octagon. Moreover, since the method for rationalizing square roots works for all multiplicities, it is conceivable that one can compute the algebraic words of higher-point three-loop MHV from those of two-loop NMHV as well. 

We have focused on the symbol so far, but it should be possible to obtain polylogarithm functions from our symbol, at least for two-loop NMHV octagons (see recent works on heptagons~\cite{Dixon:2020cnr}). Moreover, a fascinating question is if we can ``bootstrap'' octagons, based on the alphabet, first and last entries, as well as constraints from collinear limits {\it etc.} similar to the hexagon and heptagon bootstrap. A potential issue is how to implement (extended) Steinmann relations in some way, which at least naively do not apply to $n=8$ (or any multiple of $4$) due to the lack of BDS-like normalization~\cite{Alday:2009dv}. If one could resolve that issue, it may be possible to bootstrap to three loops and higher, which would be a strong test on some conjectural alphabet of octagons. A particularly simple example is given by our special class of components which are free of algebraic letters: for example the component $\chi_1 \chi_3 \chi_5 \chi_7$ of the octagon has a simple symbol with only $68$ (out of $180$) rational letters, and it would be interesting to uplift it to a weight-$4$ function (or even directly bootstrap). Given the simple relation of such components to double-pentagon integrals, such results may also shed light into these unknown Feynman integrals. 

It would be interesting to push the limit of our method based on anomaly equations even further. Higher-point three-loop NMHV and four-loop MHV amplitudes can be reached if we have the corresponding two-loop N${}^2$MHV amplitudes. The simplest one is the two-loop N${}^2$MHV octagon, which should be completely fixed by $\bar{Q}$ equations and parity; the point is not only to re-derive three-loop NMHV heptagon and four-loop MHV hexagon from first-principle computations, but also illustrate the structures of the ${\bar Q}$-method further. Of course, to go to even higher $n, k$ and loops, we would need the more general method involving solving both ${\bar Q}$ and $Q^{(1)}$ equations, which are related by parity. We leave the study of the anomaly equations and their applications to higher-loop amplitudes to the future. Finally, it is tempting to ask the following: can we formulate a question based on these anomaly equations, to which the non-perturbative S-matrix of planar ${\cal N}=4$ SYM is the (unique) answer?

\section*{Acknowledgements} We thank N. Arkani-Hamed and J. Trnka for organizing the first meeting of ``Geomplitudes'', and the participants of the meeting for comments on the results we reported. C.Z. is grateful to the Institute of Theoretical Physics, CAS for warm hospitality during this special time. This work is supported in part by Research Program of Frontier Sciences of CAS under Grant No. QYZDBSSW-SYS014 and National Natural Science Foundation of China under Grant No. 11935013.\\

\appendix


\section{The effects of $\int \epsilon\dif \epsilon \dif^{3}\chi_{n+1}$ on all rational N$^{2}$MHV Yangian invariants}\label{appa}

Here we require $1\leq i_{1}<i_{2}<\cdots <i_{10}\leq n{+}1$ and define $X:=n\wedge B$ where $\mathcal{Z}_{B}=\mathcal{Z}_{n-1}-C\tau Z_{1} $. The effects of the operation $\int \epsilon\dif \epsilon \dif^{3}\chi_{n+1}$ on NMHV Yangian invariants are known from \cite{CaronHuot:2011kk}
\begin{align} \label{A1}
    [i_{1}\,i_{2}\,i_{3}\,n\,n{+}1] \to 
    \begin{cases}
        \dif\log\frac{\langle Xi_{1}i_{2}\rangle}{\langle Xi_{2}i_{3}\rangle}\bar{Q}\log\frac{\langle \bar{n}i_{2}\rangle}{\langle \bar{n}i_{1}\rangle} +
        \dif\log\frac{\langle Xi_{2}i_{3}\rangle}{\langle Xi_{1}i_{3}\rangle}\bar{Q}\log\frac{\langle\bar{n}i_{3}\rangle}{\langle\bar{n}i_{1}\rangle} & 1<i_{1} \text{ and } i_{3}<n{-}1 \\
        \dif \log \frac{\langle Xi_{1}i_{2}\rangle}{\langle X\,n{-}2\,n{-}1\rangle}
        \bar{Q}\log\frac{\langle\bar{n}i_{2}\rangle}{\langle\bar{n}i_{1}\rangle}
        & 1<i_{1}\text{ and } i_{3}=n{-1} \\
        \dif \log \frac{\langle Xi_{2}i_{3}\rangle}{\langle X12\rangle}\bar{Q}\log\frac{\langle\bar{n}i_{3}\rangle}{\langle \bar{n}i_{2}\rangle} & 1=i_{1} \text{ and } 
        i_{3}<n{-1} \\
        \dif \log\frac{\langle X\,n{-}2\,n{-}1\rangle}{\langle X\,12\rangle}
        \bar{Q}\log\frac{\langle \bar{n}2\rangle}{\langle \bar{n} i_{2}\rangle} 
        & 1=i_{1} \text{ and } i_{3}=n{-}1
    \end{cases} \:.
\end{align}
There are 14 classes of N$^{2}$MHV Yangian invariants which can be found in~\cite{ArkaniHamed:2012nw}. One of them is the four-mass box Yangian invariants which we have elaborated in the main text. In this Appendix, we give the results of the action of the operation $\int \epsilon\dif \epsilon \dif^{3}\chi_{n+1}$ on 13 classes of rational Yangian invariants. For thoes Yangian invariants unlisted here, this operation gives zero.

\subsection*{1. $[i_{1}\,i_{2}\,(i_{2}\,i_{3})\cap(i_{4}\,n,\,n{+}1)\,(i_{2}\,i_{3}\,i_{4})\cap(n\,n{+}1)\,n{+}1]\,[i_{2}\,i_{3}\,i_{4}\,n\,n{+}1]$ and its cyclic rotations}

\subsubsection*{1.1 $[i_{1}\,i_{2}\,(i_{2}\,i_{3})\cap(i_{4}\,n,\,n{+}1),(i_{2},i_{3},i_{4})\cap(n\,n{+}1)\,n{+}1]\,[i_{2},i_{3},i_{4},n\,n{+}1]$}
\begin{align} \label{A2}
    &[i_{1}\,i_{2}\,(i_{2}\,i_{3})\cap(i_{4}\,n\,n{+1})\,(i_{2}\,i_{3}\,i_{4})\cap(n\,n{+}1)\,n{+1}] [i_{2}\,i_{3}\,i_{4}\,n\,n{+}1] \to \nonumber \\
    &\begin{cases}
        [i_{1}\,i_{2}\,i_{3}\,i_{4}\,n]\begin{pmatrix}
            \bar{Q}\log\frac{\langle n(n{-}1\,1)(i_{1}\,i_{2})(i_{3}\,i_{4})\rangle}{\langle \bar{n} i_{1}\rangle\langle i_{2}i_{3}i_{4}n\rangle}\dif \log\frac{\langle Xi_{1}i_{2}\rangle}{\langle Xi_{3}i_{4}\rangle} \\
            +\bar{Q}\log\frac{\langle \bar{n}i_{1}\rangle}{\langle\bar{n}i_{4}\rangle}\dif\log\frac{\langle Xi_{1}i_{4}\rangle}{\langle Xi_{3}i_{4}\rangle}\end{pmatrix}
            & 1<i_{1} \text{ and  }i_{4}<n{-}1 \\
        [i_{1}\,i_{2}\,i_{3}\,n{-}1\,n]\bar{Q}\log\frac{\langle\bar{n}i_{1}\rangle}{\langle\bar{n}i_{3}\rangle}\dif\log\frac{\tau}{\langle Xi_{1}i_{2}\rangle} & 1<i_{1} \text{ and }i_{4}=n{-1} \\
        [1\,i_{2}\,i_{3}\,i_{4}\,n]\bar{Q}\log\frac{\langle\bar{n}i_{4}\rangle}{\langle\bar{n}i_{2}\rangle}\dif\log\langle Xi_{3}i_{4}\rangle  & i_{1}=1 \text{ and }i_{4}<n{-1} \\
        [1\,i_{2}\,i_{3}\,n{-1}\,n] \bar{Q}\log\frac{\langle\bar{n}2\rangle}{\langle\bar{n}i_{2}\rangle}\dif\log\tau &i_{1}=1 \text{ and }i_{4}=n{-1}
    \end{cases}
\end{align}
\subsubsection*{1.2 $[i_{2}\,i_{3}\,(i_{3}\,i_{4})\cap(n\,n{+1}\,i_{1})\,(i_{3}\,i_{4}\,n)\cap(n{+}1\,i_{1})\,i_{1}]\,
[i_{3}\,i_{4}\,n\,n{+1}\,i_{1}] $}
The effect of $\int \epsilon\dif \epsilon \dif^{3}\chi_{n+1}$ on this Yangian invariant is the same as $[i_{1}(i_{3}i_{4})\cap(i_{1}i_{2}n)i_{4}\,n\,n{+1}] $ $[i_{1}\, i_{2}\, i_{3}\, i_{4}\,n] $, for which we can apply eq.\eqref{A1}.

\subsubsection*{1.3 $[i_{3}\,i_{4}\,(i_{4}\,n)\cap(n{+1}\,i_{1}\,i_{2})\,(i_{4}\,n\,n{+}1)\cap(i_{1}\,i_{2})\,i_{2}]\,[\,i_{4}\,n\,n{+1}\,i_{1}\,i_{2}] $}
The effect of $\int \epsilon\dif \epsilon \dif^{3}\chi_{n+1}$ on this Yangian invariant is the same as $[i_{1}(i_{1}i_{2})\cap(i_{3}i_{4}n)i_{4}\,n\,n{+1}]$ $[i_{1}i_{2}i_{3}i_{4}n]$, for which we can apply eq.\eqref{A1}.

\subsubsection*{1.4 $[i_{4}\,n\,(n\,n{+}1)\cap(i_{1}\,i_{2}\,i_{3})\,(n\,n{+1}\,i_{1})\cap(i_{2}\,i_{3})\,i_{3}] [i_{1}\,i_{2}\,i_{3}\,n\,n{+}1]$}

The effect of $\int \epsilon\dif \epsilon \dif^{3}\chi_{n+1}$ on this Yangian invariant is the same as in \eqref{A2}.

\subsubsection*{1.5 $[n\,n{+1}\,(n{+}1\,i_{1})\cap(i_{2}\,i_{3}\,i_{4})\,(n{+}1\,i_{1}\,i_{2})\cap(i_{3}\,i_{4})\,i_{4}]\,[i_{1}\,i_{2}\,i_{3}\,i_{4}\,n{+}1]$ }

The effect of $\int \epsilon\dif \epsilon \dif^{3}\chi_{n+1}$ on this Yangian invariant is the same as $[(n\,i_{1})\cap(i_{2}\,i_{3}\,i_{4})\,(n\,i_{1}\,i_{2})\cap(i_{3}\,i_{4})\,i_{4}\,n\,n{+}1]\,[i_{1}i_{2}i_{3}i_{4}n]$, for which we can apply eq.\eqref{A1}.

\subsubsection*{1.6 $[n{+1}\,i_{1}\,(i_{1}\,i_{2})\cap(i_{3}\,i_{4}\,n)\,(i_{1}\,i_{2}\,i_{3})\cap(i_{4}\,n)\,n] [i_{1}\,i_{2}\,i_{3}\,i_{4}\,n]$}

Directly apply eq.\eqref{A1}.

\subsection*{2. $[i_{1}\,i_{2}\,(i_{3}\,i_{4})\cap(i_{5}\,n\,n{+}1)\,(i_{3}\,i_{4}\,i_{5})\cap(n\,n{+}1)\,n{+}1]\,[i_{3}\,i_{4}\,i_{5}\,n\,n{+}1]$ and its cyclic rotations}

\subsubsection*{2.1 $[i_{1}\,i_{2}\,(i_{3}\,i_{4})\cap(i_{5}\,n\,n{+}1)\,(i_{3}\,i_{4}\,i_{5})\cap(n\,n{+}1)\,n{+}1]\,[i_{3}\,i_{4}\,i_{5}\,n\,n{+}1]$ }

For generic $i_{1}<i_{2}<i_{3}<i_{4}<i_{5}$ where $i_{1}>1$ and $i_5<n{-}1$, the operation $\int \epsilon\dif \epsilon \dif^{3}\chi_{n+1}$ gives 
\begin{align} \label{A3}
    &\quad\bar{Q}\log\frac{\langle\bar{n} (i_{1}i_{2})\cap(i_{3}i_{4}i_{5})\rangle}{\langle\bar{n}{i_{5}}\rangle\langle i_{1}i_{2}i_{3}i_{4}\rangle} [i_{1}\,i_{2}\,i_{3}\,i_{4}\,i_{5}] \,\dif\log\frac{\langle Xi_{5}(i_{1}i_{2})\cap(i_{3}i_{4}i_{5})\rangle}{\langle X i_{1}i_{2} \rangle} \nonumber \\
   & +\bar{Q}\log\frac{\langle\bar{n}i_{1}\rangle}{\langle\bar{n}i_{5}\rangle} [i_{1}\,i_{3}\,i_{4}\,i_{5}\,n]\, \dif\log\frac{\langle X i_{1}i_{5}\rangle}{\langle X i_{1}i_{2}\rangle} 
    -\bar{Q}\log\frac{\langle\bar{n}i_{2}\rangle}{\langle\bar{n}i_{5}\rangle} [i_{2}\,i_{3}\,i_{4}\,i_{5}\,n] \,\dif\log\frac{\langle X i_{2}i_{5}\rangle}{\langle X i_{1}i_{2}\rangle} \nonumber \\
  & +\bar{Q}\log\frac{\langle n(n{-1}\,1)(i_{1}i_{2})(i_{3}i_{5})\rangle}{\langle\bar{n}i_{5}\rangle\langle i_{1}i_{2}i_{3}n\rangle} \,[i_{1}\,i_{2}\,i_{3}\,i_{5}\,n] \,\dif\log\frac{\langle X i_{3}i_{5}\rangle}{\langle Xi_{1}i_{2}\rangle} \nonumber  \\
    &-\bar{Q}\log\frac{\langle n(n{-1}\,1)(i_{1}i_{2})(i_{4}i_{5})\rangle}{\langle\bar{n}i_{5}\rangle\langle i_{1}i_{2}i_{4}n\rangle} \,[i_{1}\,i_{2}\,i_{4}\,i_{5}\,n]\, \dif\log\frac{\langle X i_{4}i_{5}\rangle}{\langle Xi_{1}i_{2}\rangle} \nonumber \\
    &-\bar{Q}\log\frac{\langle n(n{-}1\,1)(i_{1}i_{2})(i_{3}i_{4})\rangle}{\langle \bar{n}i_{5}\rangle\langle i_{1}i_{2}i_{3}i_{4}\rangle} \,[i_{1}\,i_{2}\,i_{3}\,i_{4}\,n]\,\dif\log\frac{\langle Xi_{5}(i_{1}i_{2})\cap(i_{3}i_{4}n)\rangle}{\langle Xi_{1}i_{2}\rangle}\:.
\end{align}
For $i_{5}=n{-}1$, eq.\eqref{A3} reduces to
\begin{align} \label{A4}
\bar{Q}\log\frac{\langle\bar{n}i_{3}\rangle}{\langle\bar{n}i_{4}\rangle}\,[i_{1}\,i_{2}\,(i_{3}i_{4})\cap(\bar{n})\,n{-}1\,n] \, \dif\log\frac{\tau}{\langle Xi_{1}i_{2}\rangle} \:.
\end{align}
For $i_{1}=1$, eq.\eqref{A3} reduces to
\begin{align} \label{A5}
   & \bar{Q}\log\frac{\langle\bar{n}i_{2}\rangle}{\langle\bar{n}i_{5}\rangle}\biggl([1\,i_{2}\,i_{3}\,i_{4}\,i_{5}]\dif\log\langle Xi_{5}(1i_{2})\cap(i_{3}i_{4}i_{5})\rangle -\,[1\,i_{2}\,i_{3}\,i_{4}\,n]\,\dif\log\langle Xi_{5}(1i_{2})\cap(i_{3}i_{4}n)\rangle \nonumber \\
    &-[i_{2}\,i_{3}\,i_{4}\,i_{5}\,n]\,\dif\log\langle Xi_{2}i_{5}\rangle+[1\,i_{2}\,i_{3}\,i_{5}\,n]\,\dif\log\langle Xi_{3}i_{5}\rangle  -[1\,i_{2}\,i_{4}\,i_{5}\,n]\,\dif\log\langle Xi_{4}i_{5}\rangle\biggr) \:.
\end{align}
For $i_{1}=1$ and $i_5=n{-}1$, eq.\eqref{A3} reduces to
\begin{align}
    \bar{Q}\log\frac{\langle\bar{n}2\rangle}{\langle\bar{n}i_{2}\rangle} [1\,i_{3}\,i_{4}\,n{-}1\,n]\,\dif\log\tau \:.
    \label{A6}
\end{align}
\subsubsection*{2.2 $[i_{2}\,i_{3}\,(i_{4}\,i_{5})\cap(n\,n{+}1\,i_{1})\,(i_{4}\,i_{5}\,n)\cap(n{+1}\,i_{1})\,i_{1}]\,[i_{1}\,i_{4}\,i_{5}\,n\,n{+}1]$}

The effect of $\int \epsilon\dif \epsilon \dif^{3}\chi_{n+1}$ on this Yangian invariant is the same as
\begin{align}
  &\quad [i_{1}\,i_{2}\,i_{3}\,i_{4}\,n]\,[i_{1}\,i_{4}\,i_{5}\,n\,n{+}1] +
  [i_{1}\,(i_{4}\,i_{5})\cap(i_{2}\,i_{3}\,n)\,i_{5}\,n\,n{+1}]\,[i_{2}\,i_{3}\,i_{4}\,i_{5}\,n] \nonumber \\
  &-[i_{1}\,(i_{4}\,i_{5})\cap(i_{1}\,i_{2}\,i_{3})\,i_{5}\,n\,n{+}1][i_{1}\,i_{2}\,i_{3}\,i_{4}\,i_{5}] \nonumber \\
  &+[i_{1}\,(i_{4}\,i_{5})\cap(i_{1}\,i_{2}\,n)\,i_{5}\,n\,n{+}1] [i_{1}\,i_{2}\,i_{4}\,i_{5}\,n]  \nonumber \\
  &-[i_{1}\,(i_{4}\,i_{5})\cap(i_{1}\,i_{3}\,n)\,i_{5}\,n\,n{+}1]\,[i_{1}\,i_{3}\,i_{4}\,i_{5}\,n] \:, \label{A7}
\end{align}
for which we can apply eq.\eqref{A1}.

\subsubsection*{2.3 $[i_{3}\,i_{4}\,(i_{5}\,n)\cap(n{+}1\,i_{1}\,i_{2})\,(i_{5}\,n\,n{+}1)\cap(i_{1}\,i_{2})\,i_{2}] \,[i_{1}\,i_{2}\,i_{5}\,n\,n{+1}] $}

The effect of $\int \epsilon\dif \epsilon \dif^{3}\chi_{n+1}$ on this Yangian invariant is the same as $[i_{1}(i_{1}i_{2})\cap(i_{3}i_{4}n)\,i_{5}\,n\,n{+1}]$ $[i_{1}\,i_{2}\,i_{3}\,i_{4}\,n]$, for which we can apply eq.\eqref{A1}.

\subsubsection*{2.4 $[i_{4}\,i_{5}\,(n\,n{+}1)\cap(i_{1}\,i_{2}\,i_{3})\,(n\,n{+}1\,i_{1})\cap(i_{2}\,i_{3})\,i_{3}]\,
[i_{1}\,i_{2}\,i_{3}\,n\,n{+1}] $}

For $i_{1}>1$ and $i_5\leq n{-}1$, the operation $\int \epsilon\dif \epsilon \dif^{3}\chi_{n+1}$ gives 
\begin{align}
    &\biggl(\bar{Q}\log\frac{\langle\bar{n} i_{1}\rangle\langle i_{2}i_{3}i_{4}i_{5}\rangle}{\langle\bar{n}(i_{4}i_{5})\cap(i_{1}i_{2}i_{3})\rangle} \dif \log\frac{\langle X (i_{4}i_{5})\cap(i_{1}i_{2}i_{3})i_{1}\rangle}{\langle Xi_{1}i_{2}\rangle}  \nonumber \\
    &\quad +\bar{Q}\log\frac{\langle \bar{n}(i_{4}i_{5})\cap(i_{1}i_{2}i_{3})\rangle}{\langle\bar{n}(i_{1}i_{2})\cap(i_{3}i_{4}i_{5})\rangle} 
    \dif\log\frac{\langle X(i_{4}i_{5})\cap(i_{1}i_{2}i_{3})i_{3}\rangle}{\langle Xi_{1}i_{2}\rangle}\biggr) 
    \times [i_{1}\,i_{2}\,i_{3}\,i_{4}\,i_{5}] \label{A8}
\end{align}
For $i_{1}=1$ and $i_5\leq n{-}1$, eq.\eqref{A8} reduces to 
\begin{align}
    \bar{Q}\log\frac{\langle 1(i_{2}i_{3})(i_{4}i_{5})(n{-}1\,n)\rangle}{\langle\bar{n}i_{2}\rangle \langle 1i_{3}i_{4}i_{5}\rangle}\,\dif\log\langle X(i_{4}i_{5})\cap(1i_{2}i_{3})i_{3}\rangle \,[1\,i_{2}\,i_{3}\,i_{4}\,i_{5}] \label{A9} \:.
\end{align}

\subsubsection*{2.5 $[n\,n{+}1\,(i_{1}i_{2})\cap(i_{3}i_{4}i_{5})\,(i_{1}i_{2}i_{3})\cap(i_{4}i_{5})\,i_{5}]\,[i_{1}\,i_{2}\,i_{3}\,i_{4}\,i_{5}]$}

Directly apply eq.\eqref{A1}.

\subsubsection*{2.6 $ [n{+}1\,i_{1}\,(i_{2}i_{3})\cap(i_{4}i_{5}n)\,(i_{2}i_{3}i_{4})\cap(i_{5}n)\,n] \,
[i_{2}\,i_{3}\,i_{4}\,i_{5}\,n]$}

For $i_{1}\geq1$ and $i_5<n{-}1$, we can directly apply eq.\eqref{A1}. For $i_{1}>1$ and $i_5=n{-}1$, the operation $\int \epsilon\dif \epsilon \dif^{3}\chi_{n+1}$ gives 
\begin{align}
   \bar{Q}\log\frac{\langle \bar{n}i_{1}\rangle}{\langle\bar{n}i_{4}\rangle}\,[i_{2}\,i_{3}\,i_{4}\,n{-}1\,n]\,\dif\log
    \frac{\tau}{\langle X i_{1} (i_{2}i_{3})\cap(i_{4}\,n{-}1\,n)\rangle} \:. \label{A10}
\end{align}
For $i_{1}=1$ and $i_5=n{-}1$, eq.\eqref{A10} reduces to
\begin{align}
    \bar{Q}\log\frac{\langle \bar{n}2\rangle}{\langle\bar{n}i_{4}\rangle}\,[i_{2}\,i_{3}\,i_{4}\,n{-}1\,n]\,\dif\log\tau \:.
    \label{A11}
 \end{align}

\subsection*{3. $[i_{1}\,i_{2}\,i_{3}\,(i_{3}i_{4}i_{5})\cap(n\,n{+}1)\,n{+}1]\,[i_{3}\,i_{4}\,i_{5}\,n\,n{+}1]$ and its cyclic rotations}

\subsubsection*{3.1 $[i_{1}\,i_{2}\,i_{3}\,(i_{3}i_{4}i_{5})\cap(n\,n{+}1)\,n{+}1]\,[i_{3}\,i_{4}\,i_{5}\,n\,n{+}1]$}
For $i_{1}>1$ and $i_5<n{-}1$, the operation $\int \epsilon\dif \epsilon \dif^{3}\chi_{n+1}$ gives 
\begin{align}
    &{-}\bar{Q}\log\frac{\langle\bar{n}i_{1}\rangle}{\langle\bar{n}i_{2}\rangle} 
    \,[(i_{1}i_{2})\cap(\bar{n})\,i_{3}\,i_{4}\,i_{5}\,n]\,\dif\log\langle Xi_{1}i_{2}\rangle \nonumber \\
    &+\bar{Q}\log\frac{\langle\bar{n}i_{1}\rangle\langle i_{3}i_{4}i_{5}n\rangle}{\langle n(i_{1}i_{3})(i_{4}i_{5})(n{-1}\,1) \rangle} \,[i_{1}\,i_{3}\,i_{4}\,i_{5}\,n]\,\dif\log\langle Xi_{1}i_{3}\rangle \nonumber \\
    &-\bar{Q}\log\frac{\langle\bar{n}i_{2}\rangle\langle i_{3}i_{4}i_{5}n\rangle}{\langle n(i_{2}i_{3})(i_{4}i_{5})(n{-1}\,1) \rangle} [i_{2}\,i_{3}\,i_{4}\,i_{5}\,n]\,\dif\log\langle Xi_{2}i_{3}\rangle \nonumber \\
    &+\bar{Q}\log\frac{\langle\bar{n}(i_{1}i_{2})\cap(i_{3}i_{4}i_{5})\rangle}{\langle \bar{n}(i_{4}i_{5}\cap(i_{1}i_{2}i_{3})\rangle} [i_{1}\,i_{2}\,i_{3}\,i_{4}\,i_{5}] \,\dif\log\langle i_{3} (i_{1}i_{2})(i_{4}i_{5})(X)\rangle \nonumber \\
    &+\bar{Q}\log\frac{\langle\bar{n}i_{4}\rangle\langle i_{1}i_{2}i_{3}n \rangle}{\langle n(i_{1}i_{2})(i_{3}i_{4})(n{-}1\,1) \rangle} [i_{1}\,i_{2}\,i_{3}\,i_{4}\,n]\,\dif\log\langle X i_{3}i_{4}\rangle \nonumber \\
    &+\bar{Q}\log\frac{\langle n(i_{1}i_{2})(i_{3}i_{5})(n{-1}\,1)\rangle}{\langle\bar{n}i_{5}\rangle\langle i_{1}i_{2}i_{3}n\rangle}[i_{1}\,i_{2}\,i_{3}\,i_{5}\,n] \,\dif\log\langle Xi_{3}i_{5}\rangle \nonumber \\
    &-\bar{Q}\log\frac{\langle\bar{n}i_{4}\rangle}{\langle \bar{n}i_{5}\rangle}
    [i_{1}\,i_{2}\,i_{3}\,(i_{4}\,i_{5})\cap(\bar{n})\,n]\,\dif\log\langle Xi_{4}i_{5}\rangle \:. \label{A12}
\end{align}
For $i_{1}=1$ and $i_5<n{-}1$, eq.\eqref{A12} reduces to
\begin{align}
    &-\bar{Q}\log\frac{\langle \bar{n}i_{2}\rangle \langle i_{3}i_{4}i_{5}n\rangle}{\langle n(i_{2}i_{3})(i_{4}i_{5})(n{-1},1)\rangle}
    [i_{2}\,i_{3}\,i_{4}\,i_{5}\,n]\,\dif\log\langle Xi_{2}i_{3}\rangle  \nonumber \\
    &+\bar{Q}\log\frac{\langle\bar{n}i_{2}\rangle \langle 1i_{3}i_{4}i_{5}\rangle}{\langle 1(i_{2}i_{3})(i_{4}i_{5})(n{-1}\,n)\rangle}
    [1\,i_{2}\,i_{3}\,i_{4}\,i_{5}]\,\dif\log\langle i_{3}(1i_{2})(i_{4}i_{5})(X)\rangle \nonumber\\
    &-\bar{Q}\log\frac{\langle\bar{n}i_{2}\rangle}{\langle \bar{n}i_{4}\rangle} [1\,i_{2}\,i_{3}\,i_{4}\,n]\,
    \dif\log\langle Xi_{3}i_{4}\rangle+\bar{Q}\log\frac{\langle\bar{n}i_{2}\rangle}{\langle\bar{n}i_{5}\rangle}\,[1\,i_{2}\,i_{3}\,i_{5}\,n]\,\dif\log\langle Xi_{3}i_{5}\rangle \nonumber\\
    &-\bar{Q}\log\frac{\langle\bar{n}i_{4}\rangle}{\langle\bar{n}i_{5}\rangle} [1\,i_{2}\,i_{3}\,(i_{4}i_{5})\cap(\bar{n})\,n]\,\dif\log\langle Xi_{4}i_{5}\rangle \label{A13}
\end{align}
For $i_{1}>1$ and $i_5=n{-}1$, eq.\eqref{A12} reduces to
\begin{align}
    &\quad \bar{Q}\log\frac{\langle\bar{n}i_{3}\rangle}{\langle\bar{n}i_{4}\rangle}[i_{1}\,i_{2}\,i_{3}\,n{-}1\,n]\,
    \dif\log\tau-\bar{Q}\log\frac{\langle\bar{n}i_{1}\rangle}{\langle\bar{n}i_{2}\rangle}
    [(i_{1}i_{2})\cap(\bar{n})\,i_{3}\,i_{4}\,n{-}1\,n]\,\dif\log\langle Xi_{1}i_{2}\rangle \nonumber  \\
    &+\bar{Q}\log\frac{\langle\bar{n}i_{1}\rangle}{\langle\bar{n}i_{4}\rangle}\,[i_{1}\,i_{3}\,i_{4}\,n{-}1\,n]\,\dif\log\langle Xi_{1}i_{3}\rangle-\bar{Q}\log\frac{\langle \bar{n}i_{2}\rangle}{\langle\bar{n}i_{4}\rangle}R
    [i_{2}\,i_{3}\,i_{4}\,n{-1}\,n]\,\dif\log\langle Xi_{2}i_{3}\rangle \nonumber \\
    &-\bar{Q}\log\frac{\langle n(i_{1}i_{2})(i_{3}i_{4})(n{-}1\,1)\rangle}{\langle\bar{n}i_{4} \rangle \langle i_{1}i_{2}i_{3}n\rangle } [i_{1}\,i_{2}\,i_{3}\,i_{4}\,n]\,  \dif\log\langle Xi_{3}i_{4}\rangle \nonumber \\
    &+\bar{Q}\log\frac{\langle n{-}1(i_{1}i_{2})(i_{3}i_{4})(n1)\rangle}{\langle\bar{n}i_{4}\rangle \langle i_{1}i_{2}i_{3}\,n{-}1\rangle}
    [i_{1}\,i_{2}\,i_{3}\,i_{4}\,n{-}1]\,\dif\log\langle i_{3}(i_{1}i_{2})(i_{4}n{-1})(X)\rangle
\end{align}
For $i_{1}=1$ and $i_5=n{-}1$, eq.\eqref{A12} reduces to
\begin{align}
   & \bar{Q}\log\frac{\langle\bar{n}i_{2}\rangle}{\langle\bar{n}i_{4}\rangle}\biggl(
    [1\,i_{2}\,i_{3}\,n{-}1\,n]\,\dif\log\tau 
    -[i_{2}\,i_{3}\,i_{4}\,n{-}1\,n]\,\dif\log\langle Xi_{2}i_{3}\rangle  \nonumber  \\
    &-[1\,i_{2}\,i_{3}\,i_{4}\,n]\,\dif\log\langle Xi_{3}i_{4}\rangle  
    + [1\,i_{2}\,i_{3}\,i_{4}\,n{-}1]\,\dif\log\langle i_{3}(1i_{2})(i_{4}\,n{-}1)(X) \rangle \biggr)
\end{align}

\subsubsection*{3.2 $[i_{2}\,i_{3}\,i_{4}\,(i_{4}\,i_{5}\,n)\cap(n{+1}\,i_{1})\,i_{1}] \,[i_{4}\,i_{5}\,n\,n{+}1\,i_{1}] $}

The effect of $\int \epsilon\dif \epsilon \dif^{3}\chi_{n+1}$ on this Yangian invariant is the same as $[i_{2}\,i_{3}\,i_{4}\,n\,i_{1}][i_{1}\,i_{4}\,i_{5}\,n\,n{+}1]$, for which we can apply eq.\eqref{A1}

\subsubsection*{3.3 $[i_{3}\,i_{4}\,i_{5}\,(i_{5}\,n\,n{+}1)\cap(i_{1}i_{2})\,i_{2}] [i_{5}\,n\,n{+}1\,i_{1}\,i_{2}]$}

For $i_{1}>1$ and $i_5<n{-}1$, the operation $\int \epsilon\dif \epsilon \dif^{3}\chi_{n+1}$ gives 
\begin{align}
    [i_{1},i_{2},i_{3},i_{4},i_{5}]\biggl(&\bar{Q}\log\frac{\langle\bar{n}(i_{1}i_{2})\cap(i_{3}i_{4}i_{5}) \rangle}{\langle \bar{n}i_{1}\rangle\langle i_{2}i_{3}i_{4}i_{5}\rangle} \dif\log\frac{\langle Xi_{1}i_{2}\rangle}{\langle i_{5}(i_{1}i_{2})(i_{3}i_{4})(X)\rangle} \nonumber \\ 
    &+\bar{Q}\log\frac{\langle\bar{n}i_{1}\rangle}{\langle\bar{n}i_{5}\rangle}
    \dif\log\frac{\langle Xi_{1}i_{5}\rangle}{\langle i_{5}(i_{1}i_{2})(i_{3}i_{4})(X)\rangle}\biggr) \label{A16}
\end{align}
For $i_{1}>1$ and $i_5=n{-}1$, eq.\eqref{A16} reduces to
\begin{align}
    [i_{1}\,i_{2}\,i_{3}\,i_{4}\,n{-}1]\bar{Q}\log\frac{\langle\bar{n}i_{1}\rangle \langle i_{2}i_{3}i_{4}\,n{-}1\rangle }{\langle n{-}1(i_{1}i_{2})(i_{3}i_{4})(n1)\rangle}\dif\log\frac{\tau}{\langle Xi_{1}i_{2} \rangle} \:.
\end{align}
For $i_{1}=1$ and $i_5<n{-}1$, eq.\eqref{A16} reduces to 
\begin{align}
    \bar{Q}\log\frac{\langle\bar{n}i_{5}\rangle}{\langle\bar{n}i_{2}\rangle} [1\,i_{2}\,i_{3}\,i_{4}\,i_{5}]\,
    \dif\log\langle i_{5}(1i_{2})(i_{3}i_{4})(X) \rangle \:.
\end{align}
For $i_{1}=1$ and $i_5=n{-}1$, eq.\eqref{A16} reduces to 
\begin{align}
    \bar{Q}\log\frac{\langle \bar{n}2\rangle}{\langle\bar{n}i_{2}\rangle}[1\,i_{2}\,i_{3}\,i_{4}\,n{-1}]\,\dif\log\tau
\end{align}

\subsubsection*{3.4 $[i_{4}\,i_{5}\,n\,(n\,n{+}1\,i_{1})\cap(i_{2}i_{3})\,i_{3}] [n\,n{+}1\,i_{1}\,i_{2}\,i_{3}] $}
For $i_{1}>1$ and $i_5 \leq n{-}1$, the operation $\int \epsilon\dif \epsilon \dif^{3}\chi_{n+1}$ gives 
\begin{align}
    [i_{2}\,i_{3}\,i_{4}\,i_{5}\,n]\biggl(& \bar{Q}\log\frac{\langle\bar{n}i_{2}\rangle}{\langle \bar{n}i_{1}\rangle}\dif\log\frac{\langle Xi_{1}i_{2}\rangle}{\langle Xi_{1}(i_{2}i_{3})\cap(i_{4}i_{5}n)\rangle} \nonumber \\
    &+\bar{Q}\log\frac{\langle n(i_{2}i_{3})(i_{4}i_{5})(n{-}1\,1) \rangle}{\langle \bar{n}i_{2}\rangle \langle i_{3}i_{4}i_{5}n\rangle}
    \dif\log\frac{\langle Xi_{2}i_{3}\rangle}{\langle Xi_{1}(i_{2}i_{3})\cap(i_{4}i_{5}n)\rangle}\biggr). \label{A20}
\end{align}
For $i_{1}=1$ and $i_5\leq n{-}1$, eq.\eqref{A20} reduces to
\begin{align}
    [i_{2}\,i_{3}\,i_{4}\,i_{5}\,n]\bar{Q}\log\frac{\langle n(i_{2}i_{3})(i_{4}i_{5})(n{-}1\,1)\rangle}{\langle\bar{n}i_{2}\rangle\langle i_{3}i_{4}i_{5}n\rangle}\dif\log\langle Xi_{2}i_{3}\rangle \:.
\end{align}

\subsubsection*{3.5 $[i_{5}\,n\,n{+}1\,(n{+}1\,i_{1}\,i_{2})\cap(i_{3}i_{4})\,i_{4}][n{+}1\,i_{1}\,i_{2}\,i_{3}\,i_{4}]$}

The effect of $\int \epsilon\dif \epsilon \dif^{3}\chi_{n+1}$ on this Yangian invariant is the same as $[(ni_{1}i_{2})\cap(i_{3}i_{4})i_{4}i_{5}\,n\,n{+}1]$ $[i_{1}\,i_{2}\,i_{3}\,i_{4}\,n]$, for which we can apply eq.\eqref{A1}.

\subsubsection*{3.6 $[n\,n{+}1\,i_{1}\,(i_{1}i_{2}i_{3})\cap(i_{4}i_{5})\,i_{5}] [i_{1}\,i_{2}\,i_{3}\,i_{4}\,i_{5}]$}

Directly apply eq.\eqref{A1}.

\subsubsection*{3.7 $[n{+}1\,i_{1}\,i_{2}\,(i_{2}i_{3}i_{4})\cap(i_{5}n)n][i_{2}\,i_{3}\,i_{4}\,i_{5}\,n]$}

Directly apply eq.\eqref{A1}.

\subsection*{4. $[i_{1}\,i_{2}\,i_{3}\,(i_{4}i_{5}i_{6})\cap(n\,n{+}1)\,n{+}1]\,[i_{4}\,i_{5}\,i_{6}\,n\,n{+}1]$ and its cyclic rotations}

\subsubsection*{4.1 $[i_{1}\,i_{2}\,i_{3}\,(i_{4}i_{5}i_{6})\cap(n\,n{+}1)\,n{+}1][i_{4}\,i_{5}\,i_{6}\,n\,n{+1}]$}

For $i_{1}>1$ and $i_{6}<n{-}1$, the operation $\int \epsilon\dif \epsilon \dif^{3}\chi_{n+1}$ gives 
\begin{align}
    &\quad \bar{Q}\log\frac{\langle\bar{n}i_{1}\rangle}{\langle\bar{n}i_{2}\rangle}[(i_{1}i_{2})\cap(\bar{n})\,i_{4}\,i_{5}\,i_{6}\,n]\,\dif\log\frac{\langle X(i_{1}i_{2}i_{3})\cap(i_{4}i_{5}i_{6})\rangle}{\langle Xi_{1}i_{2}\rangle} \nonumber \\
    &-\bar{Q}\log\frac{\langle\bar{n}i_{1}\rangle}{\langle\bar{n}i_{3}\rangle}[(i_{1}i_{3})\cap(\bar{n})\,i_{4}\,i_{5}\,i_{6}\,n]\,\dif\log\frac{\langle X(i_{1}i_{2}i_{3})\cap(i_{4}i_{5}i_{6})\rangle}{\langle Xi_{1}i_{3}\rangle} \nonumber\\
    &+\bar{Q}\log\frac{\langle\bar{n}i_{2}\rangle}{\langle\bar{n}i_{3}\rangle}\,[(i_{2}i_{3})\cap(\bar{n})\,i_{4}\,i_{5}\,i_{6}\,n]\,\dif\log\frac{\langle X(i_{1}i_{2}i_{3})\cap(i_{4}i_{5}i_{6})\rangle}{\langle Xi_{2}i_{3}\rangle} \nonumber\\
    &+\bar{Q}\log\frac{\langle \bar{n}i_{4}\rangle}{\langle\bar{n}i_{5}\rangle}\,[i_{1}\,i_{2}\,i_{3}\,(i_{4}i_{5})\cap(\bar{n})\,n]\,\dif\log\frac{\langle X(i_{1}i_{2}i_{3})\cap(i_{4}i_{5}i_{6})\rangle}{\langle Xi_{4}i_{5}\rangle} \nonumber\\
    &-\bar{Q}\log\frac{\langle \bar{n}i_{4}\rangle}{\langle\bar{n}i_{6}\rangle}\,[i_{1}\,i_{2}\,i_{3}\,(i_{4}i_{6})\cap(\bar{n})\,n]\,\dif\log\frac{\langle X(i_{1}i_{2}i_{3})\cap(i_{4}i_{5}i_{6})\rangle}{\langle Xi_{4}i_{6}\rangle}\nonumber \\
    &+\bar{Q}\log\frac{\langle \bar{n}i_{5}\rangle}{\langle\bar{n}i_{6}\rangle}\,[i_{1}\,i_{2}\,i_{3}\,(i_{5}i_{6})\cap(\bar{n})\,n]\,\dif\log\frac{\langle X(i_{1}i_{2}i_{3})\cap(i_{4}i_{5}i_{6})\rangle}{\langle Xi_{5}i_{6}\rangle} \label{A22}
\end{align}
For $i_{1}=1$ and $i_6<n{-}1$, eq.\eqref{A22} reduces to
\begin{align}
    &-\bar{Q}\log\frac{\langle\bar{n}i_{2}\rangle}{\langle\bar{n}i_{3}\rangle}[(i_{2}i_{3})\cap(\bar{n})\,i_{4}\,i_{5}\,i_{6}\,n]\dif\log \langle Xi_{2}i_{3}\rangle  \nonumber \\
    &+\bar{Q}\log\frac{\langle\bar{n}i_{2}\rangle}{\langle\bar{n}i_{3}\rangle}[1\,(i_{2}i_{3})\cap(\bar{n})\,i_{4}\,i_{5}\,i_{6}]\dif\log\langle  X(1i_{2}i_{3})\cap(i_{4}i_{5}i_{6})\rangle \nonumber \\
    &-\bar{Q}\log\frac{\langle \bar{n}i_{4}\rangle}{\langle\bar{n}i_{5}\rangle}[1\,i_{2}\,i_{3}\,(i_{4}i_{5})\cap(\bar{n})\,n]\dif\log \langle Xi_{4}i_{5}\rangle \nonumber \\
    &+\bar{Q}\log\frac{\langle \bar{n}i_{4}\rangle}{\langle\bar{n}i_{6}\rangle}[1\,i_{2}\,i_{3}\,(i_{4}i_{6})\cap(\bar{n})\,n]\dif\log \langle Xi_{4}i_{6}\rangle \nonumber  \\
    &-\bar{Q}\log\frac{\langle \bar{n}i_{5}\rangle}{\langle\bar{n}i_{6}\rangle}[1\,i_{2}\,i_{3}\,(i_{5}i_{6})\cap(\bar{n})\,n]\dif\log \langle Xi_{5}i_{6}\rangle
\end{align}
For $i_{1}>1$ and $i_6=n{-}1$, eq.\eqref{A22} reduces to
\begin{align*}
    \bar{Q}\log\frac{\langle\bar{n}i_{4}\rangle}{\langle\bar{n}i_{5}\rangle}
    \biggl(& [i_{1}\,i_{2}\,i_{3}\,n{-}1\,n]\,\dif\log\frac{\tau}{\langle X(i_{1}i_{2}i_{3})\cap(i_{4}\,i_{5}\,n{-}1)\rangle} \\
    &-[i_{1}\,i_{2}\,(i_{4}i_{5})\cap(\bar{n})\,n{-}1\,n]\,\dif\log\frac{\langle Xi_{1}i_{2}\rangle}{\langle X(i_{1}i_{2}i_{3})\cap(i_{4}\,i_{5}\,n{-}1)\rangle} \\
    &+ [i_{1}\,i_{3}\,(i_{4}i_{5})\cap(\bar{n})\,n{-}1\,n]\,\dif\log\frac{\langle Xi_{1}i_{3}\rangle}{\langle X(i_{1}i_{2}i_{3})\cap(i_{4}\,i_{5}\,n{-}1)\rangle} \\
    & -[i_{2}\,i_{3}\,(i_{4}i_{5})\cap(\bar{n})\,n{-}1\,n]\,\dif\log\frac{\langle Xi_{2}i_{3}\rangle}{\langle X(i_{1}i_{2}i_{3})\cap(i_{4}\,i_{5}\,n{-}1)\rangle} \\
    &-[i_{1}\,i_{2}\,i_{3}\,(i_{4}i_{5})\cap(\bar{n})\,n]\, \dif\log\frac{\langle Xi_{4}i_{5}\rangle}{\langle X(i_{1}i_{2}i_{3})\cap(i_{4}\,i_{5}\,n{-}1)\rangle}\biggr)
\end{align*}
For $i_{1}=1$ and $i_6=n{-}1$, eq.\eqref{A22} reduces to 
\begin{align}
    &\bar{Q}\log\frac{\langle\bar{n}i_{4}\rangle}{\langle\bar{n}i_{5}\rangle}\biggl(
        -[1\,i_{2}\,i_{3}\,(i_{4}i_{5})\cap(\bar{n})\,n]\dif\log\langle Xi_{4}i_{5}\rangle
         -[i_{2}\,i_{3}\,(i_{4}i_{5})\cap(\bar{n})\,n{-}1\,n]\dif\log\langle Xi_{2}i_{3}\rangle \nonumber \\
    &\quad +[1\,i_{2}\,i_{3}\,n{-}1\,n]\dif\log\tau  + [1\,i_{2}\,i_{3}\,(i_{4}i_{5})\cap(\bar{n})\,n{-}1]\dif\log \langle X(1i_{2}i_{3})\cap(i_{4}\,i_{5}\,n{-}1)\rangle \biggr) \:.
\end{align}

\subsubsection*{4.2 $[i_{2}\,i_{3}\,i_{4}\,(i_{5}i_{6}n)\cap(n{+}1\,i_{1})\,i_{1}] [i_{5}\,i_{6}\,n\,n{+}1\,i_{1}] $ }

The effect of $\int \epsilon\dif \epsilon \dif^{3}\chi_{n+1}$ on this Yangian invariant is the same as $[i_{1}\,i_{2}\,i_{3}\,i_{4}\,n] [i_{1}\,i_{5}\,i_{6}\,n\,n{+}1]$, for which we can apply eq.\eqref{A1}.

\subsubsection*{4.3 $[i_{3}\,i_{4}\,i_{5}\,(i_{6}\,n\,n{+}1)\cap(i_{1}i_{2})\,i_{2}][i_{1}\,i_{2}\,i_{6}\,n\,n{+}1]$}

The effect of $\int \epsilon\dif \epsilon \dif^{3}\chi_{n+1}$ on this Yangian invariant is the same as $[i_{1}(i_{1}i_{2})\cap(i_{3}i_{4}i_{5})i_{6}\,n\,n{+}1]$ $[i_{1}\,i_{2}\,i_{3}\,i_{4}\,i_{5}]$, for which we can apply eq.\eqref{A1}.

\subsubsection*{4.4 $[i_{4}\,i_{5}\,i_{6}\,(n\,n{+}1\,i_{1})\cap(i_{2}i_{3})\,i_{3}][n\,n{+}1\,i_{1}\,i_{2}\,i_{3}]$}

The effect of $\int \epsilon\dif \epsilon \dif^{3}\chi_{n+1}$ on this Yangian invariant is the same as $[i_{1}i_{2}(i_{2}i_{3})\cap(i_{4}i_{5}i_{6})\,n\,n{+}1]$ $[i_{2}\,i_{3}\,i_{4}\,i_{5}\,i_{6}]$, for which we can apply eq.\eqref{A1}.

\subsubsection*{4.5 $[i_{6}\,n\,n{+}1\,(i_{1}i_{2}i_{3})\cap(i_{4}i_{5})\,i_{5}]\,[i_{1}\,i_{2}\,i_{3}\,i_{4}\,i_{5}]$}

Directly apply eq.\eqref{A1}.

\subsubsection*{4.6 $[n\,n{+}1\,i_{1}\,(i_{2}i_{3}i_{4})\cap(i_{5}i_{6})\,i_{6}][i_{2}\,i_{3}\,i_{4}\,i_{5}\,i_{6}]$}

Directly apply eq.\eqref{A1}.

\subsubsection*{4.7 $[n{+}1\,i_{1}\,i_{2}\,(i_{3}i_{4}i_{5})\cap(i_{6}n)\,n][i_{3}\,i_{4}\,i_{5}\,i_{6}\,n]$}

Directly apply eq.\eqref{A1}.

\subsection*{5. $[i_{1}\,i_{2}\,i_{3}\,i_{4}\,n{+}1][i_{4}\,i_{5}\,i_{6}\,n\,n{+}1]$ and its cyclic rotations}

One of the $R$-invariants have a smooth limit under $\mathcal{Z}_{n+1}\to\mathcal{Z}_{n}$, and the other follows the replacement rule \eqref{A1}.

\subsection*{6. $[i_{1}\,i_{2}\,i_{3}\,(i_{4}i_{5})\cap(i_{6}n\,n{+}1)\,n{+}1][i_{4}\,i_{5}\,i_{6}\,n\,n{+}1]$ and its cyclic rotations}

\subsubsection*{6.1 $[i_{1}\,i_{2}\,i_{3}\,(i_{4}i_{5})\cap(i_{6}n\,n{+}1)\,n{+}1][i_{4}\,i_{5}\,i_{6}\,n\,n{+}1]$}

For $i_{1}\geq 1$ and $i_{6}<n-1$, the operation $\int \epsilon\dif \epsilon \dif^{3}\chi_{n+1}$ gives
\begin{align}
    &\bar{Q}\log\frac{\langle \bar{n}i_{4}\rangle}{\langle \bar{n}i_{6}\rangle} [i_{1}\,i_{2}\,i_{3}\,i_{4}\,n]\,
    \dif\log\frac{\langle Xi_{4}i_{6}\rangle}{\langle Xi_{4}i_{5}\rangle}
    -\bar{Q}\log\frac{\langle \bar{n}i_{5}\rangle}{\langle \bar{n}i_{6}\rangle} [i_{1}\,i_{2}\,i_{3}\,i_{5}\,n] \,
    \dif\log\frac{\langle Xi_{5}i_{6}\rangle}{\langle Xi_{4}i_{5}\rangle} \nonumber \\
    &-\bar{Q}\log\frac{\langle \bar{n}(i_{4}i_{5})\cap(i_{1}i_{2}i_{3})\rangle}{\langle\bar{n}i_{6}\rangle \langle i_{2}i_{3}i_{4}i_{5}\rangle}
    [i_{1}\,i_{2}\,i_{3}\,i_{4}\,i_{5}] \,\dif\log\frac{\langle X(i_{4}i_{5})\cap(i_{1}i_{2}i_{3})i_{6}\rangle}{\langle Xi_{4}i_{5}\rangle} \nonumber \\
    &+\bar{Q}\log\frac{\langle \bar{n}(i_{1}i_{2})\cap(i_{4}i_{5}n)\rangle}{\langle\bar{n}i_{6}\rangle \langle i_{1}i_{2}i_{4}i_{5}\rangle}
    [i_{1}\,i_{2}\,i_{4}\,i_{5}\,n]\,\dif\log\frac{\langle X(i_{1}i_{2})\cap(i_{4}i_{5}n)i_{6}\rangle}{\langle Xi_{4}i_{5}\rangle} \nonumber  \\
    &-\bar{Q}\log\frac{\langle \bar{n}(i_{1}i_{3})\cap(i_{4}i_{5}n)\rangle}{\langle\bar{n}i_{6}\rangle\langle i_{1}i_{3}i_{4}i_{5}\rangle}
    [i_{1}\,i_{3}\,i_{4}\,i_{5}\,n]\,\dif\log\frac{\langle X(i_{1}i_{3})\cap(i_{4}i_{5}n)i_{6}\rangle}{\langle Xi_{4}i_{5}\rangle} \nonumber \\
    &+\bar{Q}\log\frac{\langle \bar{n}(i_{2}i_{3})\cap(i_{4}i_{5}n)\rangle}{\langle\bar{n}i_{6}\rangle\langle i_{2}i_{3}i_{4}i_{5}\rangle}
    [i_{2}\,i_{3}\,i_{4}\,i_{5}\,n]\,\dif\log\frac{\langle X(i_{2}i_{3})\cap(i_{4}i_{5}n)i_{6}\rangle}{\langle Xi_{4}i_{5}\rangle}  \:. \label{A25}
\end{align}
For $i_{1}\geq 1$ and $i_{6}=n-1$, eq.\eqref{A25} reduces to
\begin{align}
    \bar{Q}\log\frac{\langle \bar{n}i_{4}\rangle}{\langle\bar{n}i_{5}\rangle} [i_{1}\,i_{2}\,i_{3}\,(i_{4}i_{5})\cap(\bar{n})\,n]\,\dif\log\frac{\tau}{\langle Xi_{4}i_{5}\rangle}
\end{align}

\subsubsection*{6.2 $[i_{2}\,i_{3}\,i_{4}\,(i_{5}i_{6})\cap(n\,n{+}1\,i_{1})\,i_{1}][i_{5}\,i_{6}\,n\,n{+1}\,i_{1}] $}

For $i_{1}> 1$ and $i_{6}<n-1$, the operation $\int \epsilon\dif \epsilon \dif^{3}\chi_{n+1}$ gives
\begin{align}
    &-\bar{Q}\log\frac{\langle\bar{n}i_{1}\rangle}{\langle\bar{n}i_{5}\rangle} [i_{1}\,i_{2}\,i_{3}\,i_{4}\,i_{5}]\,\dif\log\frac{\langle Xi_{1}i_{5}\rangle}{\langle Xi_{5}i_{6}\rangle} +\bar{Q}\log\frac{\langle \bar{n}i_{1}\rangle}{\langle\bar{n}i_{6}\rangle}[i_{1}\,i_{2}\,i_{3}\,i_{4}\,i_{6}]\, \dif\log\frac{\langle \bar{n}i_{1}i_{6}\rangle}{\langle\bar{n}i_{5}i_{6}\rangle} \nonumber \\
    &+\bar{Q}\log\frac{\langle \bar{n}(i_{5}i_{6})\cap(i_{1}i_{2}i_{3})\rangle}{\langle \bar{n} i_{1}\rangle
    \langle i_{2}i_{3}i_{5}i_{6}\rangle}
    [i_{1}\,i_{2}\,i_{3}\,i_{5}\,i_{6}] \,\dif\log\frac{\langle i_{1}(i_{2}i_{3})(i_{5}i_{6})(X)\rangle}{\langle Xi_{5}i_{6}\rangle}  \nonumber \\
    &-\bar{Q}\log\frac{\langle \bar{n}(i_{5}i_{6})\cap(i_{1}i_{2}i_{4})\rangle}{\langle \bar{n} i_{1}\rangle\langle i_{2}i_{4}i_{5}i_{6}\rangle}
    [i_{1}\,i_{2}\,i_{4}\,i_{5}\,i_{6}]\, \dif\log\frac{\langle i_{1}(i_{2}i_{4})(i_{5}i_{6})(X)\rangle}{\langle Xi_{5}i_{6}\rangle}  \nonumber\\
    &+\bar{Q}\log\frac{\langle \bar{n}(i_{5}i_{6})\cap(i_{1}i_{3}i_{4})\rangle}{\langle \bar{n} i_{1}\rangle \langle i_{3}i_{4}i_{5}i_{6}\rangle}
    [i_{1}\,i_{3}\,i_{4}\,i_{5}\,i_{6}]\, \dif\log\frac{\langle i_{1}(i_{3}i_{4})(i_{5}i_{6})(X)\rangle}{\langle Xi_{5}i_{6}\rangle}  \nonumber \\
    &-\bar{Q}\log\frac{\langle \bar{n}(i_{5}i_{6})\cap(i_{2}i_{3}i_{4})\rangle}{\langle\bar{n}i_{1}\rangle\langle i_{2}i_{3}i_{4}i_{5}\rangle}
    [i_{2}\,i_{3}\,i_{4}\,i_{5}\,i_{6}]\,\dif\log\frac{\langle X(i_{5}i_{6})\cap(i_{2}i_{3}i_{4})i_{1}\rangle}{\langle Xi_{5}i_{6}\rangle} \:. \label{A27}
\end{align}
For $i_{1}> 1$ and $i_{6}=n-1$, eq.\eqref{A27} reduces to 
\begin{align}
    &\bar{Q}\log\frac{\langle \bar{n}i_{1}\rangle}{\langle\bar{n} i_{5}\rangle} 
    \biggl( [i_{1}\,i_{2}\,i_{3}\,i_{4}\,n{-}1]\, \dif\log\frac{\tau}{\langle X(i_{5}\,n{-}1)\cap(i_{2}i_{3}i_{4})i_{1}\rangle} \nonumber \\
    &-[i_{1}\,i_{2}\,i_{3}\,i_{4}\,i_{5}]\, \dif\log\frac{\langle Xi_{1}i_{5} \rangle}{\langle X(i_{5}\,n{-}1)\cap(i_{2}i_{3}i_{4})i_{1}\rangle} \nonumber\\
    &-[i_{1}\,i_{2}\,i_{3}\,i_{5}\,n{-}1]\,\dif\log\frac{\langle X(i_{5}\,n{-}1)\cap(i_{1}i_{2}i_{3})i_{1}\rangle}{\langle X(i_{5}\,n{-}1)\cap(i_{2}i_{3}i_{4})i_{1}\rangle} \nonumber\\
    &+[i_{1}\,i_{2}\,i_{4}\,i_{5}\,n{-}1]\,\dif\log\frac{\langle X(i_{5}\,n{-}1)\cap(i_{1}i_{2}i_{4})i_{1}\rangle}{\langle X(i_{5}\,n{-}1)\cap(i_{2}i_{3}i_{4})i_{1}\rangle} \nonumber \\
    &-[i_{1}\,i_{3}\,i_{4}\,i_{5}\,n{-}1]\,\dif\log\frac{\langle X(i_{5}\,n{-}1)\cap(i_{1}i_{3}i_{4})i_{1}\rangle}{\langle X(i_{5}\,n{-}1)\cap(i_{2}i_{3}i_{4})i_{1}\rangle} \biggr) \:.
\end{align}
For $i_{1}=1$ and $i_{6}<n-1$, eq.\eqref{A27} reduces to 
\begin{align}
    -[1\,i_{2}\,i_{3}\,i_{4}\,(i_{5}i_{6})\cap(\bar{n})]\bar{Q}\log\frac{\langle\bar{n}i_{5}\rangle}{\langle\bar{n}i_{6}\rangle}\dif\log\langle Xi_{5}i_{6}\rangle \:.
\end{align}
For $i_{1}= 1$ and $i_{6}=n-1$, eq.\eqref{A27} reduces to
\begin{align}
    [1\,i_{2}\,i_{3}\,i_{4}\,n{-}1]\bar{Q}\log\frac{\langle\bar{n}2\rangle}{\langle\bar{n}i_{5}\rangle}\dif\log\tau
\end{align}

\subsubsection*{6.3 $[i_{3}\,i_{4}\,i_{5}\,(i_{6}n)\cap(n{+}1\,i_{1}\,i_{2})\,i_{2}][i_{6}\,n\,n{+}1\,i_{1}\,i_{2}]$}

The effect of $\int \epsilon\dif \epsilon \dif^{3}\chi_{n+1}$ on this Yangian invariant is the same as $[i_{2}i_{3}i_{4}i_{5}n][i_{1}i_{2}i_{6}\,n\,n{+}1]$, for which we can apply \eqref{A1}.

\subsubsection*{6.4 $[i_{4}\,i_{5}\,i_{6}\,(n\,n{+}1)\cap(i_{1}i_{2}i_{3})\,i_{3}][n\,n{+}1\,i_{1}\,i_{2}\,i_{3}] $ }

For $i_{1}> 1$ and $i_{6}\leq n-1$, the operation $\int\epsilon\dif\epsilon\dif^{3}\chi_{n+1}$ gives
\begin{align}
    &-\bar{Q}\log\frac{\langle \bar{n} i_{1}\rangle}{\langle \bar{n}i_{2}\rangle}
    [(i_{1}i_{2})\cap(\bar{n})\,i_{3}\,i_{4}\,i_{5}\,i_{6}] \,\dif \log
    \frac{\langle Xi_{1}i_{2}\rangle}{\langle X(i_{1}i_{2}i_{3})\cap(i_{4}i_{5}i_{6})\rangle} \nonumber \\
    &+\bar{Q}\log\frac{\langle\bar{n}i_{1}\rangle\langle i_{3}i_{4}i_{5}i_{6}\rangle}{\langle\bar{n} (i_{1}i_{3})\cap(i_{4}i_{5}i_{6})\rangle}
    \,[i_{1}\,i_{3}\,i_{4}\,i_{5}\,i_{6}]\,\dif\log\frac{\langle Xi_{1}i_{3}\rangle}{\langle X(i_{1}i_{2}i_{3})\cap(i_{4}i_{5}i_{6})\rangle} \nonumber \\
    &-\bar{Q}\log\frac{\langle\bar{n}i_{2}\rangle \langle i_{3}i_{4}i_{5}i_{6}\rangle}{\langle\bar{n} (i_{2}i_{3})\cap(i_{4}i_{5}i_{6})\rangle}
    [i_{2}\,i_{3}\,i_{4}\,i_{5}\,i_{6}]\,\dif\log\frac{\langle Xi_{2}i_{3}\rangle}{\langle X(i_{1}i_{2}i_{3})\cap(i_{4}i_{5}i_{6})\rangle} \nonumber \\
    &+\bar{Q}\log\frac{\langle \bar{n}(i_{1}i_{2})\cap(i_{3}i_{4}i_{5})\rangle}{\langle \bar{n}(i_{4}i_{5})\cap(i_{1}i_{2}i_{3})\rangle} [i_{1}\,i_{2}\,i_{3}\,i_{4}\,i_{5}]\, \dif\log 
    \frac{\langle X(i_{1}i_{2})\cap(i_{3}i_{4}i_{5})i_{3}\rangle}{\langle X(i_{1}i_{2}i_{3})\cap(i_{4}i_{5}i_{6})\rangle} \nonumber \\
    &-\bar{Q}\log\frac{\langle \bar{n}(i_{1}i_{2})\cap(i_{3}i_{4}i_{6})\rangle}{\langle \bar{n}(i_{4}i_{6})\cap(i_{1}i_{2}i_{3})\rangle}[i_{1}\,i_{2}\,i_{3}\,i_{4}\,i_{6}]\, \dif\log 
    \frac{\langle X(i_{1}i_{2})\cap(i_{3}i_{4}i_{6})i_{3}\rangle}{\langle X(i_{1}i_{2}i_{3})\cap(i_{4}i_{5}i_{6})\rangle} \nonumber \\
    &+\bar{Q}\log\frac{\langle \bar{n}(i_{1}i_{2})\cap(i_{3}i_{5}i_{6})\rangle}{\langle \bar{n}(i_{5}i_{6})\cap(i_{1}i_{2}i_{3})\rangle}\,[i_{1}\,i_{2}\,i_{3}\,i_{5}\,i_{6}] \,\dif\log 
    \frac{\langle X(i_{1}i_{2})\cap(i_{3}i_{5}i_{6})i_{3}\rangle}{\langle X(i_{1}i_{2}i_{3})\cap(i_{4}i_{5}i_{6})\rangle} \:. \label{A31}
\end{align}
For $i_{1}= 1$ and $i_{6}\leq n-1$, eq.\eqref{A31} reduces to
\begin{align}
    &-\bar{Q}\log\frac{\langle\bar{n}i_{2}\rangle\langle i_{3}i_{4}i_{5}i_{6}\rangle}{\langle \bar{n}(i_{2}i_{3})\cap(i_{4}i_{5}i_{6})\rangle}
    [i_{2}\,i_{3}\,i_{4}\,i_{5}\,i_{6}]\,\dif\log\frac{\langle Xi_{2}i_{3}\rangle}{\langle X(1i_{2}i_{3})\cap(i_{4}i_{5}i_{6})\rangle} \nonumber \\
    &+\bar{Q}\log\frac{\langle\bar{n}i_{2}\rangle \langle 1\,i_{3}i_{4}i_{5}\rangle}{\langle\bar{n}(i_{2}i_{3})\cap(i_{4}i_{5}1)\rangle}
    [1\,i_{2}\,i_{3}\,i_{4}\,i_{5}]\,\dif\log\frac{\langle X(1i_{2})\cap(i_{3}i_{4}i_{5})i_{3}\rangle}{\langle  X(1i_{2}i_{3})\cap(i_{4}i_{5}i_{6})\rangle} \nonumber \\
    &-\bar{Q}\log\frac{\langle\bar{n}i_{2}\rangle\langle 1\,i_{3}i_{4}i_{6}\rangle}{\langle\bar{n}(i_{2}i_{3})\cap(i_{4}i_{6}1)\rangle}
    [1\,i_{2}\,i_{3}\,i_{4}\,i_{6}]\,\dif\log\frac{\langle X(1i_{2})\cap(i_{3}i_{4}i_{6})i_{3}\rangle}{\langle  X(1i_{2}i_{3})\cap(i_{4}i_{5}i_{6})\rangle} \nonumber \\
    &+\bar{Q}\log\frac{\langle\bar{n}i_{2}\rangle \langle 1\,i_{3}i_{5}i_{6}\rangle}{\langle\bar{n}(i_{2}i_{3})\cap(i_{5}i_{6}1)\rangle}
    [1\,i_{2}\,i_{3}\,i_{5}\,i_{6}]\,\dif\log\frac{\langle X(1i_{2})\cap(i_{3}i_{5}i_{6})i_{3}\rangle}{\langle  X(1i_{2}i_{3})\cap(i_{4}i_{5}i_{6})\rangle} \nonumber \\
    &-\bar{Q}\log\frac{\langle\bar{n}i_{2} \rangle}{\langle\bar{n}i_{3}\rangle} \,[1\,i_{3}\,i_{4}\,i_{5}\,i_{6}]\,\dif\log\langle  X(1i_{2}i_{3})\cap(i_{4}i_{5}i_{6})\rangle \:.
\end{align}

\subsubsection*{6.6 $ [i_{6}\,n\,n{+1}\,(i_{1}i_{2})\cap(i_{3}i_{4}i_{5})\,i_{5}][i_{1}\,i_{2}\,i_{3}\,i_{4}\,i_{5}] $}
Directly apply eq.\eqref{A1}.

\subsubsection*{6.7 $[n\,n{+1}\,i_{1}\,(i_{2}i_{3})\cap(i_{4}i_{5}i_{6})\,i_{6}][i_{2}\,i_{3}\,i_{4}\,i_{5}\,i_{6}]$}
Directly apply eq.\eqref{A1}.

\subsubsection*{6.8 $[n{+}1\,i_{1}\,i_{2}\,(i_{3}i_{4})\cap(i_{5}i_{6}n)\,n]\,[i_{3}\,i_{4}\,i_{5}\,i_{6}\,n]$}
Directly apply eq.\eqref{A1}.

\subsection*{7. $[i_{1}\,i_{2}\,i_{3}\,(i_{4}i_{5})\cap(i_{6}i_{7}i_{8})\,(i_{4}i_{5}i_{6})\cap(i_{7}i_{8})][i_{4},i_{5}\,i_{6}\,i_{7}\,i_{8}]$}
Using six-term identity \eqref{6ident}, this Yangian invariant is equal to
\begin{align*}
    \Bigl(&{-}[i_{1}\,i_{2}\,i_{3}\,(i_{4}i_{5}i_{6})\cap(i_{7}i_{8})\,i_{8}]+[i_{1}\,i_{2}\,i_{3}\,(i_{4}i_{5})\cap(i_{6}i_{7}i_{8})\,i_{8}]\\
    &+[i_{1}i_{2}(i_{4}i_{5})\cap(i_{6}i_{7}i_{8})\,(i_{4}i_{5}i_{6})\cap(i_{7}i_{8})i_{8}]
    -[i_{1}\,i_{3}\,(i_{4}i_{5})\cap(i_{6}i_{7}i_{8})\,(i_{4}i_{5}i_{6})\cap(i_{7}i_{8})\,i_{8}] \\
    &+[i_{2}\,i_{3}\,(i_{4}i_{5})\cap(i_{6}i_{7}i_{8})\,(i_{4}i_{5}i_{6})\cap(i_{7}i_{8})\,i_{8}]\Bigr)
    [i_{4}\,i_{5}\,i_{6}\,i_{7}\,i_{8}] \:,
\end{align*}
where the effect of the operation $\int \epsilon\dif \epsilon \dif^{3}\chi_{n+1}$ on each term has been given.

\subsection*{8. $[i_{1}\,i_{2}\,i_{3}\,i_{4}\,(i_{4}i_{5}i_{6})\cap(i_{7}i_{8})][i_{4}\,i_{5}\,i_{6}\,i_{7}\,i_{8}]$}
Using six-term identity \eqref{6ident}, this Yangian invariant is equal to
\begin{align*}
    \Bigl(&[i_{1}\,i_{2}\,i_{3}\,i_{4}\,i_{8}]-[i_{1}\,i_{2}\,i_{3}\,(i_{4}i_{5}i_{6})\cap(i_{7}i_{8})\,i_{8}] \\
    &+[i_{1}\,i_{2}\,i_{4}\,(i_{4}i_{5}i_{6})\cap(i_{7}i_{8})\,i_{8}] 
    -[i_{1}\,i_{3}\,i_{4}\,(i_{4}i_{5}i_{6})\cap(i_{7}i_{8})\,i_{8}] \\
    &+[i_{2}\,i_{3},i_{4}\,(i_{4}i_{5}i_{6})\cap(i_{7}i_{8})\,i_{8}]\Bigr) [i_{4}\,i_{5}\,i_{6}\,i_{7}\,i_{8}]\:,
\end{align*}
where the effect of the operation $\int \epsilon\dif \epsilon \dif^{3}\chi_{n+1}$ on each term has been given.

\subsection*{9. $[i_{1}\,i_{2}\,i_{3}\,i_{4}\,i_{9}][i_{5}\,i_{6}\,i_{7}\,i_{8}\,i_{9}]$}

Directly apply eq.\eqref{A1}.

\subsection*{10. $[i_{1}\,i_{2}\,i_{3}\,i_{4}\,(i_{5}i_{6}i_{7})\cap(i_{8}i_{9})][i_{5}\,i_{6}\,i_{7}\,i_{8}\,i_{9}]$}
Using six-term identity \eqref{6ident}, this Yangian invariant is equal to
\begin{align*}
    \Bigl(&[i_{1}\,i_{2}\,i_{3}\,i_{4}\,i_{9}]-[i_{1}\,i_{2}\,i_{3}\,(i_{5}i_{6}i_{7})\cap(i_{8}i_{9})\,i_{9}] \\
    &+[i_{1}\,i_{2}\,i_{4}\,(i_{5}i_{6}i_{7})\cap(i_{8}i_{9})\,i_{9}]
    -[i_{1}\,i_{3}\,i_{4}\,(i_{5}i_{6}i_{7})\cap(i_{8}i_{9})\,i_{9}] \\
    &+[i_{2}\,i_{3}\,i_{4}\,(i_{5}i_{6}i_{7})\cap(i_{8}i_{9})\,i_{9}] \Bigr)[i_{5}\,i_{6}\,i_{7}\,i_{8}\,i_{9}] \:,
\end{align*}
where the effect of the operation $\int \epsilon\dif \epsilon \dif^{3}\chi_{n+1}$ on each term has been given.

\subsection*{12. $[i_{1}\,i_{2}\,i_{3}\,i_{4}\,(i_{5}i_{6})\cap(i_{7}i_{8}i_{9})][i_{5}\,i_{6}\,i_{7}\,i_{8}\,i_{9}]$}

Using six-term identity \eqref{6ident}, this Yangian invariant is equal to
\begin{align*}
    \Bigl(&[i_{1}\,i_{2}\,i_{3}\,i_{4}\,i_{9}]-[i_{1}\,i_{2}\,i_{3}\,(i_{5}i_{6})\cap(i_{7}i_{8}i_{9})\,i_{9}] \\
    &+[i_{1}\,i_{2}\,i_{4}\,(i_{5}i_{6})\cap(i_{7}i_{8}i_{9})\,i_{9}]
    -[i_{1}\,i_{3}\,i_{4}\,(i_{5}i_{6})\cap(i_{7}i_{8}i_{9})\,i_{9}] \\
    &+[i_{2}\,i_{3}\,i_{4}\,(i_{5}i_{6})\cap(i_{7}i_{8}i_{9})\,i_{9}] \Bigr)[i_{5}\,i_{6}\,i_{7}\,i_{8}\,i_{9}]\:,
\end{align*}
where the effect of the operation $\int \epsilon\dif \epsilon \dif^{3}\chi_{n+1}$ on each term has been given.

\subsection*{13. $ \varphi [i_{1}i_{2}i_{3}\,(i_{4}i_{5})\cap(i_{7}i_{8}i_{9})\,(i_{4}i_{6})\cap(i_{7}i_{8}i_{9})]
[(i_{4}i_{5})\cap(i_{1}i_{2}i_{3})\,(i_{4}i_{6})\cap(i_{1}i_{2}i_{3})\,i_{7}i_{8}i_{9}]$}
where 
\[
    \varphi=\frac{\langle i_{4}i_{5}(i_{1}i_{2}i_{3})\cap(i_{7}i_{8}i_{9})\rangle\langle i_{4}i_{6}(i_{1}i_{2}i_{3})\cap(i_{7}i_{8}i_{9})\rangle}{\langle i_{1}i_{2}i_{3}i_{4}\rangle\langle i_{4}i_{7}i_{8}i_{9}\rangle\langle i_{5}i_{6}(i_{1}i_{2}i_{3})\cap(i_{7}i_{8}i_{9})\rangle}
\]
This Yangian invariant is equal to 
\begin{align*}
    &\quad [i_{4}i_{5}i_{6}\,(i_{7}i_{8})\cap(i_{9}i_{1}i_{3})\,(i_{7}i_{8}i_{9})\cap(i_{1}i_{3})] 
    [i_{7}i_{8}i_{9}i_{1}i_{3}] \\
    & -[i_{4}i_{5}i_{6}\,(i_{7}i_{8})\cap(i_{9}i_{1}i_{2})\,(i_{7}i_{8}i_{9})\cap(i_{1}i_{2})]
     [i_{7}i_{8}i_{9}i_{1}i_{2}] \\
    &+ [i_{1}i_{2}i_{3}\,(i_{4}i_{5}i_{6})\cap(i_{8}i_{9})\,i_{9}][i_{4}i_{5}i_{6}i_{8}i_{9}]
    +[i_{1}i_{2}i_{3}\,(i_{4}i_{6})\cap(i_{7}i_{8}i_{9})\,i_{9}][i_{4}i_{6}i_{7}i_{8}i_{9}] \\
    &-[i_{1}i_{2}i_{3}\,(i_{4}i_{5})\cap(i_{7}i_{8}i_{9})\,i_{9}][i_{4}i_{5}i_{7}i_{8}i_{9}]
    -[i_{1}i_{2}i_{3}\,(i_{4}i_{5}i_{6})\cap(i_{7}i_{9})\,i_{9}] [i_{4}i_{5}i_{6}i_{7}i_{9}] \\
    &+[i_{9}i_{1}i_{2}i_{3}\,(i_{4}i_{5}i_{6})\cap(i_{7}i_{8})][i_{4}i_{5}i_{6}i_{7}i_{8}] 
    -[i_{1}i_{2}i_{3}\,(i_{5}i_{6})\cap(i_{7}i_{8}i_{9})\,i_{9}][i_{5}i_{6}i_{7}i_{8}i_{9}] \\
    &-[i_{4}i_{5}i_{6}\,(i_{7}i_{8})\cap(i_{9}i_{2}i_{3})\,(i_{7}i_{8}i_{9})\cap(i_{2}i_{3})] [i_{7}i_{8}i_{9}i_{2}i_{3}] \:,
\end{align*}
where the effect of the operation $\int \epsilon\dif \epsilon \dif^{3}\chi_{n+1}$ on each term has been given.

\subsection*{13. $[i_{1},i_{2},i_{3},i_{4},i_{5}][i_{6},i_{7},i_{8},i_{9},i_{10}]$}
Directly apply eq.\eqref{A1}.

\section{Simple facts of field extension} \label{appb}

When the symbol involves algebraic letters, there is an important technical question: how to find a basis of (numerical) algebraic letters 
\[
	\{l_\alpha:=a_\alpha+b_\alpha\sqrt{c_\alpha}\,|\,
	a_\alpha,b_\alpha,c_\alpha \in \mathbb Q, 1\leq \alpha \leq k\}
\]
such that all algebraic letters are product of powers of letters in the basis and some rational 
numbers? It's difficult to find it directly because rational numbers are indefinite, so we 
first normalize algebraic letters to fix this uncertainty by introducing the \emph{norm} of 
a number in a field extension~\cite{Morandi1996}.

Suppose we have the square roots $\sqrt{c_1}, \sqrt{c_2}, \dots, \sqrt{c_n}$
in letters, where $\{c_i\}_{1\leq i\leq n}$ is multiplicative independent.
Consider the field $K=\mathbb Q(\sqrt{c_1},\sqrt{c_2},\dots,\sqrt{c_n})$. As a field extension of $\mathbb Q$,
$K$ is a $2^n$-dimensional $\mathbb Q$-vector field, each element
$a\in K$ defines a linear operator $L_a:K\to K$ by $L_a(b):=ab$, 
and we define the \emph{norm} $N(a)$ to be $\det(L_a)$. It's clear from the definition that 

(1). $N(ab)=N(a)N(b)$,

(2). $N(a+b\sqrt{c_i})=(a^2-b^2c_i)^{2^{n-1}}$,

(3). $N(a)=a^{2^n}$ if $a\in k$. \\
The main lemma used here is that $1$ and $-1$ are only possible rational numbers
with unit norm in $K$.

Since our irrational letters always have the form 
\[
l_\alpha=\frac{a_\alpha+b_\alpha \sqrt{c_{i_\alpha}}}{a_\alpha-b_\alpha\sqrt{c_{i_\alpha}}}
\]
with unit norm, if a product $\prod_\alpha l_\alpha^{n_\alpha}\in K$ is rational, 
it can only be $1$ or $-1$ according to the lemma.
Therefore, such a multiplicative relation is equivalent to a linear relation of 
$\log(|l_\alpha|)$ up to a overall sign,
\[
	\sum_{\alpha}n_\alpha\log(|l_\alpha|)=0,
\]
which is very easy to handle for computers, e.g. by the PSLQ algorithm \cite{Ferguson91apolynomial}.


More generally, for the original problem, one could look for multiplicative relations of 
the following numbers with unit norm instead
\[
\left\{L_\alpha :=\frac{(a_\alpha+b_\alpha \sqrt{c_{i_\alpha}})^2}{a_\alpha^2-b_\alpha^2c_{i_\alpha}}\in K
\right\},
\]
i.e. $L_\alpha$ is the normalized square of $l_\alpha$, or equivalently
\[
	l_\alpha = \pm \sqrt{a_\alpha^2-b_\alpha^2c_{i_\alpha}}\sqrt{L_\alpha}
\]
or 
\[
	\cdots \otimes l_a \otimes \cdots=\frac 12 (\cdots \otimes (a_\alpha^2-b_\alpha^2c_{i_\alpha})\otimes \cdots+ \cdots \otimes L_\alpha \otimes \cdots)
\]
in the symbol. It's easy to determine a basis of $\{L_\alpha\}_\alpha$ from 
linear relations $\sum_\alpha n_\alpha \log(|L_\alpha|)=0$.

\section{Details on case (iv)} \label{appc}
For case (iv), rational parameterizations \eqref{rationpara1} and \eqref{rationpara2} are not available, but one can easily find the following parameterization
\begin{equation}
    \tau= \frac{r(t+s)}{t(t+1)} \label{rationalboxpara}
\end{equation}
based on the rational point $\tau_{u}=0$, where
\begin{align*}
r&=\frac{\langle 1\,2\,3\,n\rangle\langle n{-}1\,(1\,2)\,(b{-}1\,b)\,(n\,n{-}2) \rangle}{\langle 2\,3\,n{-}1\,n\rangle\langle 1\,(b{-1}\,b)\,(n{-2}\,n{-1})\,(n\,2) \rangle} \:,  \\
s&=\frac{\langle 1\,2\,n{-}2\,n{-}1\rangle\langle 1\,b{-}1\,b\,n{-}1\rangle\langle n\,(1\,2)\,(b{-1}\,b)\,(n{-}2\,n{-}1) \rangle}{\langle n{-}1\,(1\,2)\,(b{-}1\,b)\,(n\,n{-}2) \rangle\langle 1\,(b{-1}\,b)\,(n{-2}\,n{-1})\,(n\,2) \rangle}  \:.
\end{align*}
Then, in terms of $t$,
\begin{align}
    &\int \dif^{2\vert 3}\mathcal{Z}_{n+1} (f_{2,b,n{-}1,n+1}^{+}+f_{2,b,n-1,n+1}^{-}) \nonumber \\
    &= \bar{Q}\log\frac{\langle\bar{n}\,2\rangle}{\langle\bar{n}\,n{-}2\rangle}\Biggl(\int_{t=0}^{t=\infty}
    \dif\log \frac{t+\bm{x}_{1}}{t+\bm{y}_{1}}\,[1,2,b{-}1,b,n{-}2]+\dif\log \frac{t+\bm{x}_{2}}{t+\bm{y}_{2}}\,[2,b{-}1,b,n{-}2,n{-}1] \nonumber \\
    &\quad +\dif\log\frac{t+\bm{x}_{b-1}}{t+\bm{y}_{b-1}} [1,2,b{-}1,n{-}2,n{-}1]-\dif\log\frac{t+\bm{x}_{b}}{t+\bm{y}_{b}} [1,2,b,n{-}2,n{-}1] \nonumber \\
    &\quad +\dif\log(t+s) [1,2,b{-}1,b,n{-}1]+\dif\log\frac{t+1}{t} [1,b{-}1,b,n{-}2,n{-}1]\Biggr) \label{rabox}
 \end{align}
 where
 \begin{align*}
     \bm{x}_{1}&=\frac{\langle 12\,n{-2}\,n \rangle\langle 1\,b{-1}\,b\,n{-}1\rangle}{\langle 1\,(b{-}1\,b)(n{-}2\,n{-}1)\,(n\,2)\rangle}\:, \\ 
     \bm{y}_{1}&=\frac{\langle 1\,2\,n{-2}\,n{-}1\rangle\langle 1\,b{-}1\,b\,n{-}2\rangle\langle n(12)(b{-}1\,b)(n{-}2\,n{-}1)\rangle}{\langle n{-}2\,(12)(b{-1}\,b)(n{-}1\,n)\rangle\langle 1\,(b{-}1\,b)(n{-}2\,n{-}1)(n\,2)\rangle} \:,\\
     \bm{x}_{2}&=\frac{\langle 12\,n{-}2\,n{-1}\rangle\langle n\,(1\,2)\,(b{-}1\,b)\,(n{-}2\,n{-}1)\rangle}{\langle 2\,n{-}2\,n{-1}\,n\rangle\langle 1\,(b{-}1\,b)\,(n{-}2\,n{-}1)\,(n\,2)\rangle} \:,\\
     \bm{y}_{2}&=\frac{\langle 1\,b{-}1\,b\,n{-}1\rangle\langle 2\,(b{-}1\,b)\,(n{-}2\,n{-}1)\,(1\,n)\rangle}{\langle 2\,b{-}1\,b\,n{-}1\rangle\langle1\,(b{-}1\,b)\,(n{-}2\,n{-}1)\,(n\,2)\rangle} \\
     \bm{x}_{b}&=\frac{\langle 12\,b\,n \rangle\langle 12\,n{-}2\,n{-1}\rangle\langle 1\,b{-}1\,b\,n{-}1\rangle}{\langle 12\,b\,n{-}1\rangle\langle 1(b{-}1\,b)(n{-}2\,n{-}1)(n\,2)\rangle}\:,  \\
     \bm{y}_{b} &=\frac{\langle 1\,b\,n{-}2\,n{-}1\rangle\langle n(1\,2)(b{-}1\,b)(n{-}2\,n{-}1)\rangle}{\langle b\,n{-2}\,n{-1}\,n\rangle\langle 1(b{-}1\,b)(n{-2}\,n{-1})(n\,2)\rangle} \:.
 \end{align*}
and $\bm{x}_{b-1},\bm{y}_{b-1}$ differ from $\bm{x}_{b},\bm{y}_{b}$ by the exchange of $b{-}1$ and $b$. The box integral under the parameterization \eqref{rationalboxpara} has the same form as eq.\eqref{boxint1}, but now with
\[
    \zeta=\frac{\frac{\langle 12\,n{-}1\,n\rangle\langle b{-}1\,b\,n{-}2\,n{-}1\rangle}{\langle 12\,n{-}2\,n{-}1\rangle\langle b{-}1\,b\,n{-}1\,n\rangle}t}{t+\frac{\langle 1\,b{-}1\,b\,n{-}1\rangle\langle n(1\,2)(b{-}1\,b)(n{-}2\,n{-}1)\rangle}{\langle b{-}1\,b\,n{-}1\,n\rangle\langle 1(b{-}1\,b)(n{-}2\,n{-}1)(n\,2)\rangle}}  \qquad
\bar{\zeta}=\frac{t+1}{t+\frac{\langle 12\,n{-}2\,n{-}1 \rangle \langle 1\,b{-}1\,b\,n\rangle}{\langle1\,(b{-}1\,b)\,(n{-}2\,n{-}1)\,(n\,2)\rangle}}\:.
\]
After subtracting the divergence of the $t$ integration properly, this integral can be performed without any obstacle.

\bibliographystyle{JHEP}
\bibliography{reference}

\end{document}